\documentclass[journal,10pt,onecolumn]{IEEEtran}
\usepackage{color}
\usepackage{bbm}
\usepackage{amsmath}
\usepackage{dcolumn}		           
\usepackage{bm}				           
\usepackage{esint}
\usepackage[normalem]{ulem}
\usepackage{xcolor}
\usepackage{amssymb}
\usepackage{overpic}
\usepackage{epic}
\usepackage{epstopdf}
\usepackage{soul}
\usepackage{textcomp}
\usepackage{float}
\usepackage{steinmetz}
\usepackage{cite}
\usepackage{graphicx}                  
\usepackage{epstopdf}
\usepackage{caption}
\usepackage{subcaption}
\usepackage[font=footnotesize]{caption}
\usepackage[font=footnotesize]{subcaption}
\usepackage{amsfonts}
\usepackage{array}

\usepackage{multicol}

\newcommand\redout{\bgroup\markoverwith{\textcolor{red}{\rule[.5ex]{2pt}{0.4pt}}}\ULon}

\newcommand{\Nmax}{N_{\rm max}}
\newcommand{\Dmax}{D_{\rm max}}
\newcommand{\ftbr}{\mathrm{FTBR}}

\begin{document}
\title{\textcolor{red}{High-Directivity Pulsed Beams for \\ Effective Power and Information Transfer}}

\author{\bf{Richard~W.~Ziolkowski} \\
\textnormal{Department of Electrical and Computer Engineering, The University of Arizona, Tucson, AZ 85721, USA \\
ziolkows@arizona.edu \\
\textcolor{blue}{2026-11-10}}
}

\maketitle

\date{\today}


\begin{abstract}

A vector spherical wave expansion (VSWE) is used to engineer an elemental multipole source whose
radiated field is a superdirective, extremely unidirectional (SDEU) needle beam.
With the expansion coefficients chosen to mimic the spherical-harmonic closure relation
truncated at the maximum degree $\Nmax$, it is proved analytically, and confirmed numerically
to machine precision in mirrored Python and MATLAB codes, that the element attains
Harrington's normal-directivity value $\Dmax = \Nmax^2 + 2\Nmax$ along its beam axis.
Its radiated field is linearly polarized with an azimuthally symmetric magnitude; its
back lobe vanishes identically in the far field.
The unidirectional behavior is quantified at any finite far-field radius by the closed-form
front-to-back ratio $\ftbr|_{\rm field} = 4(k_0r)^2/(\Nmax+1)$, obtained from an exact
per-degree cancellation of the electric and magnetic multipole contributions in the backward
direction. It exceeds $97$~dB already at $r = 10^2\lambda$ and rises $12$~dB per doubling of range.
Uniform circular arrays (UCAs) of these elements, from $0.1\lambda$ to $6\lambda$ in radius,
inherit the SDEU behavior. The electrically small UCA is superdirective by $24.9$~dB relative
to its aperture bound.
Driving the arrays with zero-area pulses (ZAPs) transfers the frequency-domain performance
intact into the time domain, the far-field signal being the time derivative of the excitation.
It is shown analytically and numerically that whether the radiated energy beam is narrower or wider
than the corresponding intensity beam is decided by a single property of the radiated waveform, the ratio
$\langle\omega^2\rangle/\omega_{\rm peak}^2$ of its mean-square bandwidth to the curvature of
its own peak, and not by the size of the radiating aperture. The ratio is exactly $65/54$ for
the single-cycle ZAP used to quantify the performance of the arrays; its energy beam is the narrower one.
The same criterion is shown to govern the focal spot of a near-field-focused pulse-driven array.
A custom-designed, $C^2$-continuous intensity-enhancement ZAP (IE-ZAP) is introduced that
raises the focal intensity by a factor of $193.9$ over a narrow-band reference while producing
a focal spot more than ten times smaller in both the axial and transverse directions, deeply
subwavelength with respect to its own effective wavelength. It is demonstrated that these stunning
IE results are produced simply with $\Nmax$ = 1 elements, i.e., with an array of pulse-excited
Huygens dipole sources.
These results identify the pulse shape, rather than the aperture, as the design variable that
governs localized electromagnetic energy and information delivery. They point to
distributed, independently addressable pulse-driven arrays as a practical route to it.
\end{abstract}


\section{\textcolor{red}{Introduction}}

Highly directive beams have many practical applied electromagnetics applications.
As we enter the age of Integrated Space-Air-Terrestrial Networks (ISATNs), the need for such needle beams
has greatly increased with current and anticipated future wireless performance expectations.
At higher frequencies they can mitigate the effects of low power sources dealing with large propagation losses
when the scenario warrants high EIRP (effective isotropic radiated power).
They can provide secure information (energy) transfer, i.e., low probability of intercept - LPI - signals.
They will facilitate the realization of the many anticipated and desirable Future-G communications, imaging,
sensing, and wireless power transfer (WPT) applications.

Needle beams are considered in this article from a vector spherical wave expansion (VSWE) point of view.
The VSWE analysis will follow the multipole paradigm emphasized in \cite{ziolkowski_prx_2017, ziolkowski_ojap2022, ziolkowski_UMMA_2023}. A fundamental multipole needle-beam source is developed analytically and then
evaluated numerically. It is demonstrated that the fields radiated by it achieve Harrington's maximum directivity
multipole results from his papers \cite{harrington1958gain,harrington1960effect,harrington1965antenna}.
In particular, it will be shown that superdirective, extremely unidirectional (SDEU) needle-beam performance
is achieved to any maximum multipole order by them and with arrays of them.
These SDEU systems have the potential to facilitate the many aforementioned applications that
require intimate control of electromagnetic power and information transfer. They could fulfill many
current desires associated with space-constrained mobile systems requiring high directivity.
Moreover, they would provide extreme spatial efficiency, lowering interference levels,
as well as facilitating enhanced SNR (signal-to-noise ratio) and information throughput performance.

While the initial discussions will consider the needle beams in the frequency domain, pulsed needle beams will then be demonstrated from a time-domain view point. It will be shown that the superdirective performance
obtained in the frequency domain is maintained by these time-domain needle beams in the far-field of the sources and arrays
that radiate them. In addition to those far-field results, focusing within the far-field radius of the pulse-driven arrays  can achieve extremely high intensities within very localized regions with properly designed zero-area pulse (ZAP) excitations. These superfocusing ZAP results are attractive, e.g., for a variety of medical and ground-penetrating radar (GPR) applications, where enhanced spatial focusing is advantageous to high-precision positioning and imaging.

The frequency-domain needle-beam multipole elements are introduced and developed in Secs. II and III.
The VSWE approach employed is briefly discussed and then superdirective multipole element engineering is described
in detail.
The superdirective and extreme unidirectional characteristics of the needle-beam elements are developed analytically.
It is proved that they achieve Harrington's upper bound for any choice of the maximum multipole selected to be emitted
from them. A proof of their extreme unidirectional features is also obtained.
These analytical details are confirmed numerically in Sec. III, where the directivity
patterns, beamwidths, and front-to-back ratios of the elements are tabulated.
These SDEU elements are then introduced into broadside-radiating uniform circular arrays (UCAs) in Sec. IV.
The performance characteristics of the fields emitted from several different sized ( from electrically small to large ) arrays of them are characterized. Amplitude tapering of these arrays is considered.
The performance of these needle-beam arrays is detailed and confirms their SDEU properties.
The time-domain versions of the SDEU needle-beam elements and arrays are discussed in Sec. V.
The properties of the ZAPs used to excite them are described. The time-derivative properties associated with
fields propagating away from pulse-excited finite-sized sources are emphasized.
The performance characteristics of the SDEU pulsed needle-beams emitted by the ZAP-excited elements
and by their UCAs in the presence of those derivatives are detailed.
The issue of whether the intensity or energy profiles of the pulsed beams propagating away from the array
that emitted them are more narrow or wider is considered and demonstrated analytically and numerically
to be not a result of the array size, but of the shape characteristics of the ZAPs exciting them.
Near-field focusing of the fields radiated by a ZAP-excited planar UCA of SDEU elements is also addressed.
A custom-designed ZAP is detailed that was developed to achieve significant enhancement of the intensity attained
at a prescribed focal spot centered on the axis of the array. Comparisons of different ZAP-based results
are presented. It was determined that like the far-field intensity-energy profiles of the pulsed needle beams,
the size of the intensity versus the energy spot at the focus again only depends on the shape of the
ZAP exciting the array.
Conclusions are discussed in Sec. VI.
Several appendices provide additional details of the analytical results.
The numerical results were obtained with both Python and mirrored MATLAB codes and checked extensively with a
variety of approaches.

\section{\textcolor{red}{VSWE Representation of Electromagnetic Fields}}
\vskip 0.1in

There are many variations of representing a vector spherical wave expansion (VSWE) of electromagnetic fields, e.g.,
\cite{harrington_1961_book, balanis_2012_theory, b-a-z-2025superdirective}.
The expressions employed herein simply avoid the additional material constants that appear when the VSWEs are
written to explicitly track the TE and TM modes during any analysis/derivation as is found in many references.
The consequences of my choice are simpler expressions, but with the cost of at least the noted care needed to make sure the
final results have the proper physical units, particularly when a physical problem is being considered.
My choice of VSWE expressions, written to emphasize that the standard infinite degree sum over $n$ will be
restricted to a choice of $N_{\rm max}$, is:

\begin{IEEEeqnarray}{rCl}\label{eq:full-VSWE-Electric-Field}
{\vec E}(r, \theta, \phi) &=& \sum_{n=1}^{N_{\rm max}} \, \sum_{m=-n}^{+n} \>
\left\{\, a_{\rm nm} \, {\vec N}_{nm}({\vec r}) + b_{\rm nm} \, {\vec M}_{nm}({\vec r}) \,\right\}
\hfill \label{eq:VSWE-E-field} \\
\noalign{\vskip 0.1in}
{\vec H}(r, \theta, \phi) &=& j \> \frac{1}{\eta_0} \> \sum_{n=1}^{N_{\rm max}} \, \sum_{m=-n}^{+n} \>
\left\{\, b_{\rm nm} \, {\vec N}_{nm}({\vec r}) + a_{\rm nm} \, {\vec M}_{nm}({\vec r}) \,\right\}
\hfill \label{eq:VSWE-H-field}
\end{IEEEeqnarray}
\vskip 0.1in

\noindent where the unit-less vector functions are

\begin{IEEEeqnarray}{rCl}
{\vec M}_{nm} &=&  w_n (kr) \> e^{jm\phi} \>
                        \left[\, \frac{jm}{\sin\theta} \, P_n^{|m|} (\cos\theta) \> {\hat \theta}
                                   - \frac{d}{d\theta} \, P_n^{|m|} (\cos\theta) \> {\hat \phi}  \,\right]
\label{eq:M-vector} \\
\noalign{\vskip 0.1in}
{\vec N}_{nm} &=& \frac{n(n+1)}{kr}  \> w_n (kr) \> P_n^{|m|} (\cos\theta) \>  e^{jm\phi} \> {\hat r}       \nonumber \\
                   &+& e^{jm\phi} \> \frac{1}{kr} \> \frac{d}{d(kr)} \, \left[\, (kr) \> w_n(kr) \,\right] \>
                      \left[\, \frac{d}{d\theta} \, P_n^{|m|} (\cos\theta) \> {\hat \theta}
                         + \frac{jm}{\sin\theta} \, P_n^{|m|} (\cos\theta) \> {\hat \phi} \,\right]
\label{eq:N-vector}
\end{IEEEeqnarray}
\vskip 0.1in

\noindent where the $w_n(kr)$ are spherical Bessel functions appropriate for the configuration.

Both MATLAB and mirrored Python programs were developed based on these expressions. Their efficacy was confirmed with comparisons with the fields radiated by the elemental electric (ED) and magnetic (MD) dipoles reported in \cite{b-a-z-2025superdirective}. Those studies adopted a version that emphasized the TE and TM modes
since those sources were used to excite both electric and magnetic multipoles in a dielectric spherical lens.

I consider all of the fields investigated herein to have been radiated into free space, i.e.,
with the free-space permittivity
$\varepsilon_0 = 8.8541878188 \times 10^{-12}$ Farads/meter (F/m) and
the free-space permeability $\mu_0 = 4\pi \times 10^{-7}$ Henries/meter (H/m), which give the speed of light
in free-space: $c = 1/\sqrt{\varepsilon_0 \, \mu_0}$; the free-space wavenumber:
$k_0 = \omega \, \sqrt{\varepsilon_0 \, \mu_0}$, where $\omega$ is the angular frequency
$\omega = 2\pi \, f$, $f$ being the frequency of the source(s) that creates the fields; and the
free-space wave impedance: $\eta_0 = \sqrt{\mu_0/\varepsilon_0}$.

The far-field $\theta$ and $\phi$ components of the electric field in free space written in a form that any
overall physical units are associated strictly with the VSWE expansion coefficients are confirmed to be:

\begin{IEEEeqnarray}{rCl}\label{eq:E-theta-ff}
E_{\theta}^{\rm tot, ff}(r, \theta, \phi) &=& \frac{e^{-jk_0r}}{k_0r} \> \>
\sum_{n=1}^{N_{\rm max}} \, \sum_{m=-n}^{+n} \> j^n \> e^{jm\phi} \>\>
\Bigg\{\,
a_{\rm nm} \, \left[\, \frac{d}{d\theta} \, P_n^{|m|} (\cos\theta) \,\right]
-
b_{\rm nm} \>
\left[\, \frac{m}{\sin\theta} \, P_n^{|m|} (\cos\theta) \,\right]
 \,\Bigg\}                                                                                 \hfill
\end{IEEEeqnarray}
\vskip 0.1in

\begin{IEEEeqnarray}{rLc}\label{eq:E-phi-ff}
E_{\phi}^{\rm tot, ff}(r, \theta, \phi)
&=& \frac{e^{-jk_0r}}{k_0r} \> \sum_{n=1}^{N_{\rm max}} \, \sum_{m=-n}^{+n} \> j^{n+1} \> e^{jm\phi}
\> \Bigg\{\, a_{\rm nm} \, \left[\, \frac{m}{\sin\theta} \, P_n^{|m|} (\cos\theta) \,\right]
-
b_{\rm nm} \> \left[\, \frac{d}{d\theta} \, P_n^{|m|} (\cos\theta) \,\right] \,\Bigg\}        \hfill
\end{IEEEeqnarray}
\vskip 0.1in

\noindent The total radiated power is expressed as:

\begin{IEEEeqnarray}{rCl}\label{eq:P-total}
P_{\rm tot} &=& \oiint_{S_\infty} \frac{1}{2\eta_0} \>
\left[\, |E_{\theta}^{\rm tot, ff}(r, \theta, \phi)|^2 + |E_{\phi}^{\rm tot, ff}(r, \theta, \phi)|^2 \,\right] \>
\> r^2 \sin\theta d\theta d\phi  \hfill \nonumber \\
\noalign{\vskip 0.1in}
&=& P_{\theta} + P_{\phi} = \frac{1}{\eta_0} \> \frac{2\pi}{k_0^2} \>\>
\sum_{n=1}^{N_{\rm max}} \, \sum_{m=-n}^{+n} \> \frac{n(n+1)}{2n+1} \> \frac{(n+|m|)!}{(n-|m|)!}
\> \left[\, |a_{\rm nm}|^2 + |b_{\rm nm}|^2 \,\right]
\end{IEEEeqnarray}
\vskip 0.1in

\noindent Consequently, the directivity
\begin{IEEEeqnarray}{c}\label{eq:Directivity}
D(\theta,\phi) = \lim_{r \to \infty} \> \left\{\> r^2 \>\>  \frac{\frac{1}{2\eta_0} \> \left[\,
\Big|\,  E_{\theta}^{\rm ff}(\theta,\phi) \,\Big|^2
+
\Big|\,  E_{\phi}^{\rm ff}(\theta,\phi) \,\Big|^2
\,\right]}{P_{\rm tot}/(4\pi)} \>\right\}
\end{IEEEeqnarray}
\vskip 0.1in

\noindent follows from (\ref{eq:E-theta-ff}), (\ref{eq:E-phi-ff}), and (\ref{eq:P-total}).

\subsection{\textcolor{blue}{Superdirective Multipole Element Engineering}}
\vskip 0.1in

Trying to mimic the SWE closure relation emphasized in \cite{ziolkowski_prx_2017} and given in Appendix C,
Eq. (\ref{eq:SWE-closure-relation}), I introduce the scalar coefficients:

\begin{IEEEeqnarray}{c}
\Lambda_{nm} = \frac{2n+1}{n(n+1)} \> \frac{(n-|m|)!}{(n+|m|)!}
\end{IEEEeqnarray}
\vskip 0.1in

\noindent Then, I introduce the coefficients $\alpha_{nm}$ and $\beta_{nm}$ as
\begin{IEEEeqnarray}{rCl}
a_{nm} &=& j^{-n} \> \Lambda_{nm} \>\> \alpha_{nm} \\
\noalign{\vskip 0.1in}
b_{nm} &=& - j^{-n} \> \Lambda_{nm} \>\> \beta_{nm}
\end{IEEEeqnarray}
\vskip 0.1in

\noindent which will yield a finite total power unless the new coefficients $\alpha_{nm}$ and $\beta_{nm}$ have unbounded factorials in them:

\begin{IEEEeqnarray}{rCl}
P_{\rm tot} &=& \frac{2\pi}{\eta_0} \> \frac{1}{k_0^2} \>\>
\sum_{n=1}^{N_{\rm max}} \, \sum_{m=-n}^{+n} \> \frac{2n+1}{n(n+1)} \> \frac{(n-|m|)!}{(n+|m|)!}
\> \left[\, |\alpha_{\rm nm}|^2 + |\beta_{\rm nm}|^2 \,\right]
\end{IEEEeqnarray}
\vskip 0.1in

\noindent Going further, I introduce the additional coefficients $\tau_{nm}$ and $\chi_{nm}$
\begin{IEEEeqnarray}{rCl}
\alpha_{nm} &=& \frac{1}{4\pi} \> e^{-jm \phi_s} \> \left[\, \frac{d}{d\theta} \, P_n^{|m|} (\cos\theta_s) \,\right]
\>\> \tau_{nm}                                                                 \\
\noalign{\vskip 0.1in}
\beta_{nm} &=& \frac{1}{4\pi} \> e^{-jm \phi_s} \> \left[\, \frac{m}{\sin\theta_s} \, P_n^{|m|} (\cos\theta_s) \,\right]
\>\> \chi_{nm}
\end{IEEEeqnarray}
\vskip 0.1in

\noindent Then the electric field components can be written in a form that is symmetric with respect to the source and
observation coordinate angles as

\begin{IEEEeqnarray}{rCl}\label{eq:E-theta-ff-needle}
E_{\theta}^{\rm tot, ff}(r, \theta, \phi) &=& \frac{e^{-jk_0r}}{4\pi \, k_0r} \> \>
\sum_{n=1}^{N_{\rm max}} \, \sum_{m=-n}^{+n} \> e^{jm (\phi-\phi_s)} \Lambda_{nm} \>
\Biggl\{\,  \tau_{nm} \>
\left[\, \frac{d}{d\theta} \, P_n^{|m|} (\cos\theta_s) \>\> \frac{d}{d\theta} \, P_n^{|m|} (\cos\theta) \,\right]
                         \hfill \nonumber \\
\noalign{\vskip 0.1in}
&& \qquad + \quad \chi_{nm} \> \left[\, \frac{m}{\sin\theta_s} \, P_n^{|m|} (\cos\theta_s) \,\right] \>
\left[\, \frac{m}{\sin\theta} \, P_n^{|m|} (\cos\theta) \,\right]  \,\Bigg\}                  \hfill
\end{IEEEeqnarray}
\vskip 0.1in

\begin{IEEEeqnarray}{rLc}\label{eq:E-phi-ff-needle}
E_{\phi}^{\rm tot, ff}(r, \theta, \phi) &=& j \> \frac{e^{-jk_0r}}{4\pi \, k_0r} \>
\sum_{n=1}^{N_{\rm max}} \, \sum_{m=-n}^{+n} \> e^{jm(\phi-\phi_s)} \Lambda_{nm} \>
\> \Bigg\{\, \tau_{\rm nm} \, \left[\, \frac{d}{d\theta} \, P_n^{|m|} (\cos\theta_s) \,\right]
\left[\, \frac{m}{\sin\theta} \, P_n^{|m|} (\cos\theta) \,\right] \>                      \hfill \nonumber \\
\noalign{\vskip 0.1in}
&& \qquad + \quad \chi_{nm} \> \left[\, \frac{m}{\sin\theta_s} \, P_n^{|m|} (\cos\theta_s) \,\right] \>
\left[\, \frac{d}{d\theta} \, P_n^{|m|} (\cos\theta) \,\right] \,\Bigg\}                      \hfill
\end{IEEEeqnarray}
\vskip 0.1in

First, let us determine what happens if I assume the sources of the fields are at the origin. We then
have $\theta_s = 0$ and $\phi_s = 0$. We know that with the choice of the Condon-Shortley phase for the
associated Legendre functions:

\begin{IEEEeqnarray}{rLc}
\left[\, \frac{d}{d\theta} \> P_n^{|m|}(\cos\theta) \,\right]_{\theta = 0} &=& - \delta_{|m|1} \>\> \frac{n(n+1)}{2} \hfill \label{eq:endpoint_zero_dP} \\
\noalign{\vskip 0.15in}
\left[\, \frac{P_n^{|m|}(\cos\theta)}{\sin\theta} \,\right]_{\theta = 0} &=& - \delta_{|m|1} \>\> \frac{n(n+1)}{2}   \hfill
\label{eq:endpoint_zero_Poversin}
\end{IEEEeqnarray}
\vskip 0.05in
\noindent Thus, $E_{\theta}$ becomes

\begin{IEEEeqnarray}{rCl}  \label{eq:E-theta-ff-source-evals}
E_{\theta}^{\rm tot, ff}(r, \theta, \phi) &=& \frac{e^{-jk_0r}}{4\pi \, k_0r} \> \>
\sum_{n=1}^{N_{\rm max}} \, \sum_{m=-n}^{+n} \> e^{jm \phi} \Lambda_{nm} \>
\Biggl\{\, - \tau_{nm} \> \delta_{|m|1} \> \frac{n(n+1)}{2} \>\>
\left[\, \frac{d}{d\theta} \, P_n^{|m|} (\cos\theta) \,\right]                                       \hfill \nonumber \\
\noalign{\vskip 0.1in}
&& \qquad - \quad \chi_{nm} \> \delta_{|m|1} \> \frac{n(n+1)}{2} \>\>
\left[\, \frac{m^2}{\sin\theta} \, P_n^{|m|} (\cos\theta) \,\right] \,\Bigg\}                         \hfill  \nonumber \\
\noalign{\vskip 0.1in}
&=& - 2 \> \cos\phi \> \frac{e^{-jk_0r}}{4\pi \, k_0r} \> \>
\sum_{n=1}^{N_{\rm max}} \> \Lambda_{n1} \> \frac{n(n+1)}{2}
\left\{\,  \tau_{n1} \>\> \left[\, \frac{d}{d\theta} \, P_n^{|1|} (\cos\theta) \,\right]
          + \chi_{n1} \> \left[\, \frac{P_n^{|1|} (\cos\theta)}{\sin\theta} \,\right] \,\right\}         \hfill
\end{IEEEeqnarray}
\vskip 0.1in

\noindent Furthermore, let $\tau_{nm} = \chi_{nm} = E_0$, a constant amplitude.
Therefore, we then have along the +z-axis:

\begin{IEEEeqnarray}{rCl}
E_{\theta}^{\rm tot, ff}(r, \theta=0, \phi=0) &=& E_0 \> \frac{e^{-jk_0r}}{4\pi \, k_0r} \> \>
\sum_{n=1}^{N_{\rm max}} \, \sum_{m=-n}^{+n} \> \Lambda_{nm} \>
\Biggl\{\, \delta_{|m|1} \> \left[\, \frac{n(n+1)}{2} \,\right]^2
+ \> \delta_{|m|1} \> \left[\, \frac{n(n+1)}{2} \,\right]^2 \,\Bigg\}                                   \hfill \nonumber \\
\noalign{\vskip 0.1in}
&=& 4 \, E_0 \> \frac{e^{-jk_0r}}{4\pi \, k_0r} \> \>
\sum_{n=1}^{N_{\rm max}} \,  \frac{2n+1}{n(n+1)} \> \frac{(n-|1|)!}{(n+|1|)!} \>  \left[\, \frac{n(n+1)}{2} \,\right]^2
                                                                                                      \hfill \nonumber \\
\noalign{\vskip 0.1in}
&=& 4 \, E_0 \> \frac{e^{-jk_0r}}{4\pi \, k_0r} \> \> \sum_{n=1}^{N_{\rm max}} \,
\frac{2n+1}{n(n+1)} \> \left[\, \frac{1}{n(n+1)} \,\right] \>  \left[\, \frac{n(n+1)}{2} \,\right]^2  \hfill \nonumber \\
\noalign{\vskip 0.1in}
&=& E_0 \> \frac{e^{-jk_0r}}{4\pi \, k_0r} \> \> \sum_{n=1}^{N_{\rm max}} \> (2n+1)         \hfill
\end{IEEEeqnarray}
\vskip 0.1in

\noindent The degree sum follows from a well-known one:

\begin{IEEEeqnarray}{c}
\sum_{n=1}^{N_{\rm max}} \> (2n+1) = 2 \> \sum_{n=1}^{N_{\rm max}} \> \left(\, n + \frac{1}{2} \,\right)
= 2 \> \left[\, \frac{N_{\rm max} ( N_{\rm max} + 1)}{2} + \frac{N_{\rm max}}{2} \,\right]
= N_{\rm max}^2 + 2 \, N_{\rm max}
\end{IEEEeqnarray}
\vskip 0.05in
\noindent Thus,
\begin{IEEEeqnarray}{c} \label{eq:E-theta-ff-on-axis}
E_{\theta}^{\rm tot, ff}(r, \theta=0, \phi=0) = E_0 \> \frac{e^{-jk_0r}}{4\pi \, k_0r} \>
\> \left(\,  N_{\rm max}^2 + 2 \, N_{\rm max} \,\right)
\end{IEEEeqnarray}
\vskip 0.1in

\noindent The corresponding $E_{\phi}$ component expressions follow immediately as:

\begin{IEEEeqnarray}{rLc} \label{eq:E-phi-ff-source-eval}
E_{\phi}^{\rm tot, ff}(r, \theta, \phi) &=& j \> \frac{e^{-jk_0r}}{4\pi \, k_0r} \>
\sum_{n=1}^{N_{\rm max}} \, \sum_{m=-n}^{+n} \> e^{jm \phi} \> \Lambda_{nm} \>
\> \Bigg\{\, - \tau_{\rm nm} \, \delta_{|m|1} \> \frac{n(n+1)}{2} \>
\left[\, \frac{m}{\sin\theta} \, P_n^{|m|} (\cos\theta) \,\right] \>                      \hfill \nonumber \\
\noalign{\vskip 0.1in}
&& \qquad - \quad \chi_{nm} \> m \> \delta_{|m|1} \> \frac{n(n+1)}{2} \>
\left[\, \frac{d}{d\theta} \, P_n^{|m|} (\cos\theta) \,\right] \,\Bigg\}                  \hfill \nonumber \\
\noalign{\vskip 0.1in}
&=& + 2 \sin\phi \>\>  \frac{e^{-jk_0r}}{4\pi \, k_0r} \> \sum_{n=1}^{N_{\rm max}} \> \Lambda_{n1} \> \frac{n(n+1)}{2} \>
\> \Bigg\{\, \tau_{\rm n1} \> \left[\, \frac{P_n^{|1|} (\cos\theta)}{\sin\theta} \,\right]
+ \chi_{n1} \> \left[\, \frac{d}{d\theta} \, P_n^{|1|} (\cos\theta) \,\right] \,\Bigg\}    \hfill
\end{IEEEeqnarray}
\vskip 0.1in

\noindent and, hence,
\begin{IEEEeqnarray}{rLc} \label{eq:E-phi-ff-on-axis}
E_{\phi}^{\rm tot, ff}(r, \theta = 0, \phi = 0) &=& 0                                       \hfill
\end{IEEEeqnarray}
\vskip 0.1in

\noindent Consequently, the total radiated field along the +z-axis is linearly polarized (LP). Moreover,
we also recognize that with $\tau_{nm} = \chi_{nm} = E_0$ the total far-field electric field
${\vec E}_{\rm tot}^{\rm ff}(r, \theta, \phi)$ is proportional to
$-\cos\phi \, {\hat \theta} + \sin\phi \, {\hat \phi}$
and, hence, it is linearly polarized everywhere and furthermore, its magnitude
$\left|\, {\vec E}_{\rm tot}^{\rm ff}(r, \theta, \phi) \,\right|^2$ is azimuthally symmetric.

The total power radiated (\ref{eq:P-total}) is readily evaluated by inserting the various coefficients
that I introduced into it.
Since $\theta_s = \phi_s = 0$, the magnitude of the original VSWE coefficients are:

\begin{IEEEeqnarray}{rCl}
|a_{nm}| &=& \Bigl|\, j^{-n} \> \frac{E_0}{4\pi} \> \Lambda_{nm} \> \left[\, \frac{d}{d\theta} \,
P_n^{|m|} (\cos\theta_s) \,\right]_{\theta_s=0} \> e^{-jm \phi_s} \,\Bigr|_{\phi_s = 0}               \hfill \nonumber \\
\noalign{\vskip 0.1in}
&=& \frac{|E_0|}{4\pi} \> \frac{2n+1}{n(n+1)} \> \left[\, \frac{1}{n(n+1)}  \,\right] \> \delta_{|m|1} \> \frac{n(n+1)}{2}
= \frac{|E_0|}{8\pi} \> \delta_{|m|1} \> \frac{2n+1}{n(n+1)}
\end{IEEEeqnarray}
\vskip 0.05in
\begin{IEEEeqnarray}{rCl}
|b_{nm}| &=& \Bigl|\, - j^{-n} \> \frac{E_0}{4\pi} \> \Lambda_{nm} \>
\left[\, m \frac{P_n^{|m|} (\cos\theta_s)}{\sin\theta_s} \,\right]_{\theta_s=0} \> e^{-jm \phi_s} \,\Bigr|_{\phi_s = 0}   \hfill \nonumber \\
\noalign{\vskip 0.1in}
&=&  \frac{|E_0|}{4\pi} \> \frac{2n+1}{n(n+1)} \> \left[\, \frac{1}{n(n+1)} \,\right] \> m \> \> \delta_{|m|1} \frac{n(n+1)}{2}
= \frac{|E_0|}{8\pi} \> \delta_{|m|1} \> \frac{2n+1}{n(n+1)}
\end{IEEEeqnarray}
\vskip 0.05in
Then the total radiated power is
\begin{IEEEeqnarray}{rCl}
P_{\rm tot} &=& \frac{1}{\eta_0} \> \frac{2\pi}{k_0^2} \>\>
\sum_{n=1}^{N_{\rm max}} \, \sum_{m=-n}^{+n} \> \delta_{|m|1} \> \frac{n(n+1)}{2n+1} \> \frac{(n+|m|)!}{(n-|m|)!}
\left\{\,
2 \> \Biggl|\, \frac{E_0}{8\pi} \> \frac{2n+1}{n(n+1)} \,\Biggr|^2  \,\right\}  \hfill \nonumber \\
\noalign{\vskip 0.1in}
&=& \frac{1}{\eta_0} \>\> \frac{|E_0|^2}{8\pi} \> \frac{1}{k_0^2} \>\> \sum_{n=1}^{N_{\rm max}} \> (2n+1)
= \frac{1}{2\eta_0} \> \frac{|E_0|^2}{4\pi} \> \frac{1}{k_0^2} \>\>  \left(\, N_{\rm max}^2 + 2 \, N_{\rm max} \,\right)
\hfill
\end{IEEEeqnarray}
\vskip 0.1in

\noindent Finally, the directivity (\ref{eq:Directivity}) along the +z-axis is
\begin{IEEEeqnarray}{rLc}
D(\theta &=& 0,\phi = 0) = 4\pi \> \frac{\frac{1}{2\eta_0} \> \left[\, |E_0|^2 \> \left(\, \frac{1}{4\pi} \,\right)^2
\> \frac{1}{k_0^2} \> \left(\, N_{\rm max}^2 + 2 \, N_{\rm max} \,\right)^2 \,\right]}{\frac{1}{2\eta_0} \> \frac{|E_0|^2}{4\pi} \>
\frac{1}{k_0^2} \>\> \left(\, N_{\rm max}^2 + 2 \, N_{\rm max} \,\right)} \hfill \nonumber \\
\noalign{\vskip 0.1in}
&=& 4\pi \cdot \frac{1}{4 \pi} \> \left(\, N_{\rm max}^2 + 2 \, N_{\rm max} \,\right)
= \left(\, N_{\rm max}^2 + 2 \, N_{\rm max} \,\right)                   \hfill
\end{IEEEeqnarray}
\vskip 0.1in

\noindent Consequently, in agreement with Harrington, we find the maximum directivity is $ N_{\rm max}^2 + 2 \, N_{\rm max}$
and increases as we include more multipoles into the radiated field, i.e., as $N_{\rm max}$ increases.

Now consider the results along the -z-axis, i.e., with $\theta = \pi$.
Recall again that with the Condon-Shortley phase choice:

\begin{IEEEeqnarray}{rLc}
\left[\, \frac{d}{d\theta} \, P_n^{|m|} (\cos\theta) \,\right]_{\theta = \pi}
= (-1)^{n+1} \> \frac{n(n+1)}{2} \> \delta_{|m|1}                                                                \hfill \\
\noalign{\vskip 0.15in}
\left[\, \frac{P_n^{|m|} (\cos\theta)}{\sin\theta} \,\right]_{\theta = \pi} =
- \left[\, \frac{d}{d\theta} \, P_n^{|m|}(\cos\theta) \,\right]_{\theta = \pi}
= - (-1)^{n+1} \> \frac{n(n+1)}{2} \> \delta_{|m|1}                          \hfill
\end{IEEEeqnarray}
\vskip 0.1in

\noindent Therefore, with the stipulated source location, the far-field expression
(\ref{eq:E-theta-ff-needle}) yields

\begin{IEEEeqnarray}{c}
E_{\theta}^{\rm tot, ff}(r, \theta=\pi, \phi) = -2 \> \cos\phi \> E_0 \> \frac{e^{-jk_0r}}{4\pi \, k_0r} \> \>
\sum_{n=1}^{N_{\rm max}} \> \Lambda_{n1} \> (-1)^{n+1} \> \left[\, \frac{n(n+1)}{2} \,\right]
\> \Biggl\{\, \left[\, \frac{n(n+1)}{2} \,\right] - \left[\, \frac{n(n+1)}{2} \,\right] \,\Bigg\} = 0       \hfill
\end{IEEEeqnarray}
\vskip 0.1in

\noindent and (\ref{eq:E-phi-ff-needle}) via (\ref{eq:E-phi-ff-source-eval}) trivially yields

\begin{IEEEeqnarray}{c}
E_{\phi}^{\rm tot, ff}(r, \theta=\pi, \phi) = 0                                                           \hfill
\end{IEEEeqnarray}
\vskip 0.1in

\noindent Consequently, $D(\theta = \pi, \phi) = 0$ for all $\phi$ in the $r \to \infty$ limit.
The front-to-back-ratio: FTBR = $\infty$.
The radiated field is extremely unidirectional, as well as superdirective.
The corresponding statement at a finite measurement radius, where the cancellation is no
longer exact, is developed below and in Appendix A.

\subsection{\textcolor{blue}{Discussion}}
\vskip 0.1in

Note that the presence of the ``$k_0$'' in the denominator of the Green's function term in many of my expressions make
them dimensionless except for whatever the physical units of the amplitude $E_0$ are. Consequently, the electric field components have the correct ``units'' if $E_0$ is an electric field amplitude, which means that it has the
units ``$V/m$''.
We recall that expressions of ideal ED and MD fields in general references simply have the
spherical wave (Green's function) present without any ``$k_0$'' in the denominator,
i.e., only the term: $\exp(-jk_0r)/(4\pi r)$.
One finds, for example, from those ED field expressions that
$E_{0, {\rm ED}} = jk_0 \, \eta_0 \> I_e \, \ell_e$, where $I_e \, \ell_e$ is the
current moment of the ED.
It has ``volts'' ($V$) as its physical units, while the Green's function has the units ``$1/m$'' originating
from its $1/r$ term.
The electric field radiated by the corresponding balanced MD field arises from the constant
$E_{0, {\rm MD}} = jk_0 \> I_m \, \ell_m$, where $I_m \, \ell_m = \eta_0 \> I_e \, \ell_e$.
Thus, setting $E_0 = k_0 \> E_{0, {\rm ED}}$ in the expressions herein would yield the correct units V/m for
the electric field.

Also note that the coefficients $a_{nm}$ and $b_{nm}$ correspond to the TM and TE modes, respectively.
This fact would be apparent in other VSWE formats found in the literature and explains why I obtain the maximum directivity:
$\left(\, N_{\rm max}^2 + 2 \, N_{\rm max} \,\right)$,
i.e., both the electric and magnetic multipoles are taken into account on an equal footing in my analysis.
An ED (MD) field is associated with the $a_{nm}$ coefficients and the TM modes ($b_{nm}$ coefficients and the TE modes).
Then, one obtains the maximum directivity $\left(\, N_{\rm max}^2 + 2 \, N_{\rm max} \,\right)/ 2$
for either of those restricted sets of modes.

\subsection{\textcolor{blue}{Additional Analytical Considerations}}
\vskip 0.1in

\noindent Writing $\aleph = k_0 r$ and taking $w_n (\aleph) = h_n^{(2)} (\aleph)$ as the $e^{+j\omega t}$
convention requires, the two radial-distance based functions that appear
in (\ref{eq:M-vector}) and (\ref{eq:N-vector}) can be identified as:
\begin{equation}\label{eq:AB}
A_n(\aleph) \;=\; \frac{1}{\aleph}\frac{d}{d\aleph}\!\left[\,\aleph\,h_n^{(2)}(\aleph)\,\right],
\qquad
B_n(\aleph) \;=\; h_n^{(2)}(\aleph),
\end{equation}
together with the two standard Mie angular functions
\begin{equation}\label{eq:PiT}
\Pi_n(\theta) \;=\; \frac{P_n^{1}(\cos\theta)}{\sin\theta},
\qquad
T_n(\theta) \;=\; \frac{d}{d\theta} P_n^{1}(\cos\theta)
\end{equation}

\noindent The source-angle endpoint values (\ref{eq:endpoint_zero_dP}) and (\ref{eq:endpoint_zero_Poversin})
cause the order sum (sum over $m$) to collapse onto $|m|=1$.
With $\Lambda_{n,\pm1} = (2n+1)/[n(n+1)]^2$, this yields
\begin{equation}\label{eq:an}
a_{n,\pm1} \;\equiv\; a_n \;=\; -\,\frac{E_0}{8\pi}\;j^{-n}\,
\frac{2n+1}{n(n+1)},
\qquad
b_{n,\pm1} \;=\; \mp\, a_n .
\end{equation}
Carrying out the order ($m$) sum in (\ref{eq:full-VSWE-Electric-Field}) with the order restricted to $\pm 1$,
one finds \emph{exactly, at every radius},
\begin{eqnarray}
E_r (r, \theta, \phi)  &=& 2\cos\phi\; R(\aleph, \theta),        \qquad
R(\aleph, \theta) = \sum_{n=1}^{\Nmax} a_n \> \frac{n(n+1)}{\aleph}\,B_n(\aleph)\,P_n^{|1|}(\cos\theta),             \hfill
\label{eq:Er}\\[2pt]
E_\theta(r, \theta, \phi) &=& 2\cos\phi\; S_\theta(\aleph, \theta), \qquad
S_\theta(\aleph, \theta) = \sum_{n=1}^{\Nmax} a_n \>
\left[\,A_n(\aleph) T_n(\theta) - j B_n(\aleph) \Pi_n(\theta) \,\right],                                   \hfill
\label{eq:Eth}\\[2pt]
E_\phi(r, \theta, \phi)  &=& 2j\sin\phi\; S_\phi(\aleph, \theta),  \qquad
S_\phi(\aleph, \theta)   = \sum_{n=1}^{\Nmax} a_n \>
\left[\,j A_n(\aleph) \Pi_n(\theta) + B_n(\aleph) T_n(\theta) \,\right].                                   \hfill
\label{eq:Eph}
\end{eqnarray}
Equations~(\ref{eq:Er})--(\ref{eq:Eph}) are confirmed against the unpruned
double sum of (\ref{eq:full-VSWE-Electric-Field}) to $2.6\times10^{-15}$ in both Python and MATLAB.

One interesting analytical observation is that (\ref{eq:Er})--(\ref{eq:Eph}) combine to yield:
\begin{eqnarray}\label{eq:cos2phi}
\left|\, \vec E(r, \theta, \phi) \,\right|^2 \;=\; 4 \, \left[\left(\, |R(\aleph,\theta)|^2 + |S_\theta(\aleph,\theta)|^2 \,\right) \> \cos^2\phi \;+\; |S_\phi(\aleph,\theta)|^2 \> \sin^2\phi \,\right],
\end{eqnarray}
i.e., the entire azimuthal dependence is exhausted by a constant plus a single $\cos 2\phi$ term,
with \emph{no higher azimuthal harmonic}, at any radius, near field or far.
This result provided the sharpest available confirmation that the $m = \pm 1$ bookkeeping in both the
Python and MATLAB codes was correct.
Furthermore, it is clear from (\ref{eq:Er})--(\ref{eq:Eph}) that $E_{\phi} = 0$ everywhere in the
zx-plane ($\phi = 0$) and that $E_r$ vanishes on the z-axis because $P_n^{|1|}(\pm 1) = 0$.
On the other hand, $E_{\theta}$ peaks on the z-axis precisely because the endpoint values
(\ref{eq:endpoint_zero_dP}) and (\ref{eq:endpoint_zero_Poversin}) are \emph{non-zero} there and,
being equal, add coherently degree by degree.
Taking the limit $\theta \to 0$ of (\ref{eq:Eth}) and (\ref{eq:Eph}) and exploiting the
endpoint relations detailed in Appendix A,
one obtains ${\vec E} \to 2 S_\theta(\aleph,0) \, {\hat x}$ -- the on-axis field is a
single Cartesian vector, directed along ${\hat x}$ and therefore orthogonal to the beam axis
and lying entirely in the zx-plane. Its magnitude carries no azimuthal dependence at all.
The apparent $\phi$ dependence of its individual spherical components is only that of the spherical basis itself.

Another is the fact that two different powers of $k_0r$ govern the approach to the far-field region of the source.
The \emph{field} converges as $O(1/k_0r)$ because $E_r$ converges as $n(n+1)/k_0r$.
On the other hand, the \emph{pattern}, hence, directivity converges as $O((k_0r)^{-2})$ because that
leading term cancels identically in $|{\vec E}|^2$.


\subsection{\textcolor{blue}{Front-to-Back Ratio Needle Beam at a Finite Far-field Radius}}
\vskip 0.1in

The unidirectional property of the needle beams was examined analytically and numerically.
The conventions that $\ftbr|_{\rm field}$ is a ratio of field magnitudes and
that $\ftbr|_{\rm power} = (\ftbr|_{\rm field})^2$ is a ratio of power magnitudes are followed.
Expressed in decibels, the two ratios yield the same number, i.e.,
$20\log_{10}(\ftbr|_{\rm field}) = 10\log_{10}(\ftbr|_{\rm power})$. Every decibel figure quoted
herein is that common value.

As proved in detail in Appendix A, the directivity along the negative z-axis at the finite distance $r$ is

\begin{equation}
D(\theta=\pi, \phi=0) = \frac{\Nmax(\Nmax+1)^2(\Nmax+2)}{16\,(k_0r)^4}.
\end{equation}

\noindent With the frequency-domain broadside-radiating elemental source being located at the origin
and the VSWE sum representing its field being truncated at $\Nmax$, it radiates a needle beam whose
directivity along the +z-axis is $\Dmax = \Nmax^2 + 2 \, \Nmax$. Therefore, the FTBR values at a
finite far-field radius from the origin are:
\begin{equation}
\ftbr\Big|_{\rm field} = \frac{4(k_0r)^2}{\Nmax+1},
\qquad
\ftbr\Big|_{\rm power} = \left[\frac{4(k_0r)^2}{\Nmax+1}\right]^{2},
\end{equation}

\noindent They have been verified to better than $0.05$~dB whenever $k_0 r \gtrsim \Nmax(\Nmax+1)$, and
confirmed against the full Eq.~(\ref{eq:full-VSWE-Electric-Field}) to six significant figures.
\noindent They illuminate three main consequences:
\begin{itemize}
\item The back lobe falls off as $r^{-4}$ in \emph{power}, not $r^{-2}$. Thus, the slope is
$12$~dB per doubling of $r$.
\item The FTBR increases, for example, by $80$~dB between $r=10^2\lambda$ and $r=10^4\lambda$ .
\item The needle beam is \emph{extremely} unidirectional even at a finite range, e.g.,
even at $r=10^2\lambda$ the FTBR is $97$--$108$~dB.
\end{itemize}


\subsection{\textcolor{blue}{Collateral Exact Results}}\label{sec:collateral}
\vskip 0.1in

\noindent Several other interesting features of the beams fell out of the derivations that are worth reporting
independently, since each is exact at every radius rather than asymptotic.
\vskip 0.05in

\noindent \textbf{(a) The front and back lobes are azimuthally flat -- exactly.} From
(\ref{eq:quadrature}) in Appendix A, one knows that
$S_\phi (\aleph,\theta=0) = +j \, S_\theta (\aleph,\theta=0)$
and
$S_\phi (\aleph,\theta=\pi) = - j \, S_\theta (\aleph,\theta=\pi)$.
Consequently, $|S_\theta| = |S_\phi|$ on the z-axis in both directions at \emph{every} radius,
near field as well as far field.
Therefore, $D(\pi,\phi)$ is independent of $\phi$ at \emph{every} radius, even though the
pattern as a whole is not azimuthally symmetric until the far field.
This feature has been verified to $5\times10^{-16}$.
The fact that this feature is also true in the forward direction is not obvious in advance.
It is a consequence of the per-degree ($n$) relations satisfied by the Mie angular functions:
$\Pi_n = +T_n$ at $\theta = 0$ and $\Pi_n = -T_n$ at $\theta = \pi$,
which survive the sum over the degree becausethey hold degree-by-degree.
\vskip 0.05in

\noindent \textbf{(b) The xy-plane cut is a circle only asymptotically.}
$|S_\theta| = |S_\phi|$ at $\theta = \pi/2$ only in the limit, so the equatorial
cut is flat in dB to within a ripple that falls as $(k_0r)^{-2}$. For example,
it is $5.55\times10^{-3}$~dB at $r=10^2\lambda$ for $\Nmax=20$
and $5.54\times10^{-7}$~dB at $10^4\lambda$.
\vskip 0.05in


\section{\textcolor{red}{Directivity Results}}
\vskip 0.1in

All the frequency-domain sources were excited initially with a $f$ = 3.0 GHz signal
(simply because the wavelength $\lambda = ~$0.1 m) with $E_0$ = 1.0 V/m to study the needle-element's directivity.
Three maximum multipole cases were considered, i.e., the $\Nmax$ values 5, 10, 20.
The elemental source is defined by the $a_{nm}$ and $b_{nm}$ coefficients.
As designed, their broadside direction coincides with the z-axis.

\begin{figure}[h!]
	\centering
        \includegraphics[width=0.98\columnwidth]{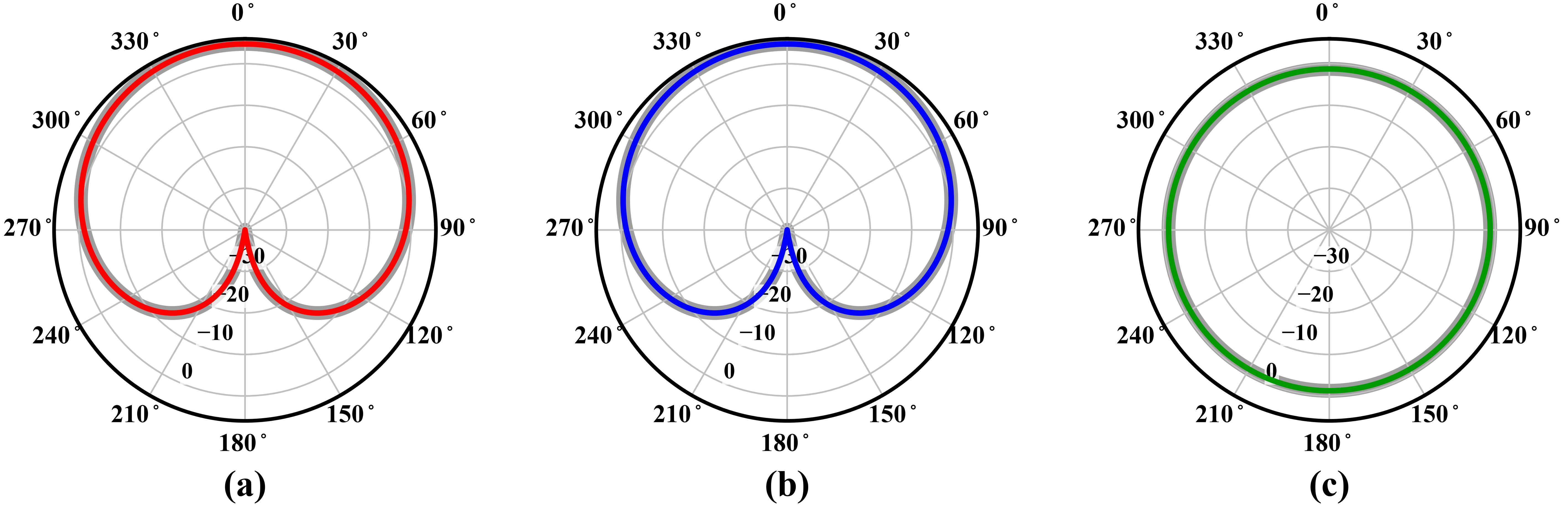}
\caption{Directivity patterns (dB) of the $\Nmax = 1$ (balanced Huygens dipole) needle element
at $f = 3.0$~GHz, evaluated at $r = 10^4\lambda$.
(a) zx-plane (red). (b) yz-plane (blue). (c) xy-plane (green).
The radial scale runs from $-40$ to $+6$~dB. The grey curve underlying each trace,
visible only where it emerges from the coloured one, is the closed-form cardioid
$D = (3/4)(1+\cos\theta)^2$; it is reproduced by both language codes to machine precision.}
	\label{fig:Fig1}
\end{figure}

As noted, the Huygens dipole element, $\Nmax = 1$, was used as prime test case for the codes.
The consequent cardioid patterns in the vertical zx and yz planes and omni-directional pattern in the
horizontal xy plane are shown in Fig.~\ref{fig:Fig1}. They illustrate the results:
$D(\theta=0,\phi=0) =$ 3.0 = 4.77 dB, HPBW = 131.06$^{\circ}$, and
$D(\pi,0) = 0.0$ (exactly, in the $r \to \infty$ limit) giving the FTBR = $\infty$.
Every beamwidth and every ``$-3$~dB'' angle quoted in this article are taken at the
\emph{exact} half-power level, $10\log_{10}(1/2) = -3.0103$~dB, rather than at a rounded
$-3.00$~dB. With that convention the value just quoted is the closed-form cardioid beamwidth
$2\arccos(\sqrt{2}-1) = 131.0604^{\circ}$ to five significant figures, which is the sharpest
available check of the beamwidth extraction itself.
%
%
%

\begin{figure}[h!]
	\centering
        \includegraphics[width=0.98\columnwidth]{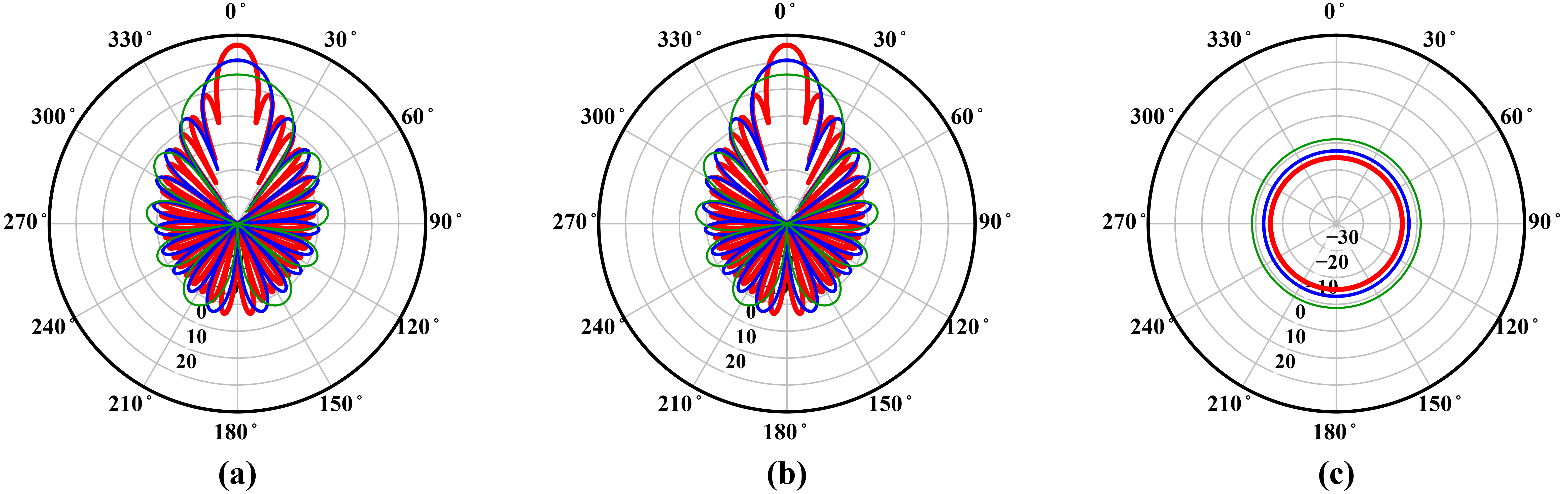}
\caption{Directivity patterns (dB) of the needle-beam source at $f = 3.0$~GHz for
$\Nmax$ = 5 (green), 10 (blue), 20 (red), evaluated at $r_{\rm meas} = 10^2\lambda$.
(a) zx-plane. (b) yz-plane. (c) xy-plane.
The radial scale runs from $-40$ to $+30$~dB. The three traces are drawn widest-first, i.e.,
$\Nmax$ = 20 beneath 10 beneath 5, with decreasing line width, so that all three remain visible
where they overlap.
The patterns at $r_{\rm meas} = 10^4\lambda$ are visually indistinguishable from these, i.e., the
needle beam is fully formed already at the shorter range; the small residual differences are
quantified in Table \ref{table:table-needle-results}.}
	\label{fig:Fig2}
\end{figure}

\begin{figure}[h!]
	\centering
        \includegraphics[width=0.98\columnwidth]{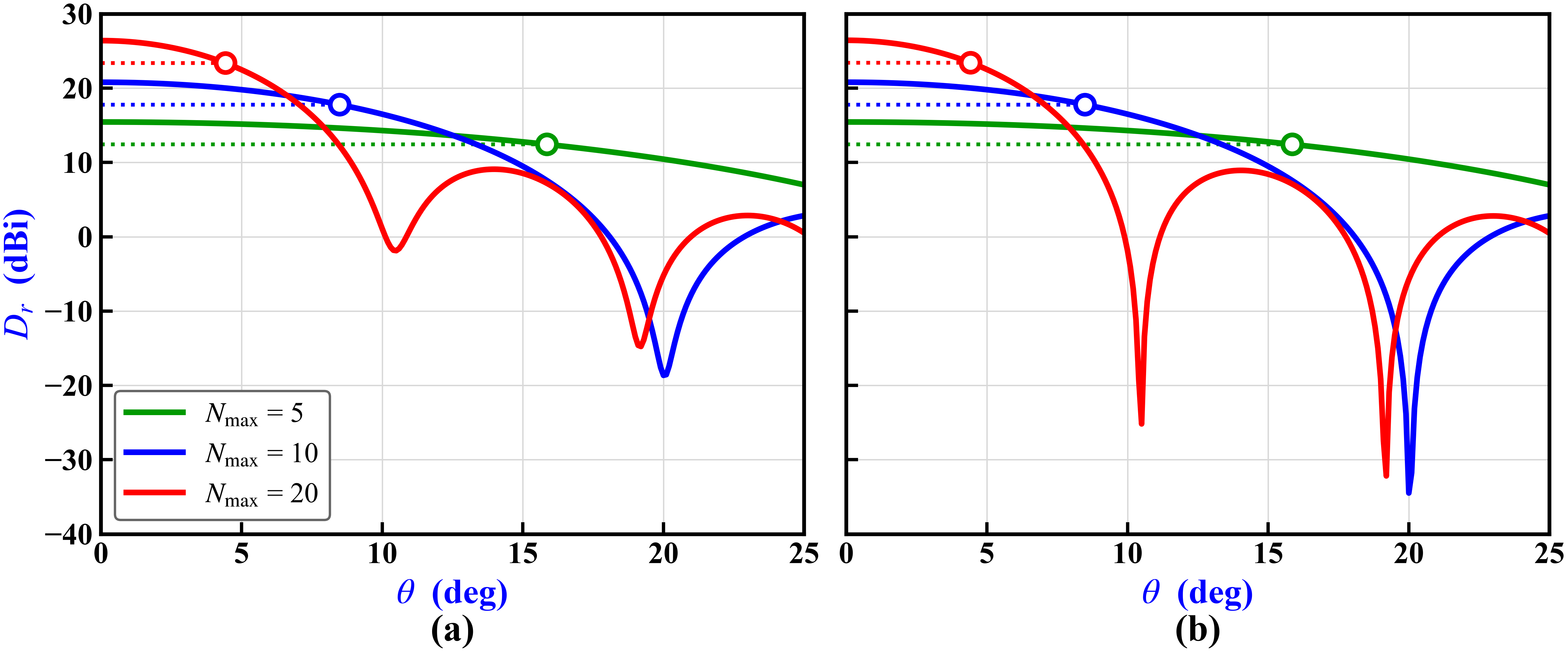}
\caption{Main beam directivities in the zx-plane as functions of the angle from the z-axis for
$\Nmax$ = 5 (green), 10 (blue), 20 (red), measured at (a) $r_{\rm meas} = 10^2\lambda$ and
(b) $r_{\rm meas} = 10^4\lambda$.
The dotted horizontal lines and the open circles mark the half-power points.}
	\label{fig:Fig3}
\end{figure}

The values obtained with the far-field code that calculated $E_{\theta}$ and $E_{\phi}$ reproduced those obtained
with the exact full VSWE code to machine precision. Moreover, it was determined explicitly to machine precision that both language codes predicted correctly that the vector functions (\ref{eq:M-vector}) and (\ref{eq:N-vector}) were divergence-free using random VSWE coefficients. This is a very strong test since it deals with the relative weights of the radial and transverse terms.
The elemental source's directivity patterns in the zx, yz, and xy planes for $\Nmax$ = 5, 10, 20 are shown
in Fig. \ref{fig:Fig2}. The main beam features are illustrated in Fig. \ref{fig:Fig3}.
%
%
%

The predicted values for several figures of merit obtained at the measurement distances
$r_{\rm meas} = 10^2\,\lambda$ and $r_{\rm meas} = 10^4\,\lambda$
are summarized in Table \ref{table:table-needle-results}.
The peak directivity measured at these distances $D_r(\theta, \phi)$ is along the z-axis. It
was determined that $D_r(0,0) = N_{\max}^2 + 2N_{\max}$ for $\Nmax$ = 5, 10, 20 at both $r_{\rm meas}$;
and $D_r(\pi,0)$ was 97--188 dB down from the peak. MATLAB and Python agreed on every printed digit.
The corresponding HPBW narrows from $31.71^{\circ} \to 16.97^{\circ} \to 8.84^{\circ}$.
The predicted unidirectional superdirective pseudo-needle beams were confirmed.
The predicted $\cos2\phi$ result was also confirmed exactly.

%
%

\begin{table*}[!h]
\vspace*{0.1in}
\caption{\textcolor{black}{Summary of $\Nmax$-Needle-element Directivity Results}}
\vspace*{0.02in}
\begin{center}
\begin{tabular}{c|ccccccc}
$\Nmax$ & $r_{\rm meas}/\lambda$ & $N^2+2N$ & $D_r(0,0)$ & $D_r(0,0)$ (dB) & HPBW (deg) & $D_r(\pi,0)$ (dB) & FTBR (dB) \\
\hline
\hline
\\[-0.5ex]
5   &  $10^2$ &  35  &  34.998  &  15.44  & 31.71  & -92.96   & 108.40
\\
5   &  $10^4$ &  35  &  35.000  &  15.44  & 31.71  & -172.96  & 188.41
\\[+0.5ex]
\hline
\\[-0.5ex]
10   &  $10^2$ &  120  & 119.91  &  20.79  & 16.97  & -82.35   & 103.14
\\
10   &  $10^4$ &  120  & 120.00  &  20.79  & 16.97  & -162.35  & 183.14
\\[+0.5ex]
\hline
\\[-0.5ex]
20   &  $10^2$ &  440  & 435.54  &  26.39  & 8.84  & -71.08   & 97.47
\\
20   &  $10^4$ &  440  & 440.00  &  26.44  & 8.84  & -151.09  & 177.52
\\[+0.8ex]
\hline
\hline
\end{tabular}
\end{center}
\label{table:table-needle-results}
\end{table*}
\vskip 0.1in


In order to plot features of the $\Nmax = 20$ needle beam on a Cartesian grid, the spherical coordinate VSWE expression of the electric field components in (\ref{eq:full-VSWE-Electric-Field}) were converted into components expressed
in Cartesian coordinates. These efforts also facilitated evaluating the needle beams radiated from points away from
the origin. Furthermore, as a consequence, they enabled the array studies to be described below.

\begin{figure}[h!]
	\centering
        \includegraphics[width=0.90\columnwidth]{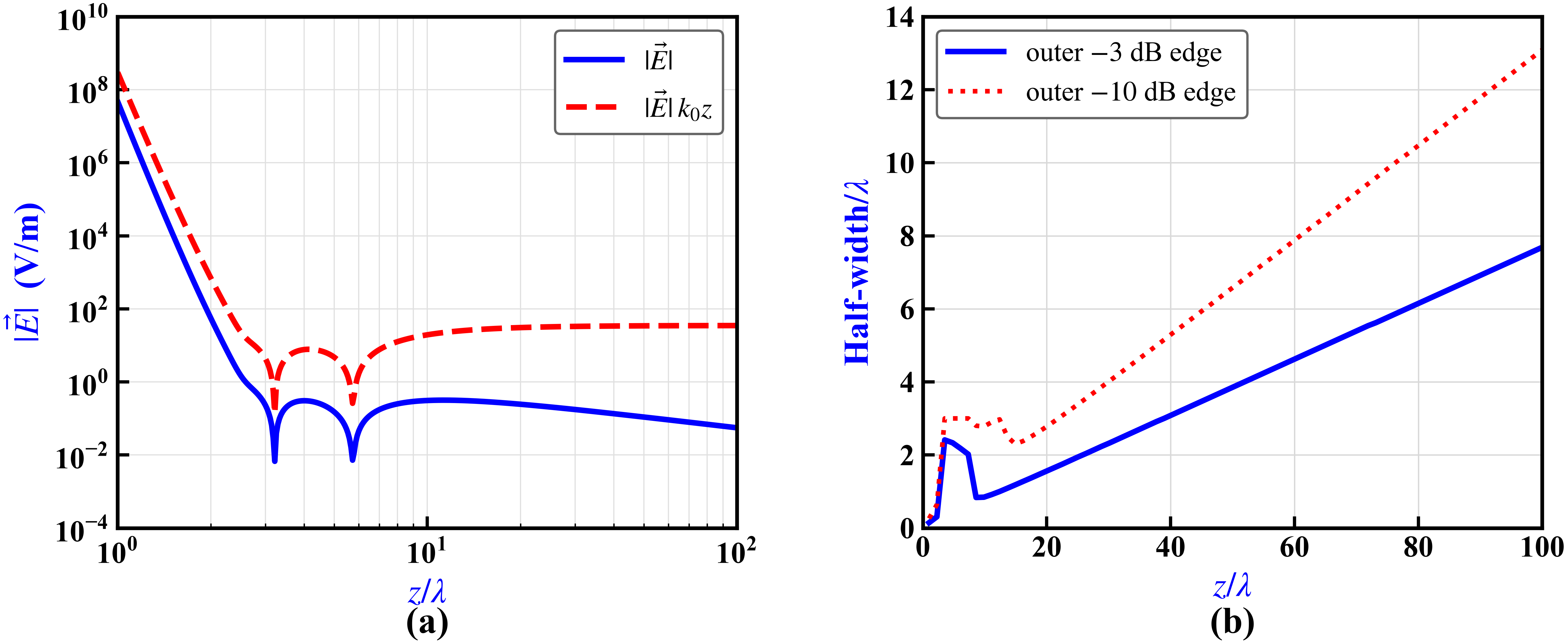}
\caption{Properties of the $\Nmax = 20$ needle beam as functions of the axial distance from the origin.
(a) On-axis total field magnitude $|\vec E|$ (blue) and its range-compensated version
$|\vec E|\,k_0z$ (red, dashed), which asymptotes to $35.0141$~V/m.
(b) Outer transverse half-width of the beam at the half-power level (blue, solid) and at the
$-10$~dB level (red, dotted), each referenced in a given $z$-slice to the maximum in that slice.}
	\label{fig:Fig4}
\end{figure}

The subsequent calculations of the on-axis magnitude $|{\vec E}| \, k_0r$ and of the transverse
half-widths of the needle beams, at the half-power and at the $-10$~dB levels, are shown in
Fig.~\ref{fig:Fig4}. Two deep on-axis nulls are located at
$z = 3.210 \, \lambda$ and $z = 5.748 \, \lambda$; they were not anticipated.
They arise from destructive interference between the longitudinal and transverse field components in the near-field region.
Clean $1/r$ decay occurs beyond $z \approx 20 \, \lambda$ (blue curve). In that far-field region the field level
$|{\vec E}|\, k_0 z$ (red dashed curve) asymptotes to $35.0141$ V/m as $z$ increases.
The width of the beam at each $z$-slice is referenced to beam's maximum in it. Since that maximum is decreasing with
increasing $z$, the width of the beam increases as the beam propagates further away from its source.

\begin{figure}[h!]
	\centering
        \includegraphics[width=0.90\columnwidth]{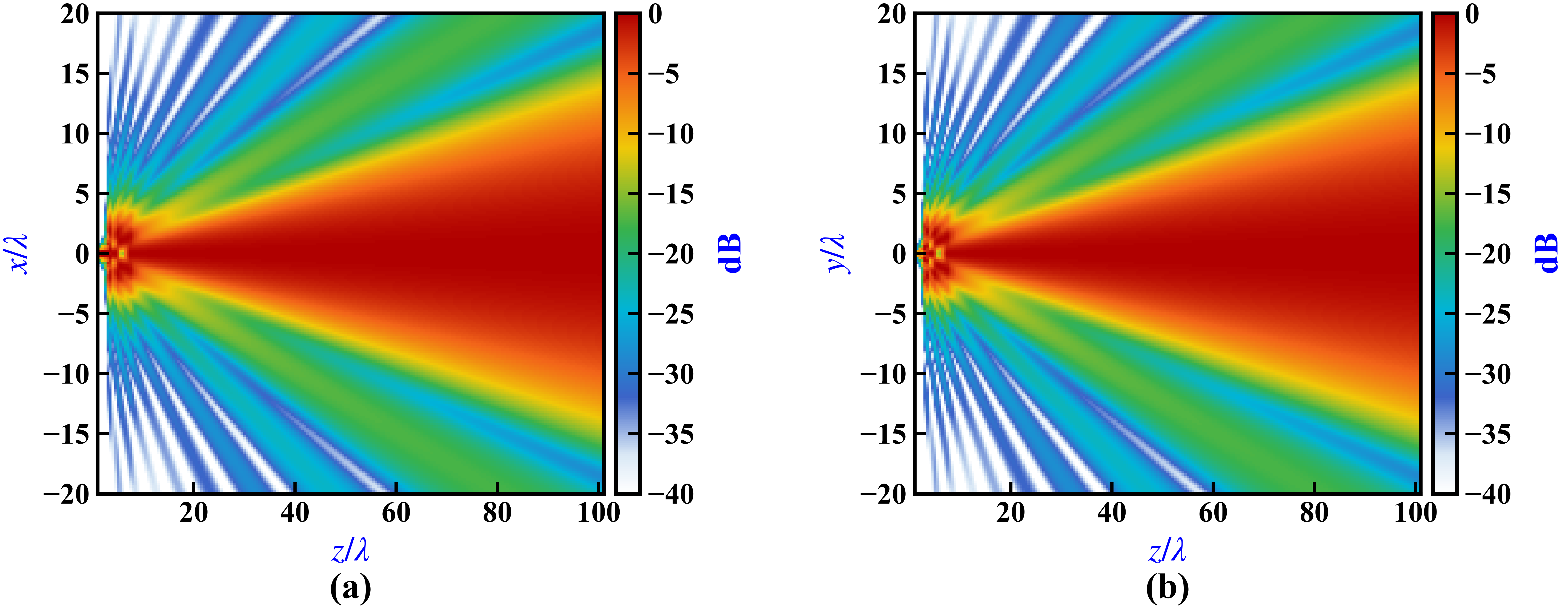}
\caption{Distribution of the total electric field magnitude of the $\Nmax = 20$ needle beam in the
(a) zx and (b) yz planes, normalized within each $z$-slice to the maximum in that slice.}
	\label{fig:Fig5}
\end{figure}

The corresponding intensity distributions of the magnitude of the total electric field
of that beam in the zx and yz planes are shown in Fig.~\ref{fig:Fig5}.
Comparing the complete field to the field arising from only its transverse components, one
finds little difference beyond the reactive zone, i.e., the $E_r$ contributions in these plots are small.
In particular, it contributes a little to the zx-plane distribution, but not at all to the yz-plane
since it is zero for $\phi = \pi/2$.
The two panels are therefore indistinguishable once $z \gtrsim 20\,\lambda$ and differ only in
the near-zone lobe structure at the left-hand edge, where $E_r$ carries up to $86\%$ of the
zx-plane field magnitude at $z = \lambda$.
To observe results across the large longitudinal distance of this plot, the color scale is referenced to the
maximum level of \emph{each z-slice}.

%
%
%
%

\section{\textcolor{red}{Frequency-domain Arrays of Superdirective Elements}}
\vskip 0.1in

Given the demonstrated exceptional performance of the elemental unidirectional superdirective multipole sources,
uniform circular arrays (UCAs) of them were then considered. The fields radiated by each element of the array
were calculated everywhere using (\ref{eq:full-VSWE-Electric-Field}) with the coefficients:

\begin{IEEEeqnarray}{rLc} \label{eq:elemental-source-coefficients}
a_{nm} &=& \frac{1}{4\pi} \> j^{-n} \> \Lambda_{nm} \> \left[\, \frac{d}{d\theta} \>
P_n^{|m|}(\cos\theta) \,\right]_{\theta_s = 0} \> \left[\, e^{-jm \phi_s} \,\right]_{\phi_s = 0}         \hfill \\
\noalign{\vskip 0.1in}
b_{nm} &=& - \frac{1}{4\pi} \> j^{-n} \> \Lambda_{nm} \>
\left[\, m \> \frac{P_n^{|m|}(\cos\theta)}{\sin\theta} \,\right]_{\theta_s = 0}
\> \left[\, e^{-jm \phi_s} \,\right]_{\phi_s = 0}                               \hfill
\end{IEEEeqnarray}
\vskip 0.1in

\noindent They are frequency-independent and, hence, dispersion-free. To achieve the largest directivity in the +z-direction, the source angles were again taken to be $\theta_s = \phi_s = 0$.
Thus, the beam emitted by each UCA is azimuthally symmetric.
Numerical integration (highly accurate Gauss-Legendre method) was employed to calculate the total power radiated:

\begin{equation}
P_{\rm tot} = \frac{r^2}{2\eta_0} \> \oiint_{\rm sphere \, of \, radius r} \>
\left[\, |E_{\theta}|^2 + |E_{\phi}|^2 \,\right] \> \sin\theta \> d\theta \, d\phi
\end{equation}

The base configuration is a UCA within a circular region of the xy-plane centered with respect to the origin
whose radius is $r_{\rm array}$.
The arrays consist of one source at the origin along with sources on rings (circles) that are concentric
with respect to the origin. The number of rings is a specified even number, $Q$. The radius of each ring is
$ar_1, ar_2, ..., ar_Q$. The source at the origin has the radius $ar_0 = 0$ for notational purposes.
In particular, the spacing between the rings was set to be uniform, i.e.,
$ar_Q = r_{\rm array}$ and, thus, $ar_{q}-ar_{q-1} = ar_Q/Q$, $q=1, ..., Q$.

Furthermore, each ring had the same even number of sources, $\xi$, uniformly distributed on it.
Thus, the spacing between each of these elements was $2\pi/\xi$. The elements on each ring were staggered
relative to those on the neighboring rings. In particular, the first source of each odd-numbered ring array:
$RA_1, RA_3, ...$, was located on the x-axis, i.e., at the angle $\phi_1 = 0$.
Then the first source of each even-numbered ring array: $RA_2, RA_4, ...$, was located at the angle $\phi_1 = \pi/\xi$.
Thus, the sources on the odd-numbered (even-numbered) rings were located at the same $\phi$ angles.
Hence, those same angle sources lay on a radial spoke from the origin through each of them.
The odd and even spokes are separated in angle by $\pi/\xi$ radians.

The vector location of each source is labeled by its ring number and its angle number, e.g., let
${\vec r}_{qp}$ for $q=1, 2, ..., Q$ and $p = 1, 2, ..., \xi$ be the location of the source so that, e.g.,
${\vec r}_{23}$ is located on the second ring with radius $2 \times \left(\, ar_Q/Q \,\right)$ and at the
angle $\phi_3 = 2 \times (2\pi/\xi) + \phi_1$.
Each element was weighted by the constant amplitude $E_0$ = 1.0 V/m. This choice changes for the amplitude
tapering considerations and for the pulse-driven arrays discussed below.
Furthermore, these array studies were restricted generally to needle elements with $\Nmax$= 10.
This choice was made because the same multipole maximum was considered for the system designs in \cite{b-a-z-2025superdirective}.
Recall that the elemental source's $\Dmax = \Nmax^2 + 2 \Nmax = $ 120 = 20.79 dB.

One main goal was to confirm that the array would radiate a unidirectional, superdirective beam in its
far field. Another was to determine if the array's needle beam has, as anticipated, attractive properties
beyond the elemental's SDEU features, e.g., narrower beam widths.
Since all of the elements of the array are identical, the total electric field in the
far-field region of the array takes on a well known form:

\begin{IEEEeqnarray}{c} \label{eq:freq-domain-far-field-of-array}
{\vec E}^{\rm ff}_{array} (\theta, \phi, \omega) = \left[\, E_0 \> \sum_{q,p} \> e^{jk_0 {\hat r} \cdot {\vec r}_{qp}} \,\right]
\>\> {\vec E}^{\rm ff}_{elem} (\theta, \phi, \omega)
= AF(\theta, \phi, \omega) \> {\vec E}^{\rm ff}_{elem}(\theta, \phi, \omega)
\end{IEEEeqnarray}
\vskip 0.1in

\noindent i.e., an array factor times the elemental multipole's far field.
Thus, we know several facts already. The array lies on the xy-plane and consists of
broadside-radiating elements with no additional phase imposed on them. The maximum
directivity will be on the z-axis where $\theta = 0$ and $\phi = 0$. This means each
phase term in the array factor is zero in that maximum direction, i.e., for all $q,p$,
$\> {\hat r} \cdot {\vec r}_{qp} = \rho_q \> \sin\theta \> \cos(\phi - \phi_{qp}) \equiv 0$.
Therefore, if the number of array elements, $K$, increases, the far-field intensity
$|{\vec E}^{\rm ff}_{array}|^2$ increases as $K^2$ times the elemental's maximum intensity.
Thus, larger broadside intensities can be attained by increasing the element density and the
$\Nmax$ of its elements.

\subsection{\textcolor{blue}{UCA Geometries}}
\vskip 0.1in

Four arrays were selected for comparison purposes and for use in the time-domain array analyses below.
All had $Q=4$, i.e., 4 outer edges demarking 4 regions.
Their sizes ranged from $r_{\rm array}$ = $0.1 \, \lambda$ to $0.564 \, \lambda$, $2.0 \, \lambda$, and $6.0 \, \lambda$.
The first choice makes the array electrically small.
The second choice has its Rayleigh distance equal to $\lambda$: $z_R = \pi r_{\rm array}^2 / \lambda \approx \lambda$.
The Rayleigh range is used herein in its beam-optics / circular-aperture sense:
$z_R = \pi r_{\rm array}^2/\lambda$,
i.e., the distance over which a beam having an initial waist $r_{\rm array}$ expands to a waist $\sqrt{2}$ times as large.
It is the natural depth scale for the near-field focusing results of Sec. V.
The third choice separates each ring by $0.5 \, \lambda$, a typical selection to minimize coupling
between physical array elements. The fourth choice is not compact, but it facilitates illustrating several additional  features tied to a larger array.

\begin{figure}[h!]
	\centering
        \includegraphics[width=0.98\columnwidth]{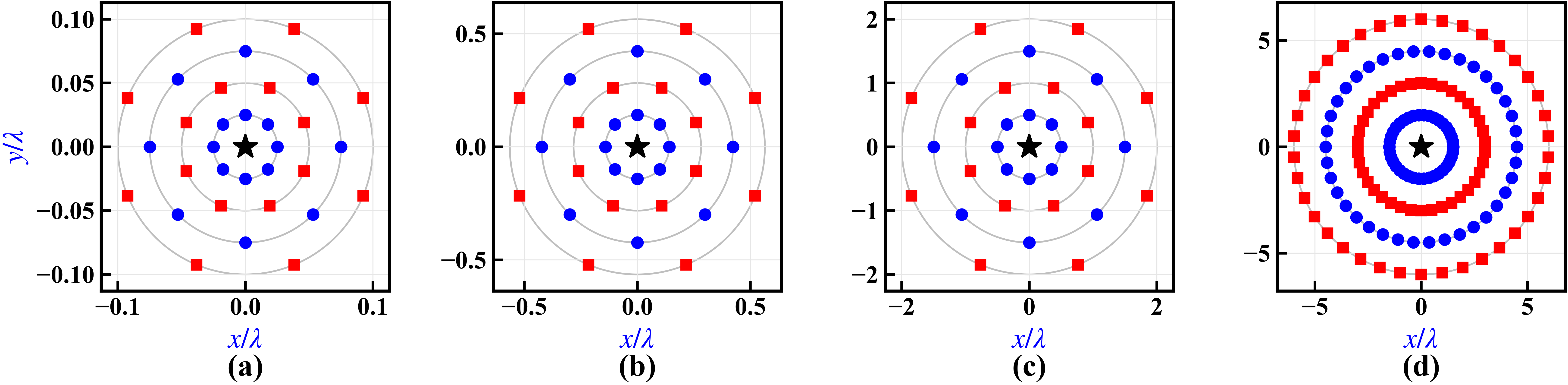}
\caption{Distribution of the elements in all four arrays:
(a) $r_{\rm array} = 0.1 \, \lambda$, (b) $0.564 \, \lambda$, (c) $2.0 \, \lambda$
(all with $Q = 4$, $\xi = 8$, $K = 33$), and (d) $6.0 \, \lambda$ ($Q = 4$, $\xi = 38$, $K = 153$).
The black star is the element at the origin; blue circles mark the odd-numbered rings and red
squares the even-numbered ones, which are rotated by $\pi/\xi$ relative to their neighbors.}
	\label{fig:Fig6}
\end{figure}

The element number on all rings was fixed at $\xi = 8$ for the three smaller arrays,
i.e., there was $45^{\circ}$ between each array element in each ring and $K$ = 33.
By maintaining a fixed number of elements in each ring, this choice had an effective $1/\rho_q$ taper of the amplitudes across the array. Other array weights will be considered below.
On the other hand, $\xi$ was increased to 38 ($K$ = 153) in the $6.0 \, \lambda$ array and avoided
an azimuthal undersampling.
It is noted that the number of elements in these arrays does not impact the directivity, rather the aperture area does.
It is also noted that the first null of the patterns of the three compact arrays occurs at $20.04^{\circ}$,
exactly where the first null of the elemental multipole source's pattern occurs: their array factors
are still too small to produce a null of their own.
Only the much larger $6 \, \lambda$ array, whose array factor does place its own first null
inside the element's main beam, breaks that degeneracy; its first null occurs at $7.96^{\circ}$.
These array configurations are illustrated in Fig.~\ref{fig:Fig6}.
%
%
%

\subsection{\textcolor{blue}{UCA Results}}
\vskip 0.1in

Several properties of the needle beams emitted from each array are listed in Table \ref{table:table-needle-prop}.
Notice that the aperture of the electrically small array is essentially uniformly illuminated with the
needle element field. Hence, it reproduces its $\Dmax$.
Recall that the broadside directivity of an array whose aperture area is uniformly illuminated is
$D_A = 4\pi {\rm Area} / \lambda^2 = (k_0 r_{array})^2$ for a circular aperture \cite{balanis_antenna_2016}.
Thus, if $\Dmax > D_A$, the array is said to be superdirective \cite{hansen1981}.

The three compact arrays are all superdirective.
The $0.1 \, \lambda$ array provides 24.9 dB of superdirectivity!
The needle element is a true small-aperture technology advance; the array result further demonstrates
the significant advantage of using multipole elements.
It is also clear that the $6 \, \lambda$ array with $\Nmax$ = 10 is barely superdirective (i.e., by only 0.35 dB);
but if the needle element with $\Nmax$ = 20 had been used, then $\Dmax$ = 1884.73 = 32.75 dB
against the same aperture bound $D_A = (k_0 r_{\rm array})^2$ = 1421.22 = 31.53 dB, i.e.,
$\Dmax/D_A$ = 1.326, which makes it superdirective by 1.23 dB.
Doubling $\Nmax$ from 10 to 20 therefore buys only 0.87 dB of additional superdirectivity in
this large array, while it buys 5.6 dB in the element itself. On the other hand, the aperture, not the element, is
already doing most of the beamforming once $r_{\rm array}$ = $6 \, \lambda$.

\begin{table*}[!h]
\vspace*{0.1in}
\caption{\textcolor{black}{Summary of Needle-beam Properties}}
\vspace*{0.02in}
\begin{center}
\begin{tabular}{c|cccccc}
$r_{\rm array}$ & $\Dmax$ & $\Dmax$ (dB) & $D_A$ (dB) &  $\Dmax/D_A$  & $ z_R = \pi r_{\rm array}^2 / \lambda$ & HPBW  \\ \hline
\hline
\\
$0.1\lambda$   &  120.97 &  20.83       & -4.04      & 306.42 = 24.86 dB   & $ 0.03   \,  \lambda$       & 16.95 $^{\circ}$
\\
$0.564\lambda$ &  142.53 &  21.54       & 10.99      &  11.35 = 10.55 dB   & $ 1.00   \,  \lambda$       & 16.29$^{\circ}$
\\
$2.0\lambda$   &  278.61 &  24.45       & 21.98      &   1.76 = 2.47  dB   & $ 12.57  \,  \lambda$       & 11.66$^{\circ}$
\\
$6.0\lambda$   & 1541.60 &  31.88       & 31.53      &   1.08 = 0.35  dB   & $ 113.10 \,  \lambda$       & 4.95$^{\circ}$
\\
\end{tabular}
\end{center}
\label{table:table-needle-prop}
\end{table*}
\vskip 0.1in

If one considers the infinite radius sphere for the directivity, each array's back null is exactly zero, i.e.,
$D(\theta=\pi, \phi=0) = 0$. Thus, each FTBR is infinite. In consideration of the SLLs, it was decided to
report the maximum backlobe level for $\theta > \pi/2$ as well.
One definition of the sidelobe level is used throughout this article and in every table:
the \emph{peak SLL} is the maximum of the total pattern over the entire sidelobe region, i.e.,
over all $\theta$ from the first null of the uniformly weighted total pattern out to
$\theta = \pi$. It is the quantity a minmax synthesis actually minimizes, and it is reported
alongside the individual near-in lobe peaks so that the two can be read against each other.

A summary of the SLLs computed for the arrays
is given in Table \ref{table:table-SLL}. The single elemental source alone has a $-17.33$ dB peak
sidelobe, at $26.92^{\circ}$, with its first null at $20.042^{\circ}$.
The near-in SLLs of the $2 \, \lambda$ array are 33 dB better from geometry alone. Moreover, the field
levels in the back hemisphere of the array are exceptionally small.
Note that the $2 \, \lambda$ array is the one case in which the peak SLL is \emph{not} the
first sidelobe. Its near-in lobes are suppressed so strongly, to $-51.05$~dB, that its worst
sidelobe, $-35.21$~dB, lies in the rear hemisphere.

\begin{table*}[!h]
\vspace*{0.1in}
\caption{\textcolor{black}{Summary of SLLs of the Arrays}}
\vspace*{0.02in}
\begin{center}
\begin{tabular}{c|ccccc}
$r_{\rm array}$ & 1st SLL                     & 2nd (dB) & 3rd (dB) & peak SLL & max level with $\theta > 90^\circ$ \\
\hline
\hline
\\
$0.1\lambda$    & $-17.41$ dB @ $26.90^\circ$ & $-23.35$ & $-26.99$ & $-17.41$ dB @ $26.90^\circ$  & $-26.94$ dB @ $163.98^\circ$ \\
$0.564\lambda$  & $-19.89$ dB @ $26.36^\circ$ & $-29.51$ & $-37.22$ & $-19.89$ dB @ $26.36^\circ$  & $-27.83$ dB @ $164.59^\circ$ \\
$2.0\lambda$    & $-51.05$ dB @ $22.97^\circ$ & $-49.01$ & $-85.18$ & $-35.21$ dB @ $169.63^\circ$ & $-35.21$ dB @ $169.63^\circ$ \\
$6.0\lambda$    & $-30.15$ dB @ $10.76^\circ$ & $-39.04$ & $-39.39$ & $-30.15$ dB @ $10.76^\circ$  & $-48.73$ dB @ $144.76^\circ$ \\
\end{tabular}
\end{center}
\label{table:table-SLL}
\end{table*}
\vskip 0.1in

\begin{figure}[h!]
	\centering
\includegraphics[width=0.98\columnwidth]{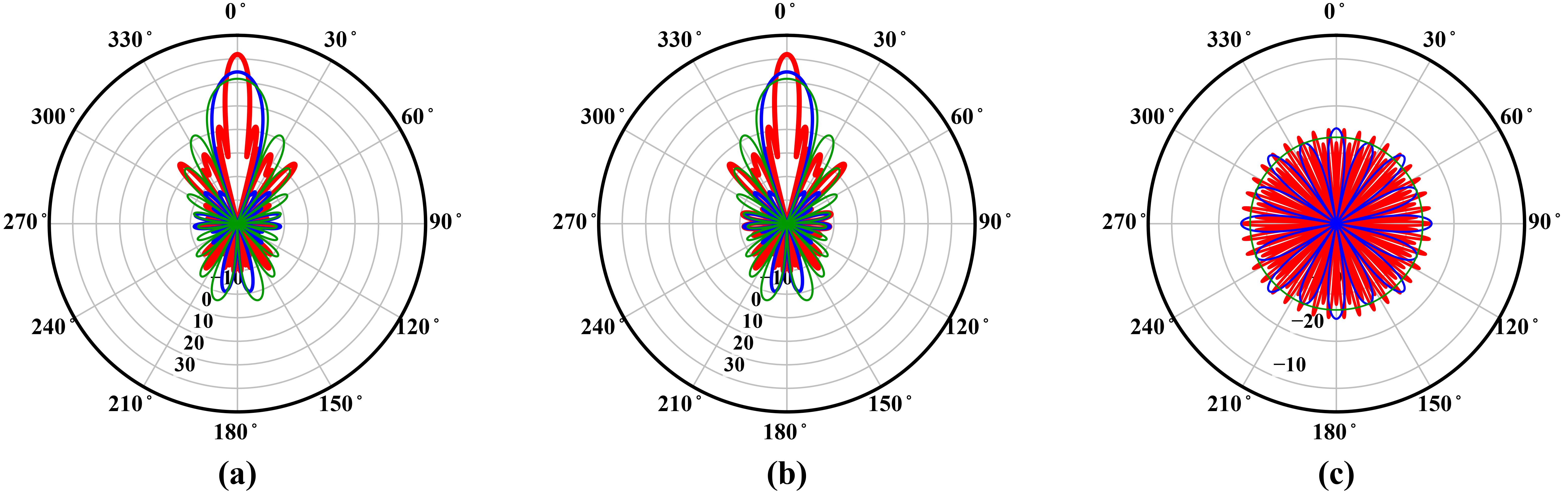}
\caption{Directivity patterns (dB) radiated by the $\Nmax = 10$ needle element
$0.564 \, \lambda$ (green), $2.0 \, \lambda$ (blue), and $6.0 \, \lambda$ (red) arrays in the
(a) zx-plane, (b) yz-plane, and (c) xy-plane, evaluated at $r_{\rm meas} = 10^4\lambda$.
The radial scale of (a) and (b) runs from $-40$ to $+40$~dB; that of (c), where the patterns lie
20--40 dB below the peak, is expanded to run from $-45$ to $-5$~dB. The $38$-fold azimuthal
ripple visible in (c) is that of the $\xi = 38$ rings of the $6 \, \lambda$ array.}
\label{fig:Fig7}
\end{figure}

\begin{figure}[h!]
	\centering
        \includegraphics[width=0.50\columnwidth]{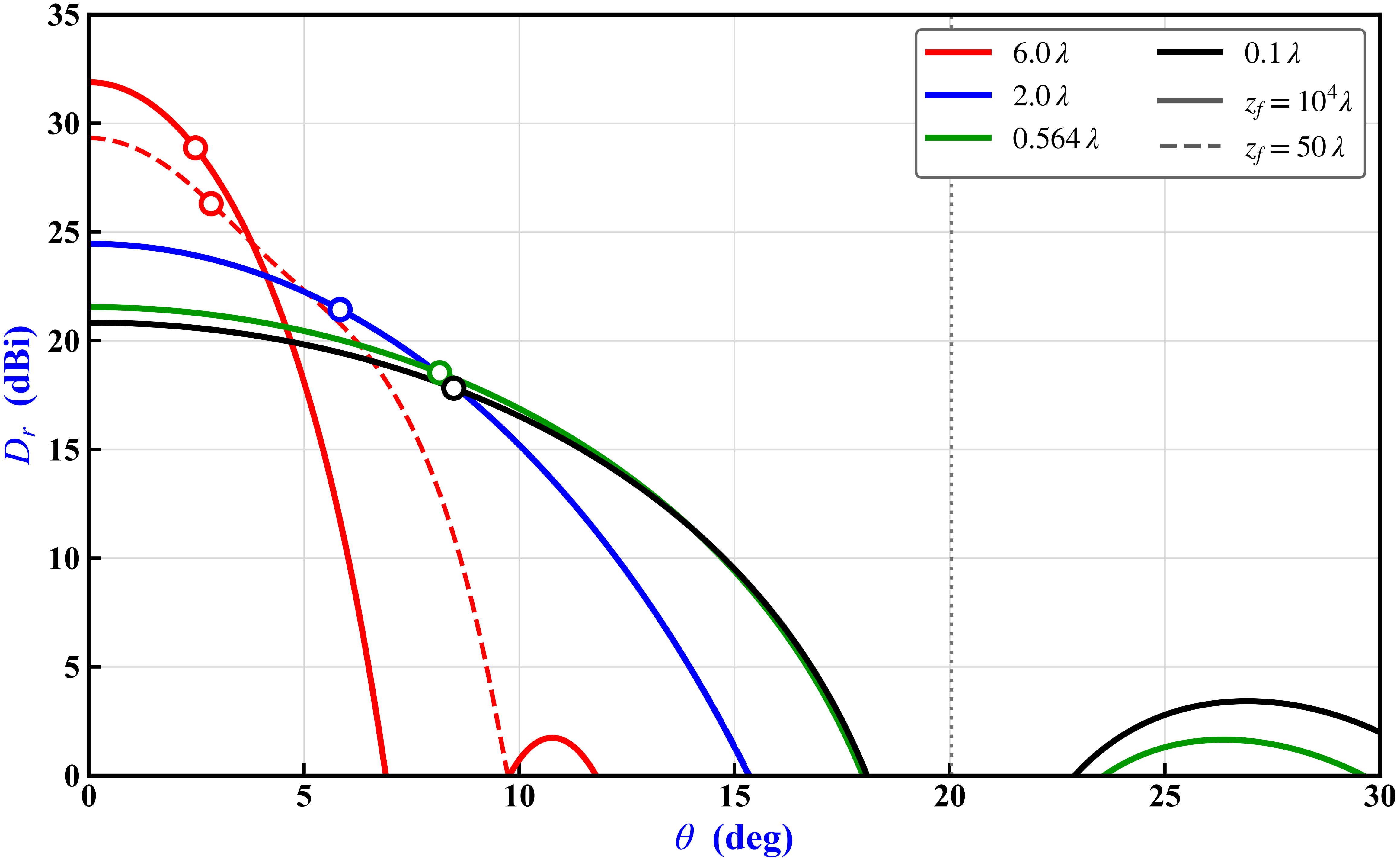}
\caption{Main beam directivities in the zx-plane as functions of the angle from the z-axis for the
$6.0 \, \lambda$ (red), $2.0 \, \lambda$ (blue), $0.564 \, \lambda$ (green) and $0.1 \, \lambda$
(black) arrays, measured at $r_{\rm meas} = 10^4 \,\lambda$ (solid) and $r_{\rm meas} = 50 \,\lambda$ (dashed).
The open circles are the half-power points; the dotted vertical line marks the $20.04^{\circ}$
first null of the $\Nmax = 10$ element.
Only the $6 \, \lambda$ array is large enough for the two measurement distances to differ
visibly: $50 \, \lambda$ lies well inside its $z_R = 113 \, \lambda$ Rayleigh range.}
\label{fig:Fig8}
\end{figure}

These results are viewed graphically in Figs.~\ref{fig:Fig7} and \ref{fig:Fig8}.
The far-field patterns of the three larger arrays in the zx, yz, and xy planes are shown in Fig.~\ref{fig:Fig7};
the $0.1 \, \lambda$ array is omitted there because its pattern is indistinguishable from that of the
$0.564 \, \lambda$ array on this scale.
The corresponding main beam results for all four arrays with the half-power points labeled are shown in Fig.~\ref{fig:Fig8}.
These results nicely illustrate that the beams radiated by the smaller arrays are already in their far field region at $r_{\rm meas} = 50 \,\lambda$ whereas the larger array is not yet ( recall that $z_R = \pi r_{\rm array}^2/\lambda$ ).
\vskip 0.1in

%
%

%
%

\subsection{\textcolor{blue}{Amplitude Tapering}}
\vskip 0.1in

To consider enhancing the needle-beam shape, in particular lowering the SLLs,
amplitude tapering across the array rings was implemented.
The VSWE coefficients of the array elements were modified into the form:
\begin{IEEEeqnarray}{rLc} \label{eq:amp-tapering-coefficients}
A_{nm,q} (\omega) &=& W_q \> E_0 \> a_{nm}
\hfill \\
\noalign{\vskip 0.1in}
B_{nm,q} (\omega) &=& W_q \> E_{0} \> b_{nm}
\hfill
\end{IEEEeqnarray}

\noindent where $W_q$ is the amplitude taper value of the q-th ring. Several traditional tapering
choices were considered and several evaluated. It was found that the natural tapering provided by
the fixed-$\xi$, uniform ring configuration and a minmax optimization of the amplitudes gave the best outcomes.
A Taylor circular taper was found to perform worse than the uniform choice. Simply with $Q=4$, there
were not enough rings to take advantage of its ``continuous aperture'' synthesis.

The performance of the arrays with the tapered weights is summarized in Table~\ref{table:table-weighted}.
While the minmax values were substantially better than many of those from the original uniform weights,
the minmax method required some negative amplitudes making the results highly sensitive to any perturbations.
The sidelobe levels reported in Table~\ref{table:table-weighted} are the peak SLLs defined
above, so its uniform-weight entries reproduce the peak-SLL column of
Table~\ref{table:table-SLL} exactly, and its uniform-weight beamwidths reproduce the HPBW column
of Table~\ref{table:table-needle-prop} exactly. The uniform and the tapered patterns are scored on
the \emph{same} sidelobe region, i.e., the one set by the first null of the uniformly weighted
pattern. Scoring each on its own first null is not a like-for-like comparison, since a taper that
narrows the main lobe moves its own null inward and is then charged for a shoulder that the
optimizer was never asked to suppress.

\begin{table*}[!h]
\vspace*{0.1in}
\caption{\textcolor{black}{Performance of Superdirective Arrays with Tapered Rings}}
\vspace*{0.02in}
\begin{center}
\begin{tabular}{c|cccc}
\hline
{\bf Fixed-$\xi$} \\
\hline
$r_{\rm array}$  & peak SLL, uniform  & peak SLL, minmax  & HPBW, uniform   & HPBW minmax    \\
\hline
$0.1\lambda$     &  -17.41 dB        & -30.55 dB        & 16.95$^\circ$   & 14.56$^\circ$  \\
$0.564\lambda$   &  -19.89 dB        & -34.47 dB        & 16.29$^\circ$   & 12.22$^\circ$  \\
$2.0\lambda$     &  -35.21 dB        & -39.55 dB        & 11.66$^\circ$   &  9.60$^\circ$  \\
\\
\hline
\\
{\bf Uniform ring density}  \\
\hline
\\
$r_{\rm array}$  & peak SLL, uniform  & peak SLL, minmax  & HPBW, uniform   & HPBW minmax    \\
\hline
$0.1\lambda$     &  -17.44 dB        & -30.55 dB        & 16.94$^\circ$   & 14.56$^\circ$  \\
$0.564\lambda$   &  -20.92 dB        & -32.99 dB        & 16.07$^\circ$   & 13.08$^\circ$  \\
$2.0\lambda$     &  -37.39 dB        & -38.29 dB        & 10.62$^\circ$   & 10.17$^\circ$  \\
\end{tabular}
\end{center}
\label{table:table-weighted}
\end{table*}
\vskip 0.1in


\section{\textcolor{red}{Pulsed beams Radiated by Time-domain Arrays of Superdirective Elements}}
\vskip 0.1in

The time-domain versions of the single multipole-element and their arrays follow readily from those
frequency domain results.
The VSWE coefficients (\ref{eq:elemental-source-coefficients}) of the multipole-element are frequency independent.
However, note that one must always remember that the free-space wavenumber is frequency dependent, i.e., $k_0 = \omega/c$.
As a consequence, the vector functions (\ref{eq:M-vector}) and (\ref{eq:N-vector}) are naturally frequency dependent.

To investigate pulsed beams radiated by these time-domain arrays in their far-field region,
one simply needs to introduce a spectrum of frequencies into the amplitudes of the array elements.
Then, one can use (\ref{eq:full-VSWE-Electric-Field}) with these modified amplitudes to calculate the
impulse response of the arrays. The time-domain needle beams are obtained with a Fourier transform of the resulting
radiated fields.

Consequently, we generalize the VSWE coefficients for the array elements to the form:
\begin{IEEEeqnarray}{rLc} \label{eq:array-coefficients}
A_{nm,q} (\omega) &=& W_q \> E_{q}(\omega) \> a_{nm} = w_q(\omega) \> a_{nm}
\hfill \\
\noalign{\vskip 0.1in}
B_{nm,q} (\omega) &=& W_q \> E_{q}(\omega) \> b_{nm} = w_q(\omega) \> b_{nm}
\hfill
\end{IEEEeqnarray}
\vskip 0.1in

\noindent where the $W_q$ would again facilitate different amplitude tapering of the rings and $E_{q}(\omega)$ is the frequency spectrum of the pulse driving the $q-th$ ring.
Herein, while we only consider all array elements being driven with the same time signal, i.e., $E_{q}(\omega) \to  E_{0}(\omega)$; nevertheless, this subtle notational generalization facilitates the discussion of
opportunities for additional applications as will be described later.

Furthermore, it must be noted that to recover known time-domain electromagnetic effects \cite{ziolkowski2026time-domain-hda} and to recover the correct units of the electric field, the frequency-dependent excitation amplitude of each array element
is taken explicitly as:
\begin{IEEEeqnarray}{c} \label{eq:current-element-amp}
E_q(\omega) = j \, k_0^2 \> \eta_0 \> J_q(\omega) \, S_e = j \> \frac{\omega}{c} \> k_0 \> F_q(\omega),
\hfill
\end{IEEEeqnarray}
\vskip 0.1in

\noindent where $F_q(\omega) \equiv \eta_0 \> J_q(\omega) \, S_e$ and
$J_q(\omega)$ is a surface current density (units $A/m$) driving each element in the q-th ring
of the UCAs at the frequency $f=\omega/2\pi$, and $S_e$ is the transverse to the z-axis surface area (units $m^2$) of each element, i.e., they lie in the xy-plane and the UCAs are broadside-radiating. Note that $J_q \, S_e$ has current-moment units. We recall that the presence of one $k_0$ cancels the $1/k_0$ leading term present in the vector functions
(\ref{eq:M-vector}) and (\ref{eq:N-vector}). The presence of the second $k_0 = \omega/c$ ensures that the signals
radiated into the far field of a finite array are a time derivative of those exciting it.
Furthermore, $F_q(\omega)$ is, to within the constant $\eta_0 S_e$, the frequency spectrum of the signal
$f_q(t)$ exciting an element in the q-th ring, obtained by Fourier transforming it. It carries the
units V-m, so that $E_q(\omega) = j k_0^2 \, F_q(\omega)$ carries the units V/m demanded of an
expansion coefficient in (\ref{eq:VSWE-E-field}), the vector functions
(\ref{eq:M-vector}) and (\ref{eq:N-vector}) themselves being dimensionless.
Again, it is emphasized that $F_q(\omega) = F(\omega)$ for $q=0, ..., Q$ -- all array elements
are excited coherently with the same time signal, $f(t)$.

The time signals to be employed here include the zero-area pulses (ZAPs) reported in \cite{ziolkowski2026time-domain-hda}.
A ZAP is a time signal whose integral over all time equals zero.
For a reader's convenience the single-cycle ZAP to be used here as one of the reference excitations and that was used for those time-domain explorations is:
\begin{IEEEeqnarray}{rCl}\label{eq:single_cycle_ZAP}
f_{\rm single}(t) = {\rm norm} \times
\begin{cases}
0 & {\rm if} \quad t < 0 \\
\left(\, \frac{t-T_L/2}{T_L/2} \,\right) \> \left[\, 1- \left(\, \frac{t-T_L/2}{T_L/2} \,\right)^2 \,\right]^3
  & {\rm if} \quad t \in [0, T_L] \\
0 & {\rm if} \quad t > T_L
\end{cases}
\end{IEEEeqnarray}
\vskip 0.1in

\noindent It is a bipolar signal, i.e., its second half is antisymmetric to its first half relative to its mid-point.
The normalization factor: ${\rm norm} = - \sqrt{7} \, (7/6)^3$. The second reference is the M-N-M multicycle
time signal \cite{ziolkowski2026time-domain-hda} that also turns on at $t=0$ and is based on a windowed sinusoidal time history.
It reaches unity after turning on smoothly in $M$ cycles, remains on for $N$ cycles, and then turns off
smoothly in $M$ cycles. Let the sinusoidal frequency be $f_0 = \omega_0 / (2 \pi)$.
The corresponding period is $T_p = 1/f_0$. Yet another ZAP will be introduced for the focusing application.

The arrays have been constructed with multipole-needle elements whose $\Nmax$ = 10; and with the taper weight
$W_q = 1.0$, they take advantage of the natural $1/\rho$ tapering from the geometry.
Then the far-field frequency domain summation of the fields radiated by each array element yields the total electric field
at each frequency:
\begin{IEEEeqnarray}{rLc} \label{eq:freq-domain-far-field-of-pulse-driven-array}
{\vec E}^{\rm ff}_{array} (r, \theta, \phi, \omega) &=& \left[\, \sum_{q,p} \> W_q \> e^{jk_0 {\hat r} \cdot {\vec r}_{qp}}
\> E_q(\omega) \,\right] \>\> {\vec E}^{\rm ff}_{elem}(r, \theta, \phi, \omega)
= j \, k_0^2 \> AF(\theta, \phi, \omega) \> F(\omega) \>\> {\vec E}^{\rm ff}_{elem}(r, \theta, \phi, \omega)
\nonumber \hfill \\
\noalign{\vskip 0.1in}
&=& \left[\, \sum_{q,p} \> e^{jk_0 {\hat r} \cdot {\vec r}_{qp}} \,\right] \>\>
\left[\, j \> \frac{\omega}{c} \> F(\omega) \> \> \frac{e^{-jk_0r}}{4 \pi r} \>\>
\overrightarrow{{\cal MEP}}_{\Nmax}(\theta, \phi) \,\right]                                                     \hfill
\end{IEEEeqnarray}
\noindent where
\begin{IEEEeqnarray}{c}\label{eq:vector-MEP}
\overrightarrow{{\cal MEP}}_{\Nmax}(\theta, \phi) = -\left[\, \cos\phi \> {\hat \theta} - \sin\phi \> {\hat \phi} \,\right] \>\> \sum_{n=1}^{\Nmax} \>\> \frac{2n+1}{n(n+1)} \left[\> \Pi_n(\theta) + T_n(\theta) \>\right]
\end{IEEEeqnarray}
\vskip 0.1in

\noindent is the vector multipole-element pattern of the electric fields radiated into the far field by the
$\Nmax$ multipole-needle element.
Recall several demonstrated facts.
The pattern's sum yields $\Dmax = \Nmax^2 + 2 \, \Nmax$ = 120 at $\theta = 0$, i.e., its on-axis value
{\it is} the element's maximum directivity.
The pattern's sum vanishes identically term-by-term at $\theta = \pi$, i.e., the FTBR of each element
and, hence, the array is {\it infinite}.
Furthermore, (\ref{eq:vector-MEP}) is frequency-independent and its vector component equals ${\hat x}$ along the z-axis.
Finally, since all of the phase terms in the array factor equal zero on the z-axis, it equals $K$,
the number of array elements.
Therefore, taking the Fourier transform of (\ref{eq:freq-domain-far-field-of-pulse-driven-array}),
again remembering that the $k_0 = \omega/c$ in (\ref{eq:current-element-amp}) carries $1/c$ with it,
the pulsed-needle beam's electric field signal that is radiated from an array and received at a far-field point
on the z-axis is the time derivative of the excitation ZAP evaluated at the retarded time $(t-z/c)$
with an amplitude that decays as $1/z$ as that distance from the array increases, i.e.,
\begin{IEEEeqnarray}{c}\label{eq:time-domain-E-field}
{\vec E}^{\rm ff}_{array} (x=0, y=0, z, t) = K \> \Dmax \>\> \frac{1}{4\pi \, z} \>\> \partial_{ct} \, f_{\rm ZAP}(t-z/c)
\>\> {\hat x}
\end{IEEEeqnarray}
\vskip 0.1in

\noindent where $\partial_{ct} \equiv (1/c) \, \partial_t$ denotes the light-time derivative in space-time.
It is instantly recognized that the SDEU properties do directly carry over to the time domain and that
the time-derivative feature appears naturally in the far field.
Furthermore, it is immediately evident that the intensity of the electric field along the z-axis
is indeed proportional to $K^2$ and, moreover, to $\Dmax^2$.
%
%
\begin{figure}[h!]
	\centering
\includegraphics[width=0.95\columnwidth]{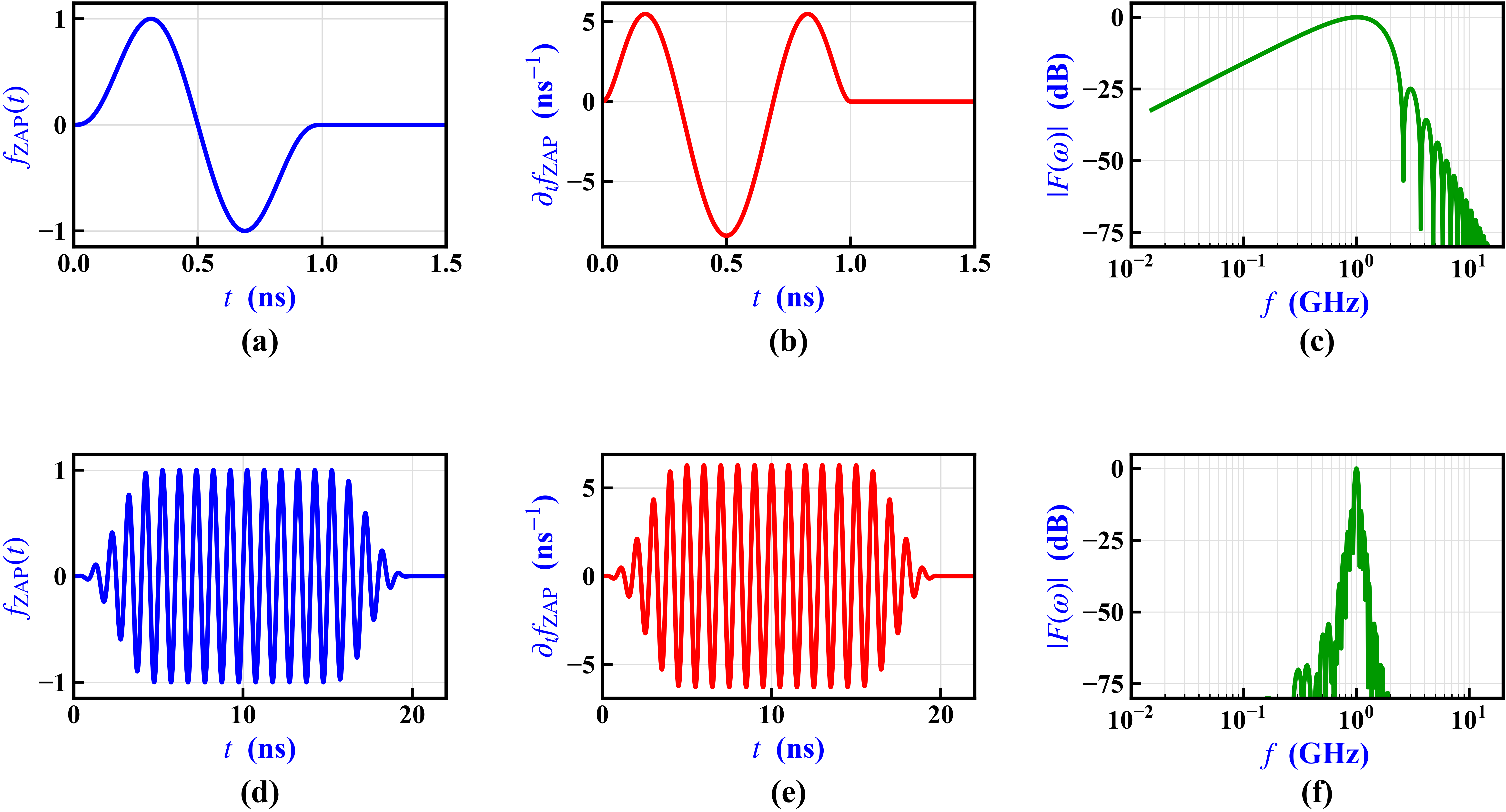}
\caption{The pulses exciting the arrays (blue), their time derivatives (red), and their
Fourier spectra (green, normalized to their own peaks).
(a)--(c) Broad bandwidth 1ns single-cycle ZAP; its spectral peak is at 1.022~GHz.
(d)--(f) Narrow-bandwidth 5-10-5 multicycle ZAP with each cycle having a 1ns period; its
spectral peak is at 1.007~GHz.}
\label{fig:Fig9}
\end{figure}
%
%
%
%
\begin{figure}[h!]
\centering
\includegraphics[width=0.90\columnwidth]{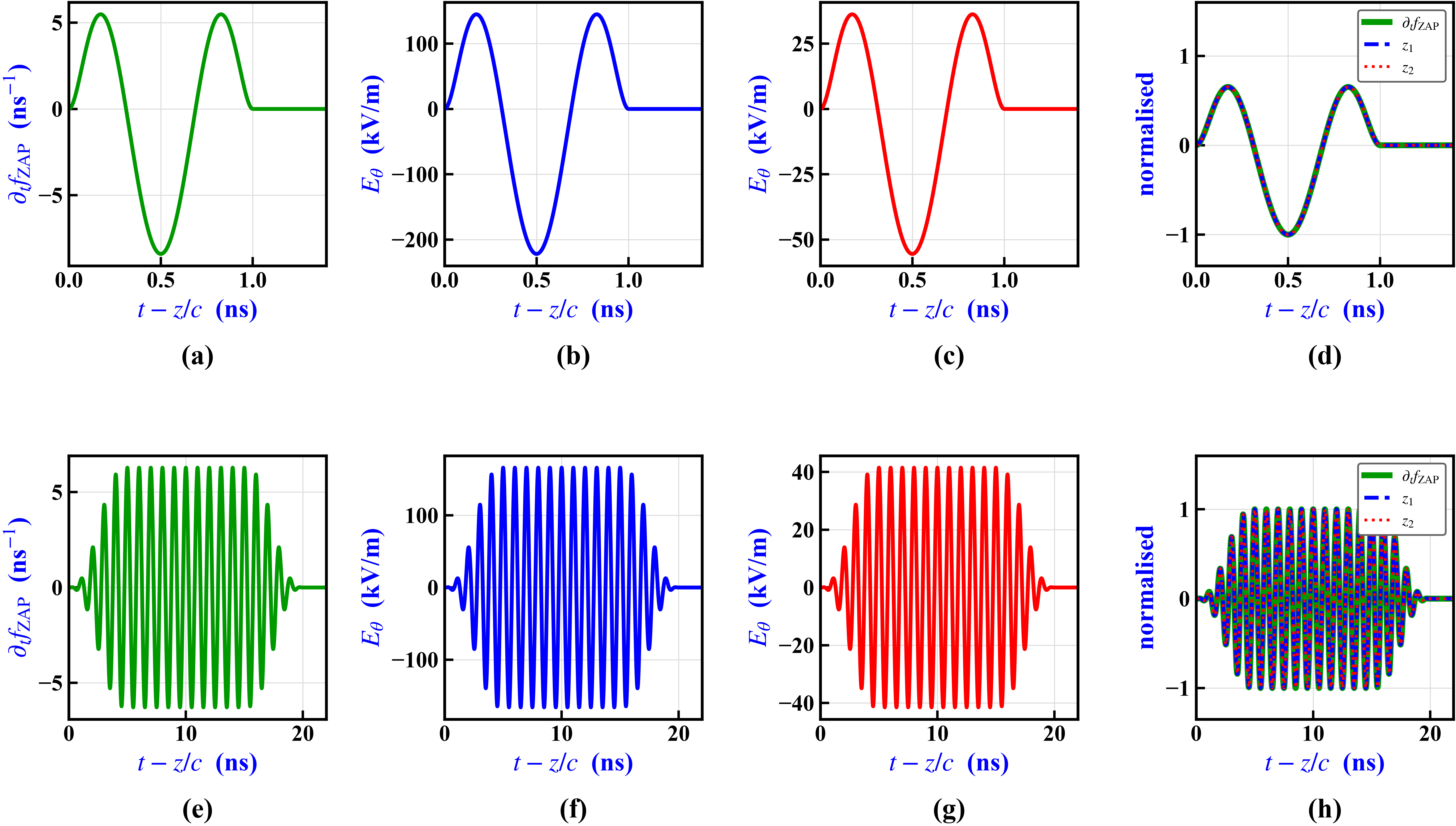}
\caption{On-axis far-field signals radiated by a $K = 33$ array excited with
the 1ns single-cycle ZAP (a--d) and with the 1ns period, 5-10-5 multicycle ZAP (e--h).
Within each row: (a), (e) the exact time derivative of the exciting ZAP (green);
(b), (f) the signal measured at $z_1 = 50 \, \lambda$ (blue);
(c), (g) the signal measured at $z_2 = 200 \, \lambda$ (red);
(d), (h) the three curves overlaid after normalization to their own peaks, where they agree to
$3.9\times10^{-16}$.}
\label{fig:Fig10}
\end{figure}
%

The arrays were excited with the broad bandwidth 1ns single-cycle ZAP and the narrow-bandwidth 5-10-5 multicycle
ZAP with each cycle having a 1ns period. The signals, their time derivatives, and their frequency spectra are
illustrated in Fig.~\ref{fig:Fig9}.
The signals measured on the z-axis at $z_1 = 50 \, \lambda$ and $z_2 = 200 \, \lambda$ that were
radiated by the $K = 33$ arrays when they were excited by those two ZAPs are shown, respectively, in
Fig.~\ref{fig:Fig10}. Because $AF(\theta = 0, \phi = 0) = K$ is independent of both frequency and
$r_{\rm array}$, the three compact arrays radiate \emph{identical} on-axis waveforms.
Thus, the single pair of panels in the figure covers all of them.
They are proportional to the time-derivative of those ZAPs (at least this is very
clear from the single-cycle subplots).
The $1/z$ decay is also clearly illustrated.
The peak of the electric field corresponding to the single-cycle ZAP (multi-cycle ZAP) excitation
at $z_2$, 55.5 kV/m, (41.5 kV/m) is 4 times smaller than it is at $z_1$, 222.0 kV/m (166.0 kV/m).

\begin{figure}[h!]
	\centering
\includegraphics[width=0.98\columnwidth]{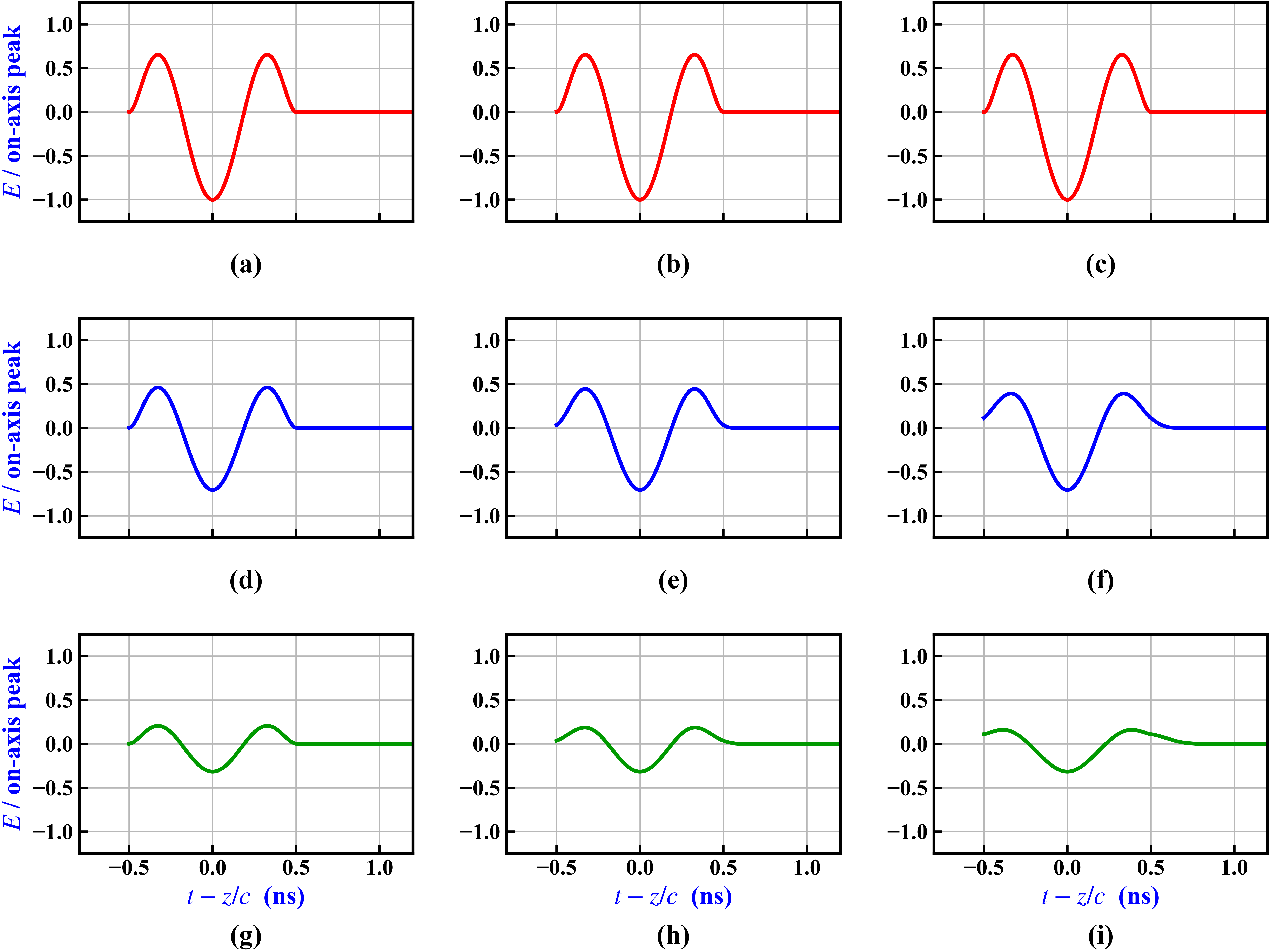}
\caption{Far-field signals generated by exciting the
$0.1 \, \lambda$ (A -- column 1: a, d, g), $0.564 \, \lambda$ (B -- column 2: b, e, h), and
$2 \, \lambda$ (C -- column 3: c, f, i) arrays with the 1ns single-cycle ZAP.
Row 1 (a--c): on-axis point (red).
Row 2 (d--f): $-3$~dB off-axis point (blue), at $8.47^{\circ}$, $7.89^{\circ}$ and $4.86^{\circ}$.
Row 3 (g--i): $-10$~dB off-axis point (green), at $14.30^{\circ}$, $13.59^{\circ}$ and $8.85^{\circ}$.
All nine panels share one vertical scale, normalized to the on-axis peak, so that both the
progressive loss of peak amplitude and the temporal stretching of the pulse can be read directly.}
\label{fig:Fig11}
\end{figure}

Because the pulsed needle beams, like any beam radiated by a finite source, will begin to spread at an angle
corresponding to the aperture size of the source and to the frequencies exciting it, their pulses away from
the beam axis will experience both diffraction {\it and} dispersion effects.
This behavior is illustrated in Fig.~\ref{fig:Fig11} for the signals
within the needle beam radiated by the $0.1 \, \lambda$, $0.564 \, \lambda$, and $2 \, \lambda$
arrays excited by the 1ns single-cycle ZAP. The signals measured on-axis, at the -3dB cone edge, and the
-10dB cone edge for each of those needle beams are presented.
While the time derivative structure is clear, the pulses are stretching and their peak values are
decreasing as the distance from the beam axis increases. This behavior becomes more clear by viewing
the intensity distribution of the pulsed beam as it is propagating away from the array that generated it.
%
%
%
%

\begin{figure}[h!]
\centering
\includegraphics[width=0.90\columnwidth]{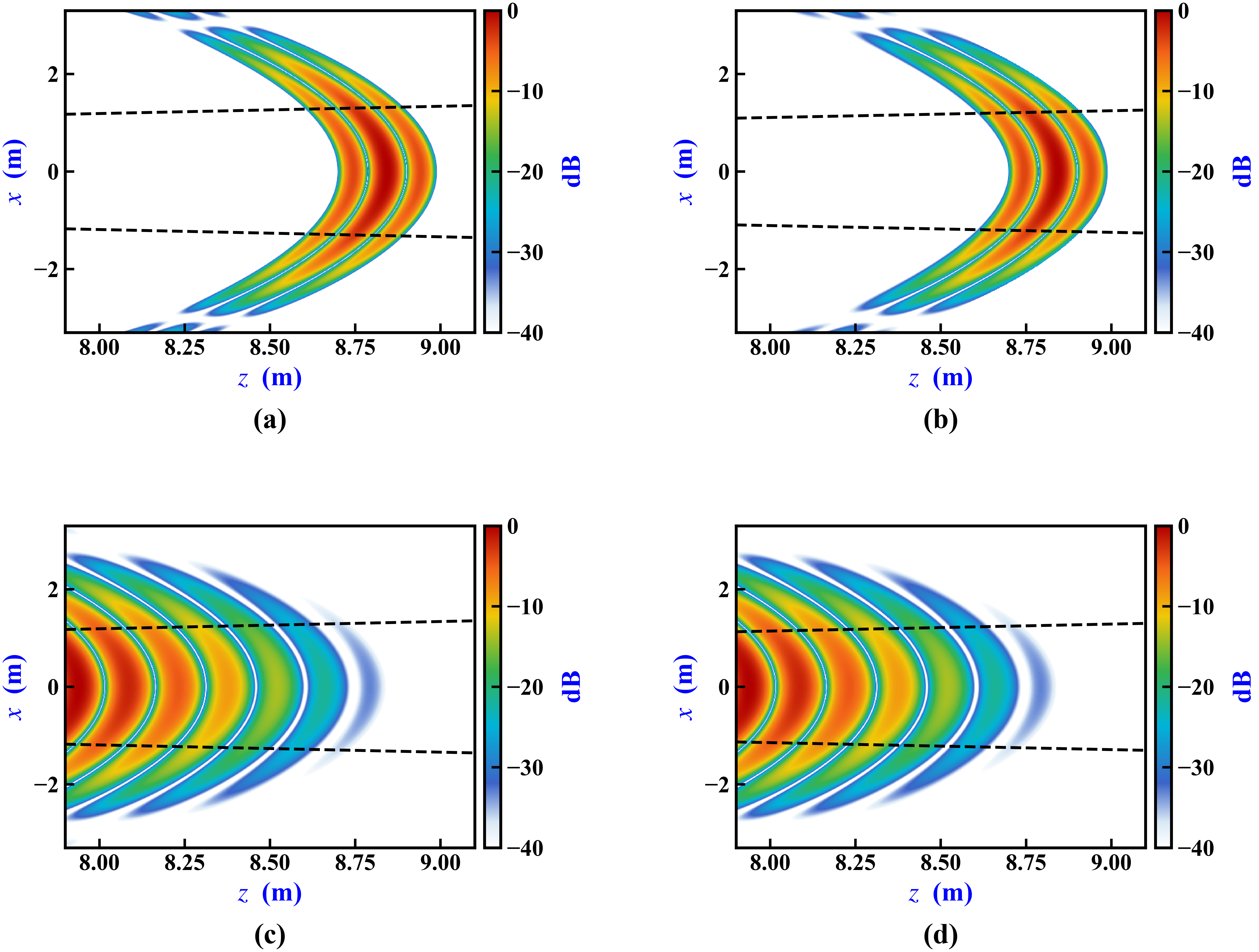}
\caption{Intensity distribution snapshots at t = 30ns in the zx-plane of the electric fields of the pulsed needle beams
excited by the 1ns single-cycle ZAP (a, b) and by the 1ns period, 5-10-5 multicycle ZAP (c, d).
The left column, (a) and (c), is the $0.1 \, \lambda$ array; the right column, (b) and (d), is the
$0.564 \, \lambda$ array. The intensity is normalized to the peak of each snapshot and is
displayed over a 40~dB window; the black dashed lines mark the $-3$~dB cone edges.}
\label{fig:Fig12}
\end{figure}

Once the fields radiated by the different sized arrays have reached their far-field regions, the widths
of their needle beams exhibit spreading. The intensity distributions of the electric field radiated by
the $0.1 \, \lambda$ and $0.564 \, \lambda$ arrays at t = 30ns are presented in Fig.~\ref{fig:Fig12}.
The 1ns single-cycle ZAP results are shown in Figs.~\ref{fig:Fig12}(a) and (b).
The -3dB cone angles defining the spread of those needle beams are, respectively, 8.47$^{\circ}$ and 7.89$^{\circ}$;
those cone edges are marked by the black dashed lines. Again, the time-derivative aspect of those radiated fields is clearly visible. The associated -10dB cone angles were, respectively, 14.30$^{\circ}$ and 13.59$^{\circ}$.
The corresponding $2 \, \lambda$ array -3dB and -10dB cone angle results were 4.86$^{\circ}$ and 8.85$^{\circ}$.
Clearly, as expected, these cone angles decrease as the array size increases.

Similarly, the 1ns period, 5-10-5 multicycle ZAP results are shown in Figs.~\ref{fig:Fig12}(c) and (d).
The -3dB cone angles defining the spread of those needle beams are, respectively, 8.47$^{\circ}$ and 8.15$^{\circ}$;
those cone edges are also marked by the black dashed lines.
In contrast to the single-cycle results, it is not possible to distinguish the time-derivative feature because
of the many cycles present. On the other hand, because the multicycle ZAP has a much narrower bandwidth,
the cone angles associated with the two arrays are not significantly different despite the arrays being
quite different in size. In contrast, the broader bandwidth single-cycle ZAP which, as a consequence, contains
many high frequencies, produced a much narrower beamwidth from the larger array.
%
%
%

\subsection{\textcolor{blue}{Properties of Intensity (Power)-Beams Versus Energy-Beams}}
\vskip 0.1in

The energy distribution of a needle beam follows from the time integral over all time of its intensity distribution.
One might expect then that their profiles may have different rates of spread.
This question was explored in \cite{ziolkowski_pulsed_gaussian_1992}.
The fields arising from a broad bandwidth, unipolar Gaussian pulse excitation of a circular aperture were considered analytically and numerically. It was demonstrated that the intensity (power) beam was considerably narrower than
the energy beam.

Exciting an array with a ZAP localizes both the intensity and energy distributions in time.
One would expect that the intensity and energy beams radiated by exciting a UCA with a narrow-band continuous wave (CW) signal or a large time-length (many cycles) multicycle ZAP would show little difference.
The single-cycle ZAP time-domain results already presented have shown that the intensity distribution becomes narrower as the UCA size increases. Larger apertures facilitate narrower intensity beams for power transfer applications.
How does a single-cycle ZAP energy needle beam behave in comparison to its intensity version?
The question is important since information transfer is known to be fundamentally a process of energy transfer and its transformation, not simply power transfer.

The distinction is the practical one for a pulsed link. A receiver that samples the field
at one instant
-- for instance, a peak detector, a threshold circuit, or a wireless-power rectifier -- responds
to the intensity beam and, therefore, to $\omega_{\rm peak}$.
A receiver that integrates over the whole pulse -- a matched filter, an energy detector, or the correlator at the heart of any
impulse-radio or ultra-wideband demodulator -- responds to the energy beam and therefore to
$\langle \omega^2 \rangle$.
The two need not have the same footprint, and the results below show that for a bipolar ZAP they do not.
Consequently, the angular sector over which a pulsed needle beam actually delivers usable \emph{information} is not, in general, the sector over which it delivers peak \emph{power}. Importantly, a system designer must then
decide which of the two the link budget is written against.

It was found numerically -- and very unexpectedly given the unipolar results -- that the energy needle beam from
a single-cycle ZAP-excited array is narrower than the corresponding intensity beam. Moreover, certain features of
the ZAP impacted the magnitude of enhancement.
The electric field result (\ref{eq:time-domain-E-field}) on the z-axis is separable into its signal and $1/r$ decay
term. Thus, the -3 dB locus of either the intensity or energy needle beam in the far-field region of the array
is a pure cone, and the corresponding cross-section at $z$ in the zx-plane
is given by $x_{-3~dB} = z \, \tan(\theta_{-3~dB})$.

The array acts on the pulse, at points away from the beam axis, as a \emph{linear filter} with
$H(\theta,\omega) = AF(\theta,\omega)/AF(0,\omega)$. Hence, its impulse response
$h(t)$ is the arrival-time distribution across the projected aperture: a non-negative,
unit-area kernel of width $2 r_{\rm array} \, \sin\theta_{\rm spread}/c$.
The off-axis waveform is therefore a pure convolutional \emph{smoothing} of the on-axis one,
$y = h * x$, where $x = \partial_t \, f_{\rm ZAP}$, with $\partial_t$ denoting a time derivative,
is the \emph{radiated} on-axis waveform.
The curvature of that signal at its own peak, $\omega_{\rm ZAP, peak}$, and its mean-square bandwidth,
$\langle\, \omega_{\rm ZAP}^2 \,\rangle$, are given by the expressions:

\begin{IEEEeqnarray}{rLc}
\omega_{\rm ZAP, peak}^2 &=& \left|\, \frac{\partial_t^2 \, x(t_{\rm ZAP, peak})}{x(t_{\rm ZAP, peak})} \,\right|  \hfill \\
\noalign{\vskip 0.1in}
\langle\, \omega_{\rm ZAP}^2 \,\rangle &=& \frac{\int_{-\infty}^{+\infty} \left|\, \partial_t \, x \, \right|^2 \> dt}
{\int_{-\infty}^{+\infty} x^2 \> dt}  \hfill
\end{IEEEeqnarray}
\vskip 0.1in

\noindent For a kernel of variance $\sigma^2$, the filter response is
$y \approx x + \frac{\sigma^2}{2} \, \partial_t^2 \, x$.
\noindent It was determined that the roll-off away from the beam-axis of the power-beam, $P_{\rm ZAP}(\theta)$,
and energy-beam, $U_{\rm ZAP}(\theta)$, occur at different rates:

\begin{IEEEeqnarray}{rLc}
\frac{P_{\rm ZAP}(\theta)}{P_{\rm ZAP}(0)} &\approx& 1-\sigma^2 \> \omega_{\rm ZAP, peak}^2                   \hfill    \\
\noalign{\vskip 0.1in}
\frac{U_{\rm ZAP}(\theta)}{U_{\rm ZAP}(0)} &\approx& 1-\sigma^2 \> \langle\, \omega_{\rm ZAP}^2 \,\rangle     \hfill
\end{IEEEeqnarray}
\vskip 0.1in

\noindent Therefore, the energy needle beam is narrower than the intensity (power) needle beam exactly when
$\langle\, \omega^2 \,\rangle > \omega_{\rm peak}^2$, i.e., when the pulse carries more high-frequency structure
overall than the sharpness of its own peak. \emph{This outcome is thus a property of the ZAP itself, not of the array}.
Equivalently, the effective duration of the ZAP is

\begin{IEEEeqnarray}{c}
T_{\rm eff}(\theta)/T_{\rm eff}(0) \approx 1-\sigma^2(\langle\, \omega_{\rm ZAP}^2 \,\rangle-\omega_{\rm ZAP, peak}^2)
\end{IEEEeqnarray}
\vskip 0.1in

\noindent Thus, the energy $U(\theta) \approx P(\theta) \cdot T_{\rm eff}$.
While a unipolar pulse's dispersive stretching would widen the energy beam,
it simply does not for the bipolar ZAP. The single-cycle ZAP's energy-beam is narrower than its power-beam.

For the single-cycle ZAP (\ref{eq:single_cycle_ZAP}) set $u = (t-T_L/2)/(T_L/2)$, so that the
radiated waveform is $x \propto \Gamma(u) \equiv \partial_u \, f_{\rm single} = (1-u^2)^2(1-7u^2)$.
Since $\partial_t = (2/T_L) \, \partial_u$, both moments below are then exact rationals in units of
$(2/T_L)^2$.
In closed form one finds
\begin{IEEEeqnarray}{rCl}
\omega_{\rm ZAP, peak}^2 &=& \big|\, \partial_u^2 \, \Gamma(0)/\Gamma(0) \,\big| \;=\; 18,
\qquad\qquad
\langle\, \omega_{\rm ZAP}^2 \,\rangle \;=\; \frac{\int_{-1}^{+1} \left(\, \partial_u \, \Gamma \,\right)^2 \, du}
                                                {\int_{-1}^{+1} \Gamma^2 \, du} \;=\; \frac{65}{3} \hfill \nonumber  \\
\noalign{\vskip 0.1in}
&& \hspace{1.0in} \frac{\langle\, \omega_{\rm ZAP}^2 \,\rangle}{\omega_{\rm ZAP, peak}^2} = \frac{65}{54} = 1.2037    \hfill
\end{IEEEeqnarray}
\vskip 0.1in

\noindent This confirms analytically that the energy-beam generated by the single-cycle ZAP is narrower than its power-beam.
For a quick comparison, consider first a Gaussian signal (GS) drive, $f_{\rm GS}(t) = e^{-t^2/\tau^2}$,
that is deliberately \emph{not} a ZAP. The waveform it radiates is the Gaussian monocycle
$x = \partial_t \, f_{\rm GS}$, for which one finds
$\omega_{\rm GS, peak}^2 = 4/\tau^2$ and $\langle\, \omega_{\rm GS}^2 \,\rangle = 3/\tau^2$.
This ratio is exactly $3/4$ -- and the ordering reverses, i.e., as found in \cite{ziolkowski_pulsed_gaussian_1992},
the power-beam is narrower than the energy-beam.
The reversal does not depend on that differentiation. Applied to the undifferentiated
Gaussian itself, for which $\omega_{\rm peak}^2 = 2/\tau^2$ and
$\langle\,\omega^2\,\rangle = 1/\tau^2$, the criterion gives the smaller ratio $1/2$ and hence
the same conclusion.
On the other hand, as expected, for a narrowband pulse with peak frequency $\omega_0$, one has $\langle\, \omega^2 \,\rangle \approx \omega_0^2 \approx \omega_{\rm peak}^2$ and the two beams coincide.

These results were confirmed numerically using the $0.564 \, \lambda$ array.
In each case the ratio $\langle\, \omega^2 \,\rangle/\omega_{\rm peak}^2$ was evaluated
directly from the sampled radiated waveform, and each value was converged in the time step.
The ratio was 0.750 for the Gaussian case -- the analytical $3/4$ --
and the energy-beam was indeed wider than the power-beam. With the 1ns period, 5-10-5 multicycle ZAP,
the ratio was 1.005, and the two beams were indistinguishable. With the 1ns single-cycle ZAP, the ratio was 1.2037 -- the
analytical $65/54$ to five significant figures -- and the energy-beam was confirmed to be narrower than the power-beam.

It is noted that the $65/54$ ratio is universal for the 1ns single-cycle ZAP.
Its time-length is $T_L =$ 1.0~ns. Its spectral peak is at 1.02~GHz. Its
-3~dB band is 0.50--1.59~GHz and its -20~dB band is 0.076--2.37~GHz.
What varies between the spreading cone angles of the arrays is \emph{how much of the $-3$~dB roll-off is
supplied by the array factor}. The multipole-element does essentially all of the beamforming for the
$0.1 \, \lambda$ array while its array factor is basically frequency-flat and contributes nothing to
the separation between the spreading of the power- and energy-beams, i.e., their spread differs by 0.05\%.
It increases with the larger arrays; the spreading differs by 1.2\% and 3.3\%, respectively, for the
$0.564 \, \lambda$ and $2  \, \lambda$ arrays.
The two statements are complementary rather than contradictory. The \emph{sign} of the
separation -- which of the two beams is the narrower one -- is fixed entirely by the pulse
through $\langle\omega^2\rangle$ versus $\omega_{\rm peak}^2$ and cannot be altered by the array.
The \emph{size} of that separation is fixed by the aperture, because it is the array factor that
supplies the dispersive smoothing kernel $h(t)$ whose variance $\sigma^2$ multiplies both
frequency moments.


\subsection{\textcolor{blue}{Focusing in the Time Domain with an Array of Superdirective Elements}}
\vskip 0.1in

Focusing of pulsed Gaussian beams with a lens was analytically and numerically considered at optical frequencies in
\cite{ziolkowski1994designer,davidson1994josaA}. Substantial intensity enhancement at the focal point was
obtained with a custom-design ZAP termed the intensity enhancement (IE) pulse.
It will be demonstrated herein that a similar result can be obtained at microwave frequencies with focusing
UCAs {\it and} a planar needle-element-based focusing linear uniform array (ULA).
Furthermore, since focusing from a conventional array is a near-field phenomena, the larger
$2 \, \lambda$ and $6 \, \lambda$ arrays gave an adequate region in which to focus pulse-shaped waveforms.
They will be used for this discussion.
It is important to note that while far-field phenomena have been the strong emphasis until this section,
there are many important applications of near-field focusing where evanescent waves and high spatial frequencies are captured before they decay. They include subwavelength imaging; microwave equivalents of near-field scanning optical microscopy (NSOM), i.e., super-resolution imaging; and localized bio-sensing.

While the original goal was to demonstrate that multipoles would also enhance focal region field levels, it was
found that this is not the case. Pulse-driven multipole-element arrays deliver the fields more effectively from
each of their elements, but the enhancement of the focal region intensity is determined solely by the pulse shape.
Comparisons are therefore made below between the IE-ZAP, multicycle ZAP, and continuous wave (CW) excitations
of the arrays.
Since UCAs of Huygens dipole elements have demonstrated superdirective performance \cite{ziolkowski-2024-EM-Science}
and higher-order multipoles contribute little to the desired effect, the arrays are assumed to be constructed with
$\Nmax$ = 1 elements. The $6 \, \lambda$ array with its rings of uniform element density effectively produced
very smooth electric field distributions over its entire focal region. In comparison, those of the
fixed-$\xi$ $2 \, \lambda$ array display characteristics associated with its geometry.
However, since the overall configurations were the same and the larger array did not impact the IE for the
comparable geometry, the $2 \, \lambda$ array results are emphasized below.

To achieve focusing from a planar array, an appropriate phase must be introduced at each element to mimic a lens.
The VSWE coefficients for the array elements are:
\begin{IEEEeqnarray}{rLc} \label{eq:array-coefficients-focus}
A_{nm,q} (\omega) &=& w_{q}(\omega) \> a_{nm}
\hfill \\
\noalign{\vskip 0.1in}
B_{nm,q} (\omega) &=& w_{q} (\omega) \> b_{nm}
\hfill
\end{IEEEeqnarray}
\vskip 0.1in
\noindent where $a_{nm}$ and $b_{nm}$ are, of course, the elemental source coefficients and the focusing weights are:

\begin{IEEEeqnarray}{rLc}
w_{q}(\omega) &=& W_q \> e^{+jk_0 \, \left(\, \sqrt{r_q^2 + z_f^2} - z_f \,\right)} \> E_q(\omega)   \hfill  \nonumber \\
\noalign{\vskip 0.1in}
&=& j \> \frac{\omega}{c} \> k_0 \> W_q \> e^{+jk_0 \, \left(\, \sqrt{r_q^2 + z_f^2} - z_f \,\right)} \> F(\omega)
\end{IEEEeqnarray}
\vskip 0.1in

\noindent where $z_f$ is the target focal point along the z-axis. The beam radiated by the array remains azimuthally symmetric. The exponential term accounts for beam focusing at the specified z-distance, $z_f$.
Furthermore, no additional taper across the aperture of the array was considered, i.e., $W_q$ = 1.0.
As already noted, in contrast to the far-field efforts, the focal point will essentially be in the near/mid-range regions of the array. The total electric field is again obtained as the straightforward superposition of the full VSWE radiated fields from each array element.

The IE-ZAP was custom-designed to be a $C^2$ function borrowing from aspects of the smooth window portion of the
 multicycle ZAPs and the large time derivative aspect of the $UWB_1$ ZAP employed in \cite{ziolkowski1994designer,davidson1994josaA}. It is defined as:

\begin{IEEEeqnarray}{c}
f_{\rm IE-ZAP}(t)=
\begin{cases}
0                                                & \text{ for }  \quad t \le 0                                 \\
-( 10.0 \, x_1^3 - 15 \, x_1^4 + 6.0 \, x_1^5 )  & \text{ with } \quad x_1 = t / T_1
                  \quad \text{ for }  0 \le t \le T_1                                                          \\
-1 + 2 \, (35 \, x_2^4 - 84 \, x_2^5 + 70 \, x_2^6 - 20 \, x_2^7) & \text{ with }
                  \quad x_2 = \frac{(t-T_1)}{(T_2-T_1)} \quad \text{ for } T_1 \le t \le T_2                   \\
1 - ( 10.0 \, x_3^3 - 15 \, x_3^4 + 6.0 \, x_3^5 )    & \text{ with }
                  \quad x_3 = \frac{t-T_2}{T_3-T_2}  \quad \text{ for }  T_2 \le t \le T_3                     \\
0                                                & \text{ for }  \quad t \ge T_3
\end{cases}
\end{IEEEeqnarray}
\vskip 0.1in

\noindent While the original $UWB_1$ ZAP was also a three region function, it was only $C^1$ at the times
$T_1$ and $T_2$ when it transitioned from Region 1 into Region 2 and from Region 2 into Region 3, respectively.
The increase in smoothness of this IE-ZAP redesign fosters an advantageous, completely symmetric signal within
the middle Region 2.
In fact, the septic smoothstep of Region 2 is actually $C^3$, while the quintic smoothsteps of Regions 1 and 3
are only $C^2$. The overall smoothness is thus limited by the latter -- the wings of the pulse -- not by the spike.
While we opted not to attain $C^3$ overall, the enhancement reported below would actually increase approximately 3\%
if we had.
The time constants were determined with a numerical parameter study to achieve an IE factor of at least 100.
This goal was reached and very nicely surpassed with $\delta = 0.05 \times 10^{-9}$ s = 50 ps,
$T_1 = \zeta \, \delta$, $T_2 = T_1 + \delta$, and $T_3 = T_2 + \zeta \, \delta$, where $\zeta = 66.5$.
Thus, the total time of the pulse was $T_L = T_3 = (2\zeta+1) \, \delta$ = 6.70 ns, and
its spatial length is, of course, $c \, T_L \simeq $ 2.01 m.

\begin{figure}[h!]
\centering
\begin{subfigure}[t]{0.49 \columnwidth}
      \centering
      \includegraphics[width=0.8\columnwidth]{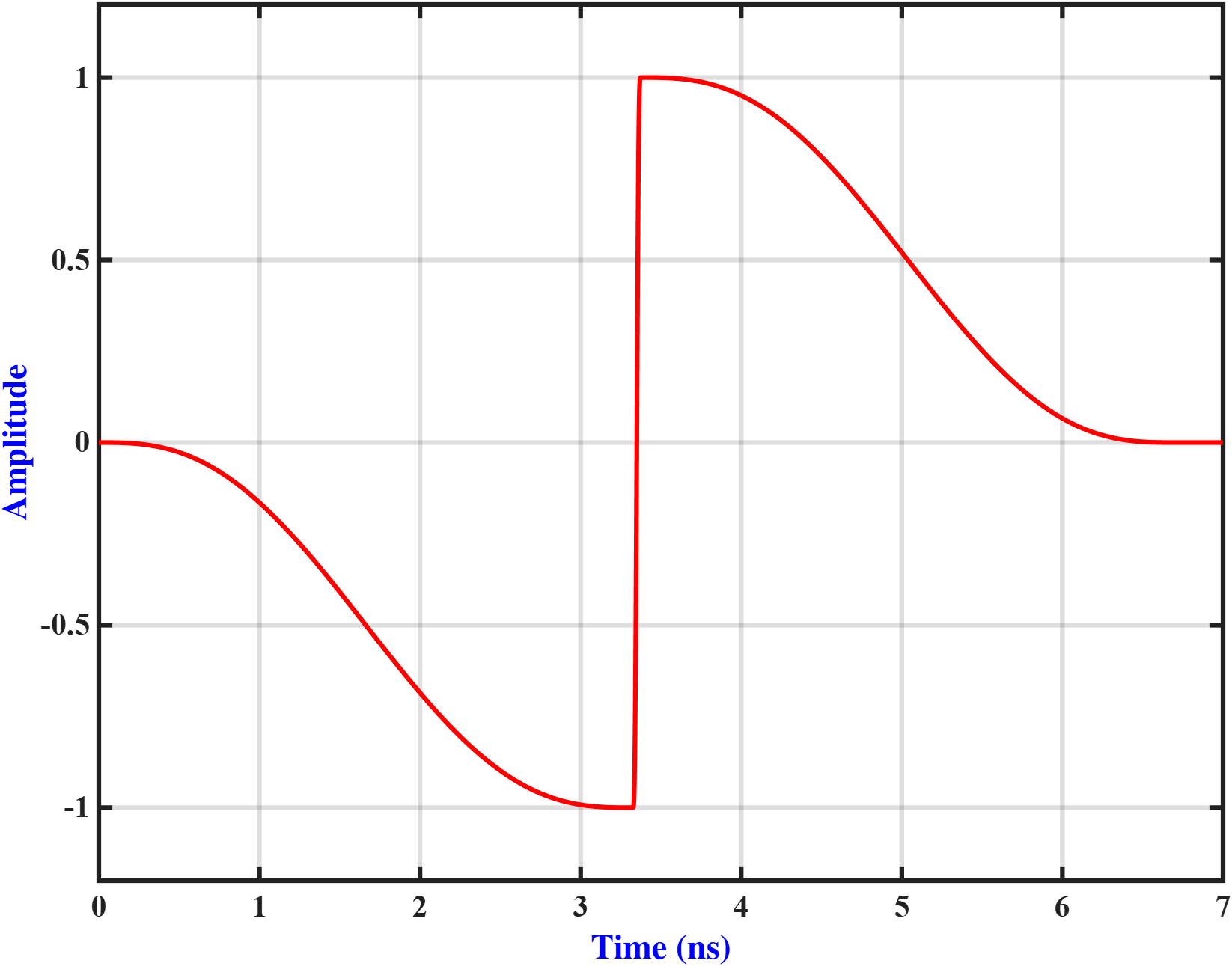}
      \caption{}
\end{subfigure}
\begin{subfigure}[t]{0.49 \columnwidth}
      \centering
      \includegraphics[width=0.8\columnwidth]{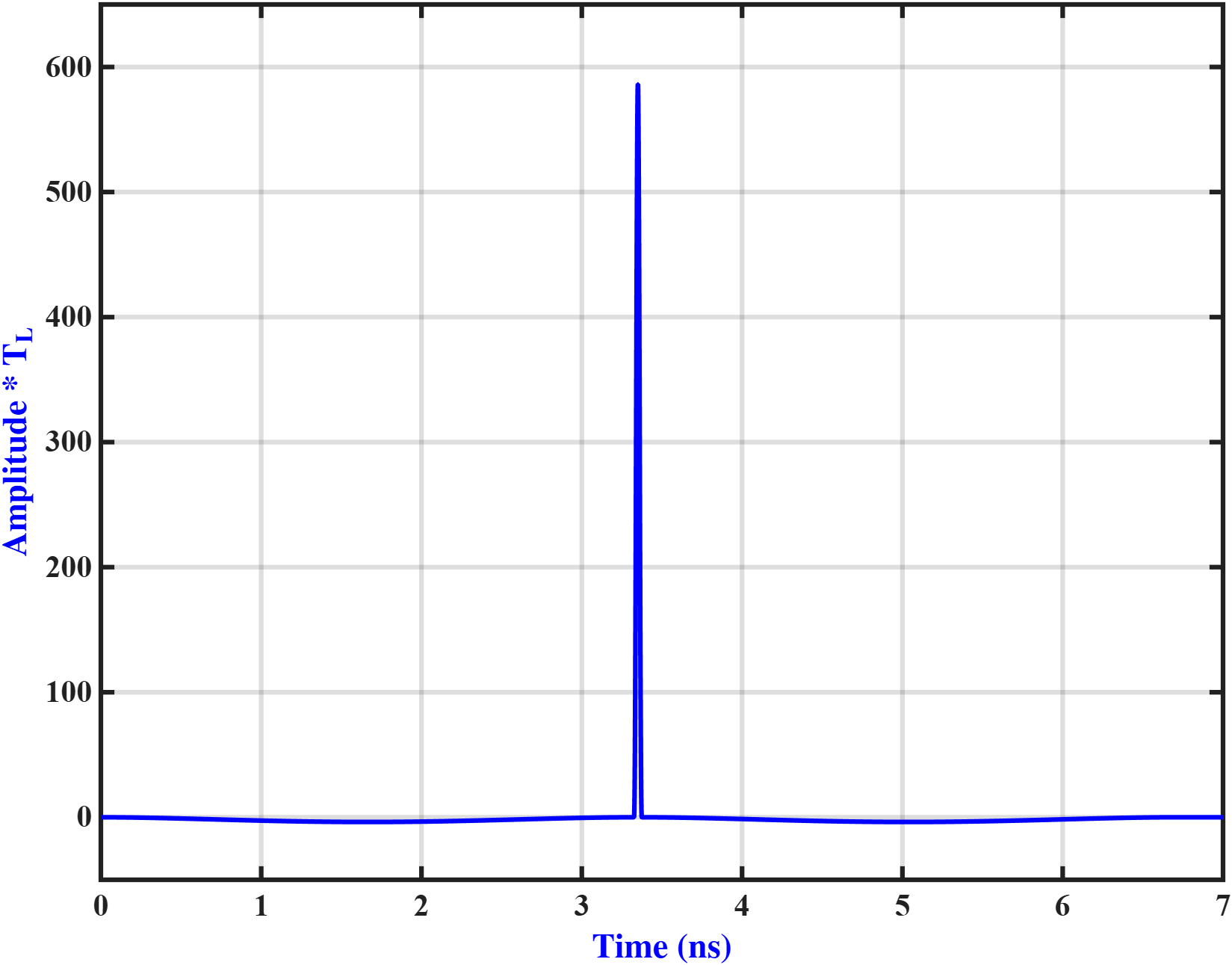}
      \caption{}
\end{subfigure}
\caption{The ZAP pulses employed in the time-domain focal enhancement results.
(a) IE-ZAP and (b) the time derivative of the IE-ZAP.}
\label{fig:Fig13}
\end{figure}

\begin{figure}[h!]
\centering
\begin{subfigure}[t]{0.49 \columnwidth}
      \centering
      \includegraphics[width=0.8\columnwidth]{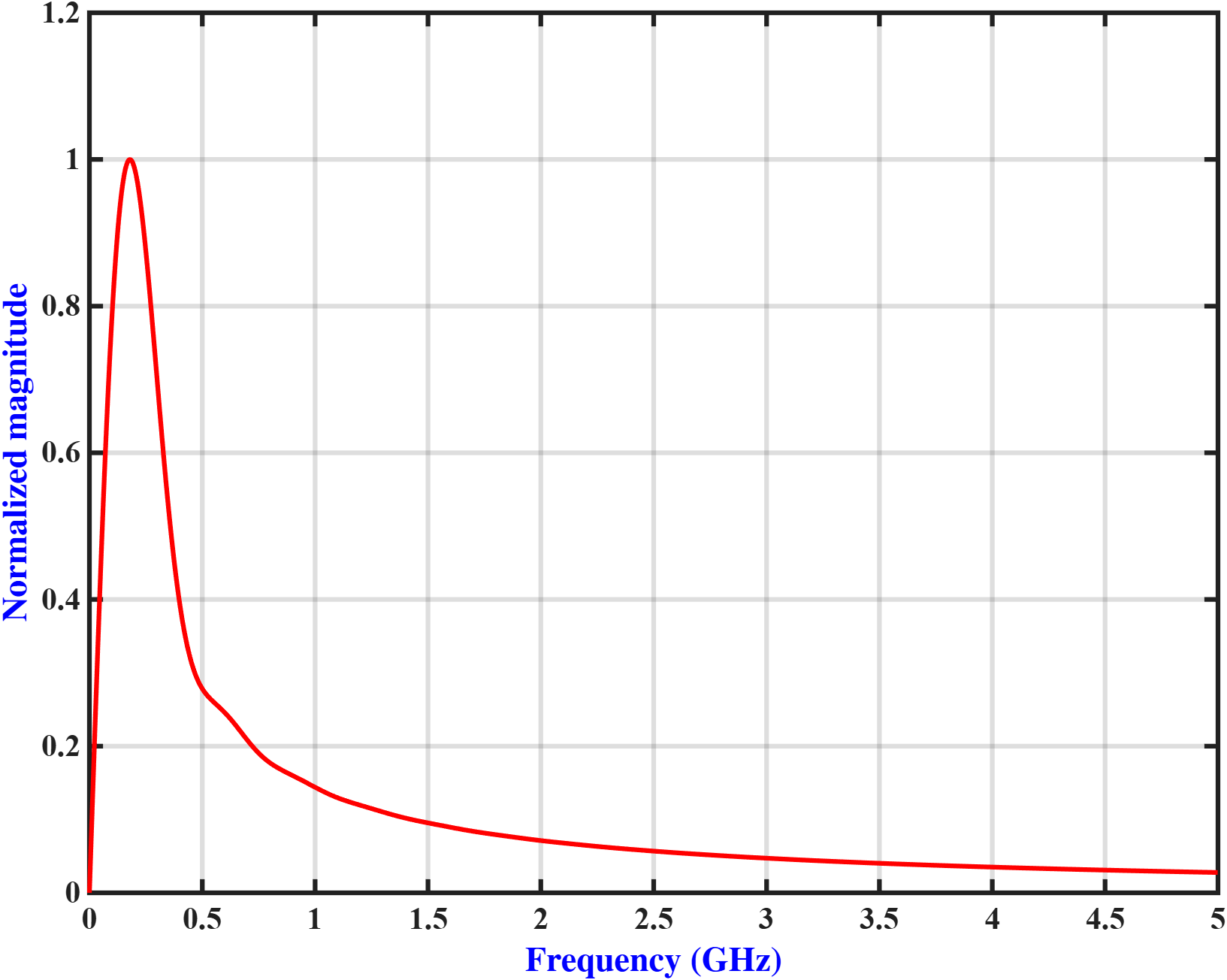}
      \caption{}
\end{subfigure}
\begin{subfigure}[t]{0.49 \columnwidth}
      \centering
      \includegraphics[width=0.8\columnwidth]{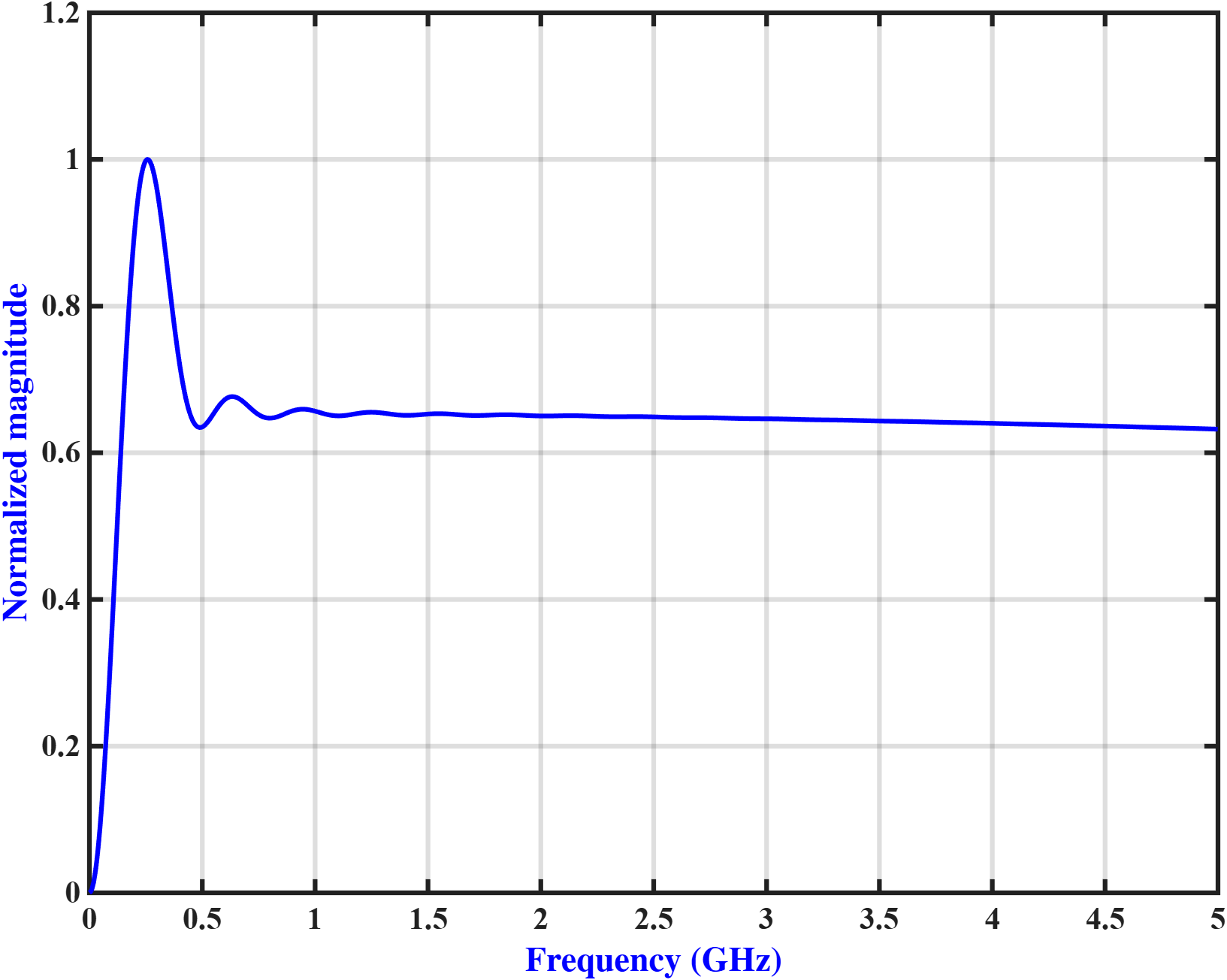}
      \caption{}
\end{subfigure}
\caption{Frequency spectra of the signals in Fig.~\ref{fig:Fig13}.
(a) Spectrum of the IE-ZAP and (b) spectrum of the time derivative of the IE-ZAP.}
\label{fig:Fig14}
\end{figure}

The time history of the IE-ZAP and its derivative are shown in Fig.~\ref{fig:Fig13}. The frequency spectra of both
signals are shown in Fig.~\ref{fig:Fig14}. The smooth appearance of the IE-ZAP is quite a contrast to its time derivative.
The turn-on (Region 1) and turn-off (Region 3) portions of the pulse follow from the window function in the multicycle ZAP.
The spike is a consequence of the fast transition in Region 2 of the IE-ZAP (i.e., its central region).
Its beginning and ending portions were designed specifically to match the end and beginning functions and their derivatives at the transition from Region 1 to Region 2 at $T_1$ and at the transition from Region 2 to Region 3 at $T_2$, respectively.
The abundance of higher frequency components apparent in the derivative's spectrum is directly associated with the derivative spike. The lower frequency contributions arise from the beginning and ending ``wings'' of the signal.
The spike itself is the septic smoothstep of Region 2. Its peak slope is available in closed
form, $\max |\partial_t f_{\rm IE\mbox{-}ZAP}| = 2 \, (35/16)/\delta = 87.5$~ns$^{-1}$, and this
single number, together with the effective frequency of the radiated pulse, accounts for the entire intensity enhancement
reported below.

%
%

%
%

The effective angular frequency, $\omega_{\rm rad}$, of the fields radiated by an array excited with a UWB signal
was determined in \cite{ziolkowski_IAPDA_1992}:

\begin{IEEEeqnarray}{c}
\omega_{\rm rad}^2 = \frac{ \int_{-\infty}^{+\infty} \> \left|\, \partial_t \, f(t) \,\right|^2 \> dt }
{\int_{-\infty}^{+\infty} \> \left|\, f(t) \,\right|^2 \> dt}
\end{IEEEeqnarray}
\vskip 0.1in

\noindent It is identical to the mean-square bandwidth $\langle\, \omega^2 \,\rangle$ introduced in the
intensity-versus-energy beam discussion above. While it was applied there to the radiated waveform, it is applied
here in a similar context to the exciting pulse in relation to the waveform that reaches the focus.
It is repeated here for convenience and for direct comparison with the terms defined in \cite{ziolkowski1994designer,davidson1994josaA}.
In particular, the IE factor at the focus achieved by the time-derivative features of the fields radiated by an array
excited with the IE-ZAP, $f_{\rm IE-ZAP}$, is:

\begin{IEEEeqnarray}{c} \label{eq:IE-factor}
\Upsilon_{\rm int, IE-ZAP} = \frac{\max_t \, \left| \, \partial_t \, f_{\rm IE-ZAP}(t) \,\right|^2}
{\max_t \, \left| \, f_{\rm IE-ZAP}(t) \,\right|^2} \bigg/ \omega_{\rm rad,~IE-ZAP}^2                       \hfill
\end{IEEEeqnarray}
\vskip 0.1in

\noindent where $\max_t$ denotes finding the maximum in time.
It thus compares the squared peak slope of the exciting pulse with the squared peak slope of a sinusoid
having the same effective frequency $\omega_{\rm rad}$ and the same peak amplitude.
As written, it is manifestly dimensionless; it equals unity for that CW
reference; and it is invariant to the amplitude normalization of $f_{\rm IE\mbox{-}ZAP}$.

\noindent The IE factor was originally introduced to quantify and emphasize the enhancement the intensity
attains at the focus as a consequence of the time derivative associated with time-domain radiated fields.

From the IE-ZAP design process it was learned that the independent control of the spike and other features of
its spectrum rests in its {\it shape factor}:

\begin{IEEEeqnarray}{c}\label{eq:kappa}
\kappa \;\equiv\; \Delta\int_{T_1}^{T_2}\Bigl|\, \partial_t \, f_{\rm IE-ZAP} (t) \,\Bigr|^{2} \> dt
\end{IEEEeqnarray}
\vskip 0.1in

\noindent where $\Delta = \delta = T_2 - T_1$. It characterizes the shape of Region 2's spike
in terms of its time extent and its energy.
The total energy content of the IE-ZAP is:
\begin{IEEEeqnarray}{c}\label{eq:total-energy-IE-ZAP}
{\cal E}_{\rm IE-ZAP} = \int_{-\infty}^{+\infty} \> \left| \, f_{\rm IE-ZAP}(t) \,\right|^2 dt
\end{IEEEeqnarray}
\vskip 0.1in

\noindent which is determined almost entirely by its ``wings''. Importantly,
it was found that its effective angular frequency is approximately:
$\omega_{\rm rad,~IE-ZAP}^2 \simeq \kappa / ( {\cal E}_{\rm IE-ZAP} \, \Delta )$,
{\it and} this approximation is 99.35\% accurate.
The smoothstep wings of the pulse carry 98.8\% of its energy, ${\cal E}_{\rm IE-ZAP}$.
The energy of the spike in Region 2, $\kappa/\Delta$, carries 99.3\% of the total derivative energy.
Consequently, these two ``knobs'' are very nearly independent of one another;
the shape factor emphasizes them.
It is bounded below since Region 2 must carry the function $f$ defined over it from $-1$ to $+1$ in the
time $\Delta$. The Cauchy--Schwarz inequality gives
$\kappa = \Delta \int_{T_1}^{T_2} |\partial_t f|^2 dt \ge \left(\int_{T_1}^{T_2} \partial_t f \, dt\right)^2 = 4$,
with equality only for a straight-line transition.
The $f_{\rm IE-ZAP}$ excitation attains $\kappa = 4 \, (140)^2 \, B(7,7) = 6.527$, well above that floor.
The IE at fixed pulse energy and fixed spike width is bestowed by the combination
$\left(\, \Delta \, \max_t |\partial_t f_{\rm IE-ZAP}| \,\right)^2/\kappa$, precisely the ratio of
the peak to the mean power of the derivative within Region 2.
It is emphasized that the floor is not the design target. The enhancement is rewarded by that ratio.
The very same straight-line transition minimizes it rather than maximizing it.
Smoothness in the middle region is therefore profitable.

The septic smoothstep pays a penalty in $\kappa$ (Region 2's time width $\Delta$ times the energy
in the IE-ZAP's derivative there) as it rises 63\% from 4 to 6.527.
Nevertheless, it more importantly raises the square of the normalized peak slope from 4 to
$(35/8)^2$ = 19.14, a factor of 4.785, for a net gain of 19.14/6.53 = 2.93.
This modest value is a multiplier, not an increment.

Consider the same pulse with the same $\delta$, the same $\zeta$, and the same peak amplitude,
but with a straight line in Region 2.
This $C^0$ variant has the effective frequency $\omega_{\rm rad}/2\pi$ = 0.884 GHz. When referenced
with respect to it, its IE is 51.88. Relative to the 1.0 GHz reference, it is only 40.5.
Consequently, the septic smoothstep collects its reward in both currencies.
Graded against each pulse's effective CW sinusoid, it collects a net gain of 2.96
when the two IE factors are evaluated exactly rather than through
the approximation $\omega_{\rm rad}^2 \simeq \kappa/({\cal E}_{\rm IE-ZAP} \, \Delta)$.
This means the septic smoothstep's IE is 2.96 $\times$ 51.88 = 153.61.
Similarly, when graded against the fixed 1.0 GHz reference, its IE collects the
full squared peak-slope factor making its IE = 4.785 $\times$ 40.5 = 193.9.
This outcomes occurs because a ratio taken at a fixed reference frequency never
involves $\omega_{\rm rad}$ and, hence, never involves $\kappa$ at all.
Region 2's septic smoothstep therefore supplies this 79\% enhancement.
Thus, $C^3$ smoothness in Region 2, instead of a straight-line transition, yields a very profitable reward.
Even so, a steep spike and a small $\kappa$ are not independent goals to be traded against one another;
the single figure of merit is their ratio.

The effective frequency of the IE-ZAP was calculated to be $\omega_{\rm rad, IE-ZAP} / 2\pi$ = 1.1236 GHz.
The IE factor $\Upsilon_{\rm int, IE-ZAP}$ was then determined for several of the $\zeta$ values in $f_{\rm IE-ZAP}(t)$ using the full VSWE calculation of the fields and compared to those obtained with the 5-10-5 1ns period, multicycle ZAP (1.00 GHz peak frequency) and a CW signal at 1.124 GHz.
It was found that
$\Upsilon_{\rm int, IE-ZAP} \simeq 2.298 \, \zeta + 0.795$
regardless of the $\Nmax$ elemental multipole element used in any of the UCAs.
This linear law follows directly from (\ref{eq:IE-factor}), i.e., the numerator, $(87.5$~ns$^{-1})^2$,
is set by $\delta$ alone and is independent of $\zeta$, whereas ${\cal E}_{\rm IE-ZAP}$ -- and
hence $1/\omega_{\rm rad}^2$ -- grows linearly with the length $\zeta \delta$ of the wings.
The $\zeta$ = 66.5 choice yields $\Upsilon_{\rm int, IE-ZAP} = 153.61$,
which significantly surpasses the desired target enhancement and is approximately an order of magnitude
larger than the original IE numerical results in \cite{davidson1994josaA}. The reason for the latter
is not a better pulse, but the fact that the microwave time scales treated here make the 50 ps spike
affordable to thoroughly resolve, whereas the computing resources at the time could not.

Because it was determined that the $\Nmax$ = 1 elemental multipole element array was an effective choice,
the full VSWE field comparison results were obtained with the $2 \, \lambda$ array of them.
The focal point was set at $z_f = 8 \, \lambda$. Thus, the f-number (ratio of the focal
distance to the diameter of the aperture) was $f/\# = 8\lambda/(4\lambda) = 2.0$.
The comparisons of the full VSWE calculation of the electric field intensity at the focus of the IE-ZAP generated fields
and of those generated by the multicycle ZAP and CW signal reference cases gave
$\Upsilon_{\rm int, IE-ZAP}$ = 193.9 and 153.61, respectively, the latter being identical to the predicted value.
Thus, the large intensity enhancement produced by an IE-ZAP-excited array at a focal point in its near field at
microwave frequencies is confirmed.

Note that the two IE numbers answer two different questions and both are quoted throughout.
The definition (\ref{eq:IE-factor}) is referred to a CW sinusoid at
$\omega_{\rm rad}/2\pi =$ 1.124 GHz of the same peak amplitude.
It is that comparison which the closed form predicts, and the simulation reproduces as $153.61$.
The larger value, $193.9$, is the realized enhancement measured against the 5-10-5 multicycle ZAP (1.0 GHz),
i.e., against a physically radiatable narrow-band drive rather than against an idealized sinusoid.
It is the larger of the two simply because the multicycle reference itself delivers less at the focus
than an ideal CW drive of the same peak amplitude. Its windowed turn-on and turn-off, as well as its slightly
lower center frequency, work against it.
The time-domain calculations reported below use the 5-20-5 version of that multicycle drive instead because of
timing issues at the focus, and they return the same $193.9$ to four figures.
The insensitivity is itself a result, i.e., both windowed multicycle
ZAPs peak at 1.0 GHz and both are already excellent stand-ins for a CW signal at that frequency.

\begin{figure}[h!]
\centering
\begin{subfigure}[t]{0.49 \columnwidth}
      \includegraphics[width=0.8\columnwidth]{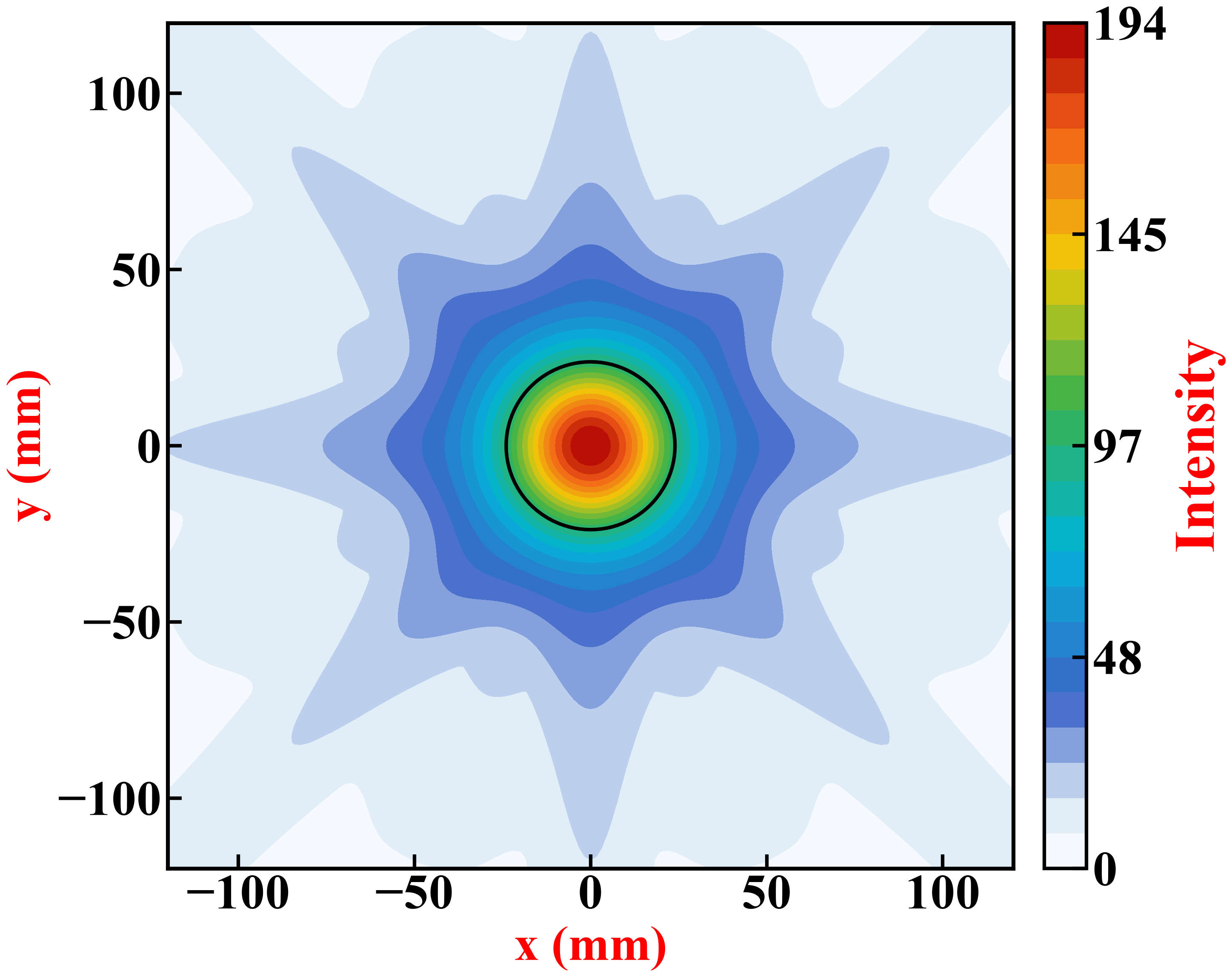}
      \caption{}
\end{subfigure}
\begin{subfigure}[t]{0.49 \columnwidth}
      \centering
      \includegraphics[width=0.8\columnwidth]{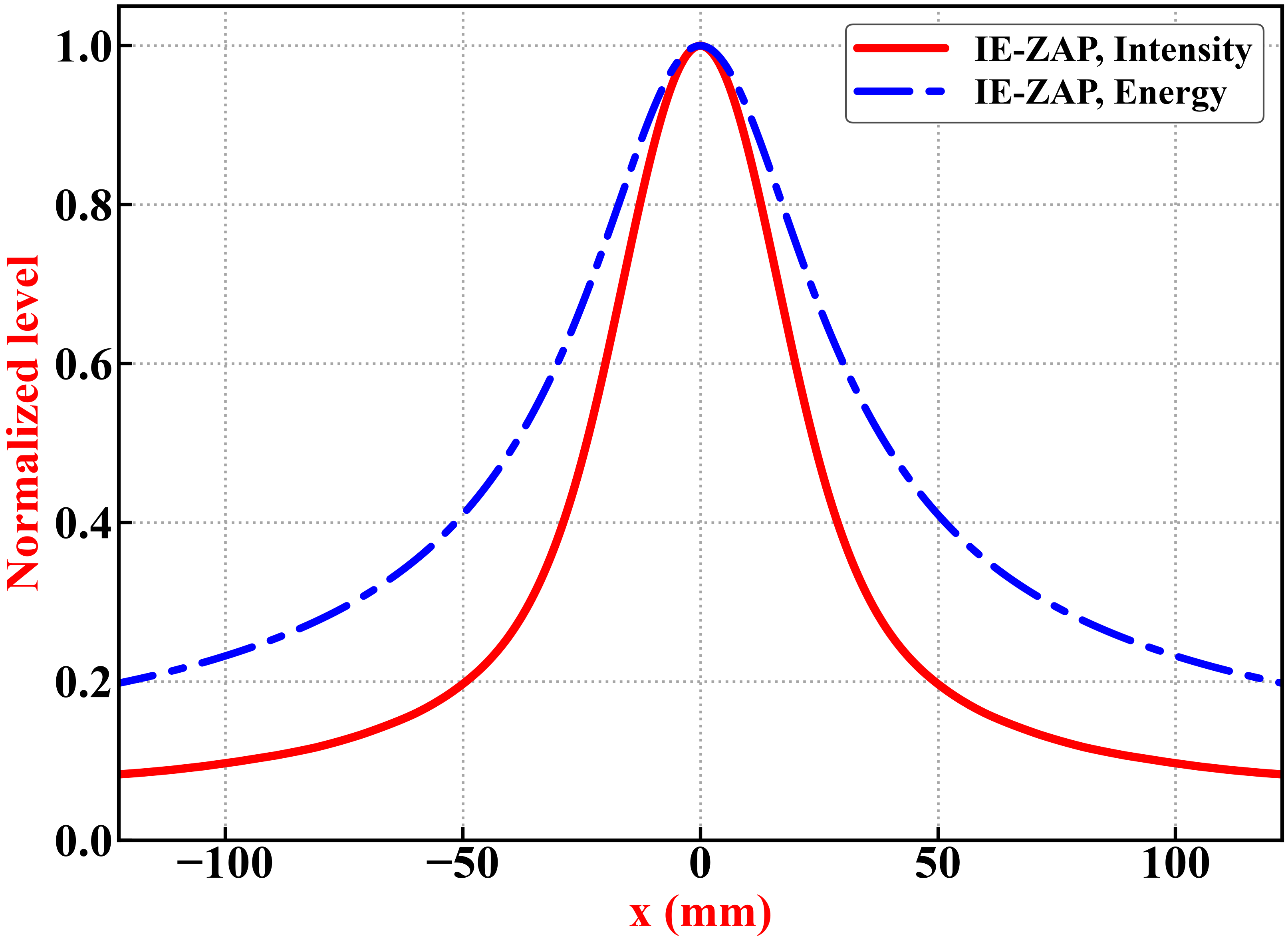}
      \caption{}
\end{subfigure}
\caption{The focal spot at $z=z_f$ produced by the IE-ZAP excited $2 \, \lambda$ UCA.
(a) Intensity distribution normalized to the corresponding 5-20-5 peak value. The black
circle is the -3~dB (relative to the maximum) contour whose radius is 23.965 mm = $\lambda_{IE-ZAP}/11.13$.
(b) x-axis profiles of that intensity distribution and the corresponding energy distribution.}
\label{fig:Fig15}
\end{figure}

The VSWE-calculated intensity distribution (relative to the 5-20-5 ZAP, 1.0 GHz reference) is shown in the
focal plane $z=z_f$ in Fig.\ref{fig:Fig15}(a).
The black circle is the -3~dB contour relative to the maximum, $=193.96$.
It surrounds the core of the focal spot, which is azimuthally symmetric.
On the other hand, the overall eight-pointed star shapes of the intensity distribution arise from the staggered,
fixed-$\xi$ = 8 number of elements on each ring of the array (i.e., the 45$^{\circ}$ separation between the elements).
The x-axis cuts of that focal plane distribution and the corresponding energy distribution are shown in
Fig.\ref{fig:Fig15}(b) normalized to their individual maximum value. With $\lambda_{\rm eff} =$ 266.82 mm,
the FWHM of the intensity spot is 47.93 mm = $\lambda_{\rm eff}$ / 5.567;
the FWHM of the energy spot is wider at 77.80 mm = $\lambda_{\rm eff}$ / 3.43.
Their ratio, 1.623, matches the predicted value (see below).
Note that the radius of the core intensity distribution is highly subwavelength,
i.e, the HWHM is 23.965 mm =  $\lambda_{\rm eff}$ / 11.13.
%
%
%

The same IE values were obtained with the $6 \, \lambda$ array with the focal point selected to have the same f/\# = 2.0,
i.e., with $z_f = 24 \, \lambda$.
On the other hand, the overall intensity and energy distributions in the focal plane were completely azimuthally symmetric.
This feature arises from the element density being the same on each ring ($\xi =$ 38, $K$ = 153).
Both examples give a profound demonstration that the IE effect associated with a
time-derivative focus is tied intimately to the IE-ZAP pulse shape and not the array size nor the number of elements,
as well as the fact that it is maintained even for compact arrays.
Nevertheless, it is noted that if the focal point is taken closer, for example, to the larger array,
the intensity focal spot can be made smaller. With the f/\# set to 0.67, i.e., with $z_f = 8 \, \lambda$,
the $6 \, \lambda$ UCA produced a 2.5 times smaller intensity spot in the focal plane, i.e.,
its FWHM was 19.08 mm = $\lambda$ / 13.98.

A totally independent confirmation of the IE effect was facilitated by using the direct time-domain calculational approach reported in \cite{ziolkowski2026time-domain-hda}. A Cartesian version of Eq. (42) in \cite{ziolkowski2026time-domain-hda} was used to calculate the time-domain electric fields radiated by each z-directed ED and $-$y-directed MD combination (cardioid maximum in the +x-direction), i.e., by each HDA in the array:
\begin{IEEEeqnarray}{rLc}\label{eq:TD_total_field}
E^{\rm total}_{z,n} (x, y, z, t) &=& -\frac{1}{4\pi \, R_n} \> \eta_0 \> [\, \partial_{ct} \, i_n(t-t_{R_n}) \> \ell \,] \> FF_{\rm term}
- \frac{1}{4\pi \, {R_n}^2} \> \eta_0 \>  \left[\, i_n(t-t_{R_n}) \> \ell \,\right] \> MR_{\rm term}            \hfill \nonumber \\
\noalign{\vskip 0.1in}
&& \qquad - \frac{1}{4\pi \, {R_n}^3} \> \eta_0 \> \left[\, c \, \ell \> H(t-t_{R_n}) \>
\int_{0}^{t-t_{R_n}} \> i(\tau) \> d\tau \,\right]
\end{IEEEeqnarray}
\vskip 0.1in

\noindent where $H(\tau)$ is the Heaviside function, i.e., $H(\tau) = 0$ for $\tau \le 0$ and $H(\tau) = 1$ for $\tau > 0$.
The retarded time term $t_{R_n} = R_n/c$ is the travel time from the $n$-th source to the observation point.
Because the fields were calculated in the xy-plane, the distance
$R_n = \left[\, (x-x_n)^2 + (y-y_n)^2 \,\right]^{1/2}$.
The Cartesian coordinate terms:

\begin{IEEEeqnarray}{rLc}
FF_{\rm term} &=& \left[\, \frac{(x-x_n)^2 + (y-y_n)^2}{R_n^2} \,\right]
   + \left[\, \frac{(x-x_n)}{R_n} \, \frac{(x_{\rm max} - x_n)}{R_{{\rm max}, n}} \,\right]
   + \left[\, \frac{(y-y_n)}{R_n} \, \frac{(y_{\rm max} - y_n)}{R_{{\rm max}, n}} \,\right]                            \hfill \\
\noalign{\vskip 0.1in}
MR_{\rm term} &=& 1 +
   \left[\, \frac{(x-x_n)}{R_n} \, \frac{(x_{\rm max} - x_n)}{R_{{\rm max}, n}} \,\right]
 + \left[\, \frac{(y-y_n)}{R_n} \, \frac{(y_{\rm max} - y_n)}{R_{{\rm max}, n}} \,\right]                              \hfill
\end{IEEEeqnarray}
\vskip 0.1in

A 9-element one-dimensional (1D) uniform linear array (ULA) along the y-axis of these x-pointing HDAs was constructed numerically in MATLAB. The elements were located (in meters) at
$-0.60$, $-0.45$, $-0.30$, $-0.15$, 0, 0.15, 0.30, 0.45, 0.60,
which mimics the 4-ring, $r_{\rm array} = 2 \, \lambda$ configuration, the separation between elements being approximately half of the effective wavelength of the IE-ZAP.
The focal point was selected to be along the x-axis at $x_f$ = 1.8 m. Referred to the 1.0~GHz wavelength,
$\lambda_{\rm 1~GHz}$ = 0.3 m,
this is a focal distance of $6 \, \lambda_{\rm 1~GHz}$.
With the aperture diameter being $4 \, \lambda_{\rm 1~GHz}$, the f-number $f/\#$ = 6/4 = 1.5.
The focus terms in the frequency-domain VSWE coefficients become a time delay for each element for this pulse-excited
array -- equal to the retarded time from each source to the focal point minus the time from the center element to it:
$t_{\rm delay, i} = [\, R_{\rm f, i} - x_f \,] / c
= [\, (\, x_f^2 + y_i^2 \,)^{1/2} - x_f \,] / c$, where $y_i = -0.60 + (i-1)*0.15$ for $i=1,2, ...,9$.
Because the array is symmetric about the x-axis, only 4 time delays are actually involved.

As noted above, a 5-20-5 1ns, multicycle ZAP was taken as the comparison case.
While the peak of its spectrum is also 1.0 GHz, it is even more narrow band than the 5-10-5 version.
Thus, it is an even closer embodiment of the 1.0 GHz CW reference.
More importantly, its longer constant amplitude section leaves ample margin for every element to be delivering
that section to the focal point simultaneously. In particular, the outermost elements of the larger array
considered below stand nearly 10.78 wavelengths from its focal point and the arrival times of their
radiated signals differ from the center element's by 0.77 $T_p$. Even with the imposed time delays to compensate,
the ten constant amplitude cycles of the 5-10-5 pulse did not comfortably cover the spread in arrival times
from all of the array elements. Roughly, only 4 of those cycles would be present from the outer elements
to help maintain the entire focal spot. Thus, the 5-20-5 ZAP was used since it guaranteed that the reference peak
against which the enhancement is measured is a genuine steady-state value and not a transient one.

\begin{figure}[h!]
	\centering
        \includegraphics[width=0.50\columnwidth]{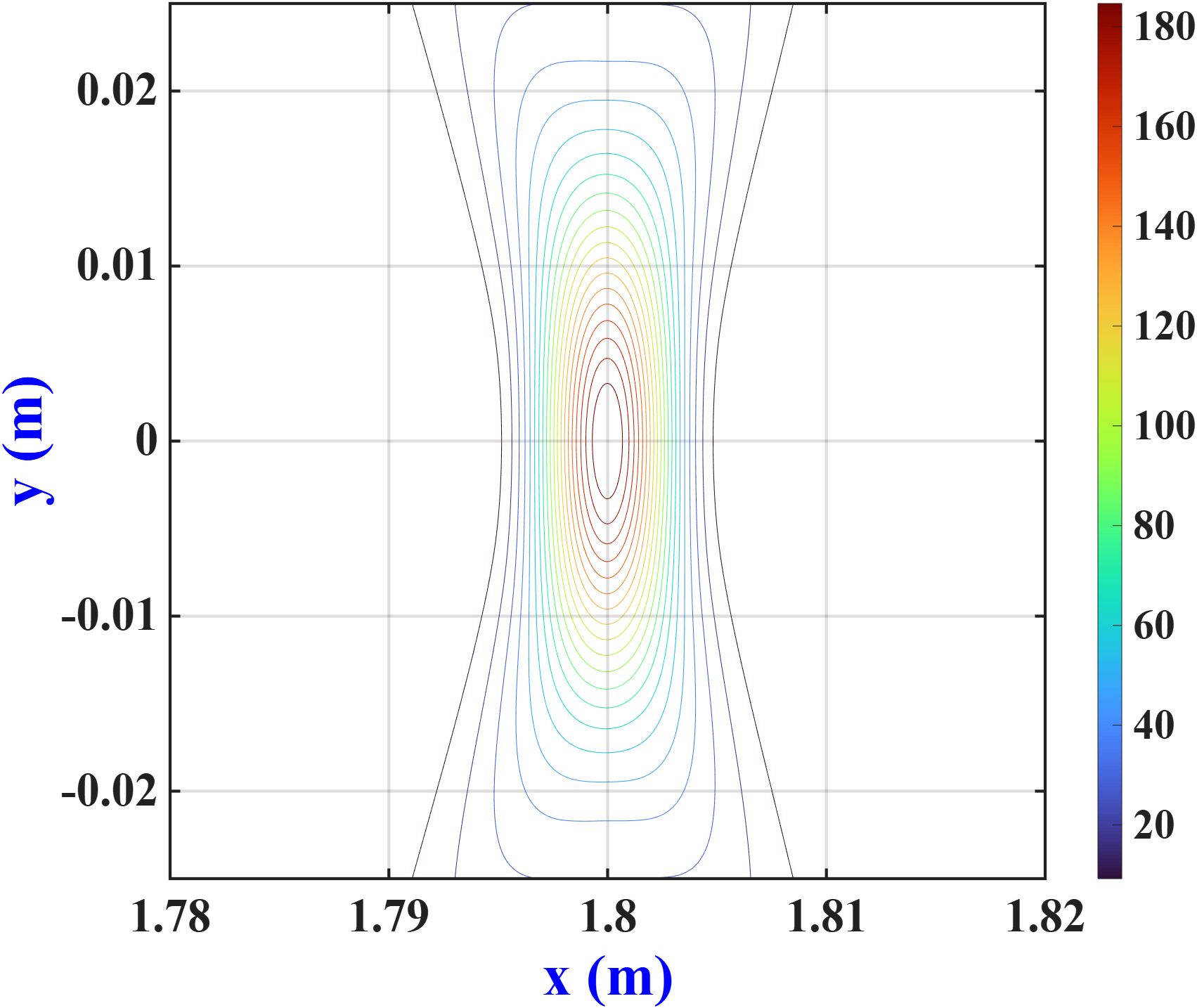}
\caption{Electric-field intensity distribution achieved in the focal region around the focal point $x_f$ = 1.8 m
with the IE-ZAP-excited 9 HDA element ULA. The contour values are normalized to the peak intensity
produced at the same point by the same array excited with the 5-20-5 multicycle ZAP reference; the
peak contour is therefore the IE factor itself, $\Upsilon_{\rm IE-ZAP} = 193.9$.}
	\label{fig:Fig16}
\end{figure}

\begin{figure}[h!]
\centering
\begin{subfigure}[t]{0.49 \columnwidth}
      \includegraphics[width=0.8\columnwidth]{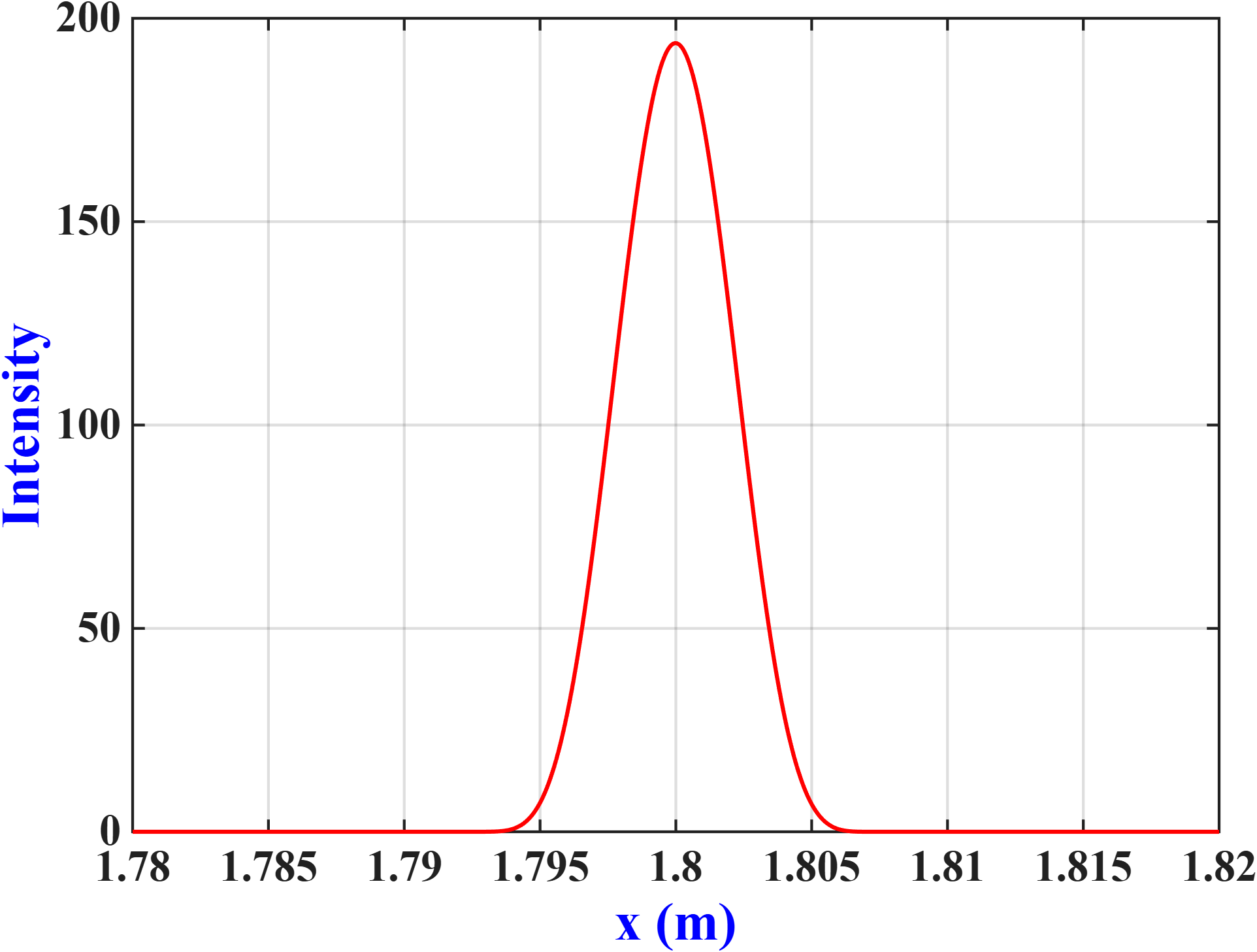}
      \caption{}
\end{subfigure}
\begin{subfigure}[t]{0.49 \columnwidth}
      \centering
      \includegraphics[width=0.8\columnwidth]{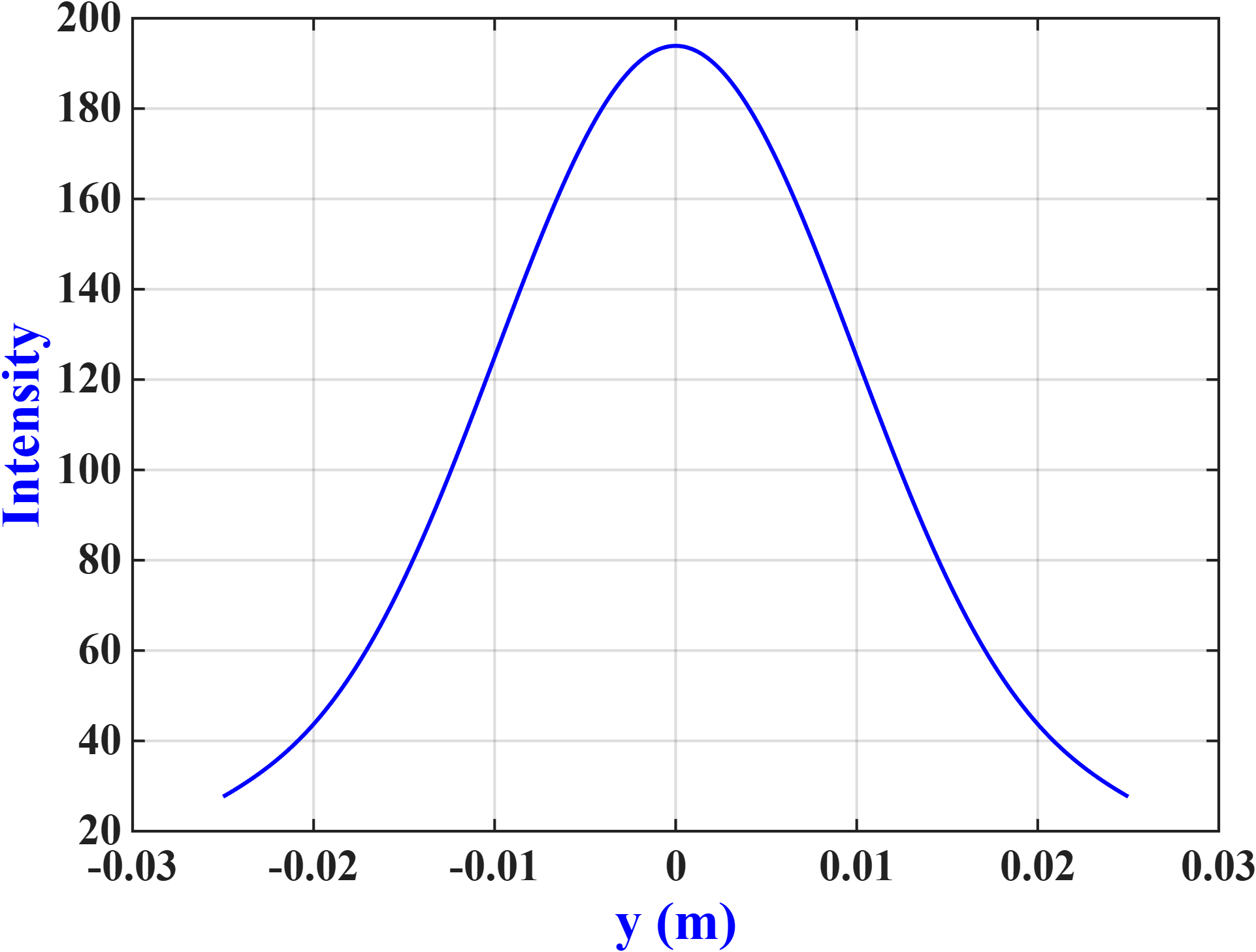}
      \caption{}
\end{subfigure}
\caption{Cross-sections of the focal-region intensity distribution of Fig.~\ref{fig:Fig15}.
(a) Cut along the propagation direction, the x-axis, through $y = 0$; $FWHM_x$ = 5.06 mm $\simeq \lambda_{\rm eff} / 52.73$.
(b) Transverse cut along the y-axis through $x = x_f$; $FWHM_y$ = 25.41 mm $\simeq \lambda_{\rm eff} / 10.50$.}
\label{fig:Fig17}
\end{figure}

The electric-field intensity in the focal region and its x- and y-cross-sections
predicted for the IE-ZAP are shown in Figs.~\ref{fig:Fig16} and \ref{fig:Fig17}.
Their peak values (clearly seen to be at the focal point $(x, y) = (x_f, 0)$ ) are normalized to the peak value produced by the reference array at the same focal point. The IE factor is $\Upsilon_{\rm IE-ZAP} = 193.9$.
This 1D ULA value is a perfect match to the VSWE results for the two-dimensional (2D) planar $2 \, \lambda$ and $6 \, \lambda$
UCAs, which carry $K = 33$ and $K = 153$ elements rather than only $9$.

Another aspect of the time-derivative facilitated intensity enhancement is the narrowing of the focal spot.
The classic diffraction limit for focusing a CW field is a spot whose FWHM is $\simeq 0.5 \lambda/{\rm NA}$,
i.e., $\lambda/2$ for a numerical aperture ${\rm NA} \to 1$ system. The arrays used here have larger f-numbers:
$f/\# = 1.25$ to $2.0$. Thus, the honest spot-size benchmark is not $\lambda/2$, but a narrow-band reference spot
computed with the identical array. Furthermore, this reference spot is itself a useful check.
The transverse FWHM of $366$ mm quoted below for the 5-20-5 drive of the 9-element ULA is within $3\%$ of the classical uniform-line-aperture value:
$0.886 \, \lambda_{\rm eff} \, x_f/L = 355$ mm
for its $L = 1.2$ m aperture and $x_f = 1.8$ m.
The narrow-band case is therefore behaving exactly as diffraction theory says it must, and it is only the
IE-ZAP that departs from it.

The effective wavelength for the IE-ZAP is $\lambda_{\rm eff} = c/(\omega_{\rm rad,~IE-ZAP}/2\pi) =$ 266.81 mm.
From Fig.~\ref{fig:Fig17}(a), the FWHM of the focal spot in the propagation direction (along the x-axis) is
$FWHM_x$ = 5.06 mm $\simeq \lambda_{\rm eff} / 52.73$, which is highly subwavelength.
From Fig.~\ref{fig:Fig17}(b), the FWHM of the focal spot in the transverse y-direction is
$FWHM_y$ = 25.41 mm $\simeq \lambda_{\rm eff} / 10.50$, which is also very subwavelength.
In contrast, the corresponding properties of the 5-20-5 ZAP focal spot are
$FWHM_x$ = 76.16 mm $\simeq \lambda_{\rm eff} / 3.50$ and $FWHM_y$ = 366.07 mm $\simeq 1.37 \lambda_{\rm eff}$.
The IE-ZAP focal spot is more than 10 times narrower in both directions.
This outcome is very attractive for a variety of near-field imaging applications.

%
%
%
%
%
%

A larger version of the 9-element 1D ULA was also considered. The elements were located (in meters) at
$-1.20$, $-0.90$, $-0.60$, $-0.30$, 0, 0.30, 0.60,  0.90,  1.20, twice the length of the original version. Thus,
it now mimics a 4 ring, $r_{\rm array} \simeq 4 \, \lambda$ configuration, the separation between elements being approximately the IE-ZAP's effective wavelength. Its Rayleigh distance is $z_R \simeq 50 \lambda \sim$ 15 m.
The focal point was set at the near-field distance $x_f$ = 3.0 m, $\approx 10 \, \lambda$.
Therefore, the $f/\#$ = 3/2.4 = 1.25 was quite similar to the smaller array version.
The same IE factor, $\Upsilon$ = 193.9, was obtained in comparison to the same array excited with the 5-20-5 1ns period,
multicycle ZAP. As the VSWE results indicated for the different 2D arrays, the size of the array had no impact on the
IE outcome; the pulse shape controls its value, not the array size.
Consequently, because this configuration was a distributive array with widely separated array elements,
these results indicate that the {\it coherent arrival of the time-synchronized fields} radiated by a distributive array excited with a custom-designed waveform having an intense time-derivative feature can produce a
substantial enhancement of the intensity at a prescribed near-field focal point.

Another interesting consequence of our studies is the fact that the size of the intensity and energy focal regions,
i.e., their width at half-maximum, follows the same rule as the FWHM of the intensity- and energy-beams in the
far-field of the array. Determination of which beam has the smaller on-axis width of its focal spot also follows
directly from the on-axis temporal waveform's ratio $\langle \omega^2 \rangle / \omega_{\rm peak}^2$. If
$\langle \omega^2 \rangle > \omega_{\rm peak}^2$, then the energy spot size is narrower than the intensity spot size.

The one point that must be kept firmly in view when this criterion is applied is that the waveform
$g$ for which it is evaluated is the one that appears \emph{in the focal region}, which is the time
derivative of the pulse that drives the array, and not the drive itself.
Writing the drive and its focal-region waveform as a pair, $f \to g = \partial_u f$ with $u$ the normalized time of
(\ref{eq:single_cycle_ZAP}), three exactly solvable cases settle the question of smaller or larger.
They are collected here. Every moment quoted below was confirmed in exact rational arithmetic.
\vskip 0.05in

\noindent \textbf{(i) The single-cycle ZAP:}
$f = u(1-u^2)^3 \to g = (1-u^2)^2(1-7u^2)$.
This $g$ is the three-lobed waveform of Fig.~\ref{fig:Fig9}(b), and, as already established,
$\omega_{\rm peak}^2 = 18$ and $\langle \omega^2 \rangle = 65/3$, so
$\langle \omega^2 \rangle/\omega_{\rm peak}^2 = 65/54 = 1.204 > 1$: its energy spot is the narrower one.
\vskip 0.05in

\noindent \textbf{(ii) Any drive whose focal waveform is a single smooth hump:}
Taking $g = (1-u^2)^n$ as the representative family, one finds $\omega_{\rm peak}^2 = 2n$ and
$\langle \omega^2 \rangle = n(4n+1)/[2(2n-1)]$, so that
\begin{IEEEeqnarray}{c}\label{eq:single-hump-law}
\frac{\langle \omega^2 \rangle}{\omega_{\rm peak}^2} \;=\; \frac{4n+1}{4(2n-1)} ,
\end{IEEEeqnarray}
which decreases monotonically from $3/4$ at $n = 2$ to $1/2$ as $n \to \infty$, i.e., it is below
unity for every $n \ge 2$. (It exceeds unity only at $n = 1$, where $\partial_u g$ does not vanish
at $u = \pm 1$ and the hump is consequently not smooth.) Region 2 of the IE-ZAP is exactly this
case with $n = 3$. Its drive there is the septic smoothstep whose derivative is
$\partial_x(35x^4 - 84x^5 + 70x^6 - 20x^7) = 140 \, x^3(1-x)^3$. Hence, with $x = (1+u)/2$, it is
proportional to $(1-u^2)^3$. Thus, $\omega_{\rm peak}^2 = 6$, $\langle \omega^2 \rangle = 39/10$, and
(\ref{eq:single-hump-law}) gives $13/20 = 0.650$. Including the long Regions 1 and 3 wings of the
full IE-ZAP lowers this a little closer to $0.645$.
\vskip 0.05in

\noindent \textbf{(iii) A bipolar focal waveform is not by itself sufficient:}
If the drive is taken to be the time integral of the single-cycle ZAP, i.e., the single hump
$f = (1-u^2)^4$, its focal waveform $g = u(1-u^2)^3$ is bipolar, yet
$\omega_{\rm peak}^2 = 49/3$ (evaluated at its peak, $u = \pm 1/\sqrt{7}$) and
$\langle \omega^2 \rangle = 27/2$ give $\langle \omega^2 \rangle/\omega_{\rm peak}^2 = 81/98 = 0.827 < 1$.
Thus, its energy spot is still the wider one. It takes the \emph{three}-lobed $g$ of case (i), whose
central lobe is sharper than its own envelope, to push the ratio above unity.
\vskip 0.05in

The IE-ZAP therefore falls squarely in case (ii). The mechanism is the physical one, i.e.,
a single smooth hump owns no feature sharper than its own peak. Thus, its mean-square bandwidth can
never exceed the curvature at that peak; and, hence, it does not yield an energy spot narrower than its intensity spot.
In distinct contrast, even though the IE factor of the ULA excited with
the 1ns single-cycle ZAP is only 1.3073 against the CW sinusoid
at that pulse's own $\omega_{\rm rad}/2\pi$ = 1.170~GHz (in agreement with the predicted 1.31) and,
analogously, only 1.79 against the 5-20-5 multicycle ZAP,
its energy spot is nevertheless smaller than its intensity spot.
As Fig.\ref{fig:Fig9} indicates, its time derivative is a three-lobed function whose central lobe
is in fact sharper than the envelope of the three lobes taken together. This feature is what pushes
its ratio above unity; and, hence, its energy spot is narrower.

In analogy with a pulsed beam's FWHM, it is the \emph{ordering} of the two spots, i.e., which of the
energy and intensity spots is the narrower, that is independent of the radius of the aperture forming
them, of the focal distance and, hence, of the $f$-number.
As the narrow-band reference case above confirms, the absolute size of either spot of course
scales with the aperture and the focal distance in the usual diffraction-limited way.
It is only the ordering that is decided elsewhere.
Therefore, whether the aperture is a ring or a filled disc, the ordering is determined strictly by the
temporal waveform's shape.

It is important to be explicit about what the IE factor is, and is not, a statement about.
Given a number as large as 193.9, it is the first question a reader should ask.
Substituting $\omega_{\rm rad}^2 \simeq \kappa/({\cal E}_{\rm IE\mbox{-}ZAP} \, \Delta)$ into
(\ref{eq:IE-factor}) and recognizing that $f_{\rm IE-ZAP}(t)$ is normalized to unity, it can be written as:
\begin{IEEEeqnarray}{rLc}\label{eq:IE-decomposition}
\Upsilon_{\rm int, IE\mbox{-}ZAP} &=&
\left[\, \max_t \, \left| \, \partial_t \, f_{\rm IE-ZAP}(t) \,\right| \,\right]^2 \>\>
\left[\, \frac{{\cal E}_{\rm IE\mbox{-}ZAP}}
             { \int_{-\infty}^{+\infty} \> \left|\, \partial_t \, f_{\rm IE-ZAP}(t) \,\right|^2 \> dt } \,\right]
\hfill \nonumber \\
\noalign{\vskip 0.1in}
&\;\simeq\;&
\frac{1}{\kappa}
\left[\, \frac{\Delta \, \max_t |\, \partial_t \, f_{\rm IE\mbox{-}ZAP} \,|}
              {\max_t |\, f_{\rm IE\mbox{-}ZAP} \,|} \,\right]^{2}
\left[\, \frac{{\cal E}_{\rm IE\mbox{-}ZAP}}{\Delta \, \max_t |\, f_{\rm IE\mbox{-}ZAP} \,|^{2}} \,\right]
\hfill
\end{IEEEeqnarray}
\vskip 0.1in

\noindent The first line emphasizes the peak of the time-derivative spike and the ratio of the
total energy of the pulse and its derivative.
It is read directly as a statement about the time derivative of the IE-ZAP and its spike, which is how the
original IE pulse \cite{ziolkowski1994designer,davidson1994josaA} was conceived.
The second line has been rewritten as a consequence of the values of those energies given above to facilitate
introducing the shape factor and grouping the terms into dimensionless combinations  to address
more quantitatively here what the IE factor tells us.
Its first bracket is the normalized peak slope of the spike. As was emphasized, the septic
smoothstep significantly enhances its value and depends only on Region 2.
Taken together with $1/\kappa$, it describes the pure spike shape.
The second bracket is the pulse energy measured in units of the spike width and, again,
is set almost entirely by the wings.

The IE factor is therefore a comparison made at fixed \emph{peak}
excitation amplitude -- the constraint that applies when a transmitter is limited by breakdown or
by amplifier compliance rather than by average power. At that fixed peak amplitude it does
grow in direct proportion to the pulse energy: ${\cal E}_{\rm IE\mbox{-}ZAP}$. Thus, lengthening the wings
of the pulse raises the energy and lowers $\omega_{\rm rad}$ in the same proportion. This behavior
is precisely the origin of the observed linear law in $\zeta$.
Therefore, the enhancement is bought with pulse duration, not with peak field.
Conversely, the only remaining freedom at fixed pulse energy and fixed spike width is the spike itself, and (\ref{eq:IE-decomposition}) reduces to
$\Upsilon \propto (\Delta \max_t |\partial_t f|)^2 / \kappa$.
The design task that remains is then a single one: maximize
$(\Delta \max_t |\partial_t f|)^2/\kappa$, the peak-to-mean power ratio of the derivative within Region 2.
However, steepening the spike and lowering $\kappa$ are not two separate goals to be balanced;
it is only their ratio that the enhancement follows.

\subsection{\textcolor{blue}{Discussion}}
\vskip 0.1in

The reason for excitement about the IE results stems from the many applied time-domain electromagnetic systems
that exist and would find them beneficial.
A personal keen interest stems from the types of 1D and 2D time-domain distributed arrays considered herein, e.g., those with larger than the typical half-wavelength separation between the array elements. For example, the 9-element,
$4 \, \lambda$ ULA had a full wavelength between its elements. The fixed-$\xi$, with every other ring rotated by $45^{\circ}$, $2 \, \lambda$ array had distances between elements greater than a $\lambda/2$. In fact, the distances between nearest-neighbor ring 3 and 4 elements were approximately 0.93 $\lambda$. Nevertheless, the ZAP-excited arrays demonstrated that appropriately timed coherent custom-designed signals could produce extremely large localized intensities, which corresponds to producing extremely large localized electric fields in a designated spot in their near-field.
Such fields would be of particular interest to near-field imaging systems because
the subsequent fields scattered by objects in the targeted spot would be quite large as well.

From an information-theoretic point-of-view, the element's directivity
concentrates the transmitted power into a solid angle that shrinks as $4\pi/(\Nmax^2+2\Nmax)$.
Thus, the signal-to-noise ratio available to a receiver on the beam axis at fixed radiated power
rises by the same factor while the interference delivered to every other direction falls by it.
Since $C = B \log_2 (1 + {\rm SNR})$, there is a direct increase in the capacity of the intended link
and a direct decrease in the capacity of any unintended one.
The front-to-back ratio $4(k_0r)^2/(\Nmax+1)$ makes this statement quantitative in the one direction
where it can be evaluated in closed form. Near-field focusing adds a further degree of freedom, that of
localizing in \emph{range} as well as in angle \cite{nf_comm_2023, nf_focus_2023}.
Consequently, the region within which the transmitted information is recoverable becomes a compact volume
rather than a ray.

The pulsed results sharpen this picture in a way that is specific to a time-domain link.
The intensity beam and the energy beam do not have the same footprint. Which of the two a receiver
sees is decided by the receiver, not by the transmitter. For instance, a peak detector or a rectifier is scored
against the intensity beam and, hence, against $\omega_{\rm peak}$.
On the other hand, a matched filter or a correlating impulse-radio demodulator is scored against the energy beam
and, hence, against $\langle \omega^2 \rangle$.
For the single-cycle ZAP the energy footprint is the narrower of the two and, hence, the angular sector over
which the link closes for information is strictly smaller than the sector over which it closes for power.
A pulsed WPT and communication system sharing one aperture would therefore have two different coverage maps
drawn on the same beam; and the waveform, through the single ratio $\langle \omega^2 \rangle/\omega_{\rm peak}^2$,
is the design variable that sets the difference between them.

This electromagnetic engineering of information creation and delivery has many remote sensing applications.
An example would be a drone-based array in which each element would emit an appropriate applications-designed ZAP.
In GPR systems, which are well-known for their use of pulsed beams, different frequencies have greater penetration depths
in different soil types.
Custom-designed ZAPs applied to excite each drone antenna could facilitate a large distributed array at
safe heights above the ground that produces an intense localized field at interesting depths, for instance, to achieve the
safe detonation of buried ordnance.
On the other hand, the large scattered field responses associated with those intense fields
carry information about what is present in the target area. The swarm could be reconfigured in real time
to attain multiple views of the same scene. With on-board cognitive receivers, enhanced target detection and
identification could be achieved.

The same is true for ULA or UCA microwave or ultrasound arrays for medical applications.
Custom-designed ZAPs could facilitate enhanced signal penetration at interesting depths in a body.
Intense localized fields using an IE-ZAP could then be used to probe a small targeted region in a
body for cancer detection and its physical destruction without heating the surrounding tissues since the energy
of each beam would be initially localized to it and then once all the beam fields are in the target area
coherently combined, this array field would create a very large intensity while the energy would be less since it would be
spread over a larger area.
On the other hand, the array excitations could be switched to a simpler single-cycle ZAP and then
the more highly localized energy beams could be used to cause significant heating (thermal energy transfer)
in a small portion of the high intensity target area to, e.g., stimulate drug release or initiate particular chemical reactions.
Many of these and other applications await further pulse-driven array understanding.

Bessel beams, a frequency domain solution of the Helmholtz equation, and several other related structured electromagnetic beams,
e.g., OAM (orbital angular momentum) or vortex beams, have become a popular topic with applications to near-field
information transfer, particularly at millimeter-wave, terahertz and optical frequencies
\cite{rao-2024-BB-review, bliokh2023roadmap, forbes2020structured, structured_light_2021, CityUHK-2022-THz-BB, CityUHK-2023-THz-BB}.
Their time-domain counterpart, localized waves (LWs, more recently also referred to as STWPs - Space-Time Wave Packets),
are (coupled) space-time solutions of the wave and Maxwell's equations. They have been used to form a variety of
structured acoustic and electromagnetic beams \cite{hernandez-2008-LWs, hernandez-2013-non-diff-waves}.
In contrast to Bessel beams, several classes of the LWs are finite-energy diffraction-resistant beams
with many appealing characteristics for localized energy transfer
\cite{ziolkowski-1989-localized, ziolkowski_LW_physics_1991}.
One is their extended near-to-far field distances, which were verified
experimentally with acoustic arrays \cite{ziolkowski-1989-evidence, ziolkowski-1990-verification}.
It was further demonstrated that their electromagnetic versions could be radiated from independently addressable pulse-driven arrays \cite{ziolkowski_IAPDA_1992}, i.e., an array whose elements are excited with the many pulses that are directly correlated to those that form the spatial structure of the exact solution.
Looking into the future, it will be interesting to determine whether or not electromagnetic SDEU needle-element arrays excited with such finite-energy LWs would radiate pulsed beams that have analogous, enhanced performance characteristics and to explore
if the consequent localized information (energy) transfer could be maintained to yet even greater Rayleigh distances.
The broader programs of structured light \cite{structured_light_2021} and of engineered
electromagnetic pulses \cite{pulses_review_2025}
supply a large catalogue of candidate excitations, and the criterion
$\langle \omega^2 \rangle/\omega_{\rm peak}^2$ established herein is a one-line test that can be applied
to any of them to evaluate their relevance for any specific application scenario before a single field is computed.

There is also increasing awareness that it is crucial to employ near-field and far-field WPT to
power batteryless sensor networks associated with early-warning systems to mitigate potential natural (and human-caused) disasters such as forest fires.
Changing batteries in remote sensors is costly, time consuming, and often not
physically possible once they are deployed.
As was noted, time-domain near-field focusing adds localization in range as well as in angle.
A SDEU system built from several needle-beam elements can therefore cover several prescribed compact volumes
rather than just several rays.
Its needle beams could be simply pointed at or steered to point at the sensor nodes in such networks to power them.
In this manner, the SDEU elements, either from a base station type or aerial type of platform,
would be acting as an effective wireless power distribution system \cite{nf_comm_2017}.
The frequency-domain counterpart of this idea, Bessel-beam launchers engineered for near-field
WPT, has already been demonstrated \cite{nf_wpt_2021, nf_wpt_2024}.
The pulsed near-field focusing described above is its time-domain analogue, with the IE-ZAP supplying
an intensity at the target volume that a narrow-band launcher of the same peak drive amplitude cannot reach.
The subwavelength focal spot reported in Sec.~V, $\lambda_{\rm eff}/52.7$ axially and
$\lambda_{\rm eff}/10.5$ transversely, belongs to the same near-field super-resolution family
\cite{NF_superresolution1992, NF_superresolution2018review}, but it is obtained without any
subwavelength structure in the aperture. Consequently, the resolution is supplied entirely by the temporal
waveform.

We have successfully developed several variations of MMA-excited (mixed-multipole antenna as the source of a)
spherical dielectric lens antennas \cite{b-a-z-2025superdirective}.
With a 10-layered lens whose radius and, hence, radiating aperture is only $0.4 \, \lambda$,
multipoles up to degree 10 have been excited and radiated.
The MMAs are a combination of independently addressable electrically small electric and magnetic dipole antennas.
While the overall SDEU systems have currently narrow bandwidths, their overall efficiencies are quite high.
Progress on a proof-of-concept 3-layer prototype is nearing the fabrication and testing phase.
We are quite confident that this needle-beam element will efficiently radiate at least a combination of dipole-quadrupole-octopole fields because several electrically small, electric MMA prototypes have been tested
successfully, confirming their attractive performance characteristics \cite{ziolkowski_superD_compact_2025}.
Furthermore, we have developed and tested a wide variety of highly efficient Huygens dipole antennas
\cite{ziolkowski2022chapter}. Despite being electrically small systems, large bandwidths have been achieved \cite{ziolkowski_bandwidth_2024, ziolkowski_VP_HDA_2025, ziolkowski_bandwidth_2025, Hrabar-2025-Huygens}.
Arrays of these $\Nmax$ = 1 unidirectional MMAs are available to design a measurement campaign to test the
near-field pulsed-beam focusing concepts described herein.

It is worth stating plainly what this article does and does not settle about realizability,
since the elemental source has been engineered here at the level of its multipole coefficients
rather than at the level of a structure that radiates them. Every superdirective multipole
distribution confined to a small volume pays some level of Chu-type penalty, i.e.,
the stored reactive energy and, hence, the radiation $Q$ of the degree-$n$ term of a source
of radius $a$ rises steeply once $k_0 a < n$. Thus, the fractional bandwidth of an $\Nmax$-multipole needle
element decreases and its sensitivity to ohmic loss increases as $\Nmax$ is increased \cite{ziolkowski_bandwidth_2024}.
The $\Nmax = 20$ element used for the near-field maps of Sec.~III is, therefore, currently an idealization whose
realization in a finite volume is a separate and hard problem. On the other hand, the $\Nmax = 10$ element
used for every array result herein is squarely within what the layered-lens MMA route above has already
demonstrated.
Furthermore, two features of the present design mitigate the penalty rather than aggravating it.
The coefficients (\ref{eq:elemental-source-coefficients}) are frequency independent.
Therefore, the element itself introduces no dispersion, and the entire bandwidth limitation is the one imposed by
the radiating structure.
Moreover, the array results show that the multipole order and the aperture size are complementary
rather than competing knobs, e.g., the $6 \, \lambda$ array gains only 0.87 dB in
going from $\Nmax = 10$ to $\Nmax = 20$. Consequently, a system that can afford a reasonable
aperture size has no reason to push $\Nmax$. On the other hand, a space-constrained system buys 24.9 dB of
superdirectivity from the element at $r_{\rm array} = 0.1 \, \lambda$ and that is precisely the case in which paying
any $Q$ penalty is worthwhile.

\section{\textcolor{red}{Conclusions}}
\vskip 0.1in

A fundamental multipole needle-beam source has been engineered from a VSWE approach. Its coefficients
were chosen to mimic the spherical-harmonic closure relation truncated at the maximum degree $\Nmax$.
It was proved analytically, and confirmed numerically to machine precision with mirrored Python
and MATLAB codes, that this element radiates a superdirective, extremely unidirectional needle beam.
It was demonstrated to any degree $\Nmax$ that its broadside directivity attains
Harrington's normal-directivity value $\Dmax = \Nmax^2 + 2 \Nmax$ exactly.
Its far field is linearly polarized with an azimuthally symmetric magnitude, and its
corresponding back lobe vanishes identically.

Two exact results emerged along the way that were not anticipated.
The azimuthal dependence of $|{\vec E}|^2$ is exhausted by a constant plus a single $\cos 2\phi$
term at every radius, near field as well as far, because the source-angle endpoint values collapse
the order sum onto $|m| = 1$.
Moreover, the front and back lobes are azimuthally flat exactly, at every radius, because the relations
$\Pi_n = +T_n$ at $\theta = 0$ and $\Pi_n = -T_n$ at $\theta = \pi$ hold degree by degree and, therefore,
survive the sum over the degree $n$.
The extreme unidirectionality was then quantified at a practical finite, rather than infinite, measurement radius.
Because $b_{n1} = -a_{n1}$, the TM and TE contributions of each single degree annihilate one another in
the backward direction to leading order. What survives is only the failure of the radial-distance based functions
$A_n$ and $jB_n$ to be exactly equal and opposite at finite $k_0r$.
This per-degree structure, together with a family of exact integer identities proved in Appendix B,
gives the major figures of merit in closed form:
$D(\pi,0) = \Nmax(\Nmax+1)^2(\Nmax+2)/[16 (k_0r)^4]$
and
$\ftbr|_{\rm field} = 4(k_0r)^2/(\Nmax+1)$.
The back lobe therefore falls as $r^{-4}$ in power rather than $r^{-2}$, i.e., it falls 12 dB per doubling
of range. The FTBR already exceeds 97 dB at $r = 10^2\lambda$ for every case examined.
It is worth remarking that the extreme nature of the FTBR actually \emph{degrades} as the directivity is raised:
doubling $\Nmax$ costs the FTBR between 5 and 6 dB at fixed range.
Thus, a superdirective needle is the harder, not the easier, beam in which extreme unidirectionality is maintained at
finite ranges.

It was further demonstrated that uniform circular arrays of these elements inherit their behaviors.
The electrically small $0.1 \, \lambda$ array is superdirective by 24.9 dB relative to the uniformly illuminated
aperture bound $(k_0 r_{\rm array})^2$. This result is a direct statement of what a multipole element
buys a space-constrained system.
Amplitude tapering across the rings lowers the sidelobes by a further 4 to 15 dB, although the minmax weights
that achieve the most require negative amplitudes and are correspondingly sensitive to perturbation.

The primary time-domain outcome of this article was the demonstration that driving those arrays with
zero-area pulses carries the frequency-domain SDEU performance into the time domain intact.
The far-field signal on the beam axis is the time derivative of the exciting ZAP, reproduced to
$4 \times 10^{-16}$ at every range examined.
Moreover, the $-3$~dB locus of their pulsed beams is an exact cone.

Another central time-domain result is its answer to the question of whether the radiated energy beam is
narrower or wider than the corresponding intensity beam. It was determined explicitly that the ordering
is decided by the pulse and not by the aperture/array radiating it.
Moreover, it was shown that the array acts off axis as a non-negative, unit-area convolutional smoothing kernel.
Thus, the two beams roll off as $1 - \sigma^2 \omega_{\rm peak}^2$ and $1 - \sigma^2 \langle \omega^2 \rangle$,
respectively, where $\sigma^2$ is the variance of that kernel.
The ordering of the beams was found to be fixed by the single dimensionless ratio
$\langle \omega^2 \rangle / \omega_{\rm peak}^2$
of the radiated on-axis waveform, i.e., of the time derivative of the exciting one.
The aperture (array radius) enters only through $\sigma^2$, i.e., it sets how large the separation is,
never which way the ordering goes.
For the single-cycle ZAP that ratio is exactly 65/54, so its energy beam is the narrower one.
The reverse of the ordering was found for a unipolar Gaussian excitation, for which the ratio is 3/4.
For a CW signal, the ratio is, of course, 1.0, and the beams coincide.

The same criterion was then shown to govern the intensity-energy ordering of the
focal spot of a near-field-focused, pulse-driven array. It was demonstrated that all of its
aperture geometry -- the radius, the focal distance, the $f$-number, and even
whether the aperture is a ring or a filled disc -- cancels out and is not part of the conclusion.

A custom-designed, $C^2$-continuous IE-ZAP excitation was developed to achieve significant
intensity-enhancement at a focal point in a pulse-driven array's near field.
An IE factor, $\Upsilon_{\rm int, IE\mbox{-}ZAP}$, was introduced to guide and optimize its design.
Unexpectedly, it also facilitated an explanation of that ordering.
It was emphasized that this IE factor depends on the squared peak slope of the ZAP measured against its
effective angular frequency $\omega_{\rm rad}$.
Its complete independence of $\Nmax$ and the size of the array was also demonstrated.
Moreover, it was obtained in closed form as the linear law
$\Upsilon_{\rm int, IE\mbox{-}ZAP} \simeq 2.298 \, \zeta + 0.795$.
The IE factor, through this closed form, also makes explicit what the enhancement costs are.
At fixed \emph{peak} excitation amplitude, the IE effect is paid for in the pulse's duration and, hence,
in the energy radiated rather than in the peak field.

The IE-ZAP was customized as a 6.70 ns, three-region function: a $C^2$ quintic smoothstep turn-on and turn-off,
each lasting $\zeta \, \delta$ = 3.325 ns with $\zeta =$ 66.5 and $\delta$ = 50 ps,
that surround a $\delta$-length septic smoothstep center which is $C^3$ at its own edges.
Those quintic wings set the smoothness of the pulse as a whole at $C^2$, one order better than the
$C^1$ of the original $UWB_1$ design in \cite{ziolkowski1994designer,davidson1994josaA}.
Edge regularity was found to govern only the asymptotic decay rate of the spectrum,
and the asymptotic tail carries a negligible share of the radiated energy.
The linear law predicted $\Upsilon_{\rm int, IE\mbox{-}ZAP} =$ 153.61, well beyond the initial goal of 100.
The VSWE calculation of the focal-region fields reproduced that value to five significant figures
against the same array excited by a CW signal at $\omega_{\rm rad, IE\mbox{-}ZAP}/2\pi$ = 1.124 GHz.

The shape factor, $\kappa$, of the IE-ZAP was introduced to help engineer and understand the
IE factor's value. It is defined by Region 2's time width $\Delta$ and the energy in the IE-ZAP's
derivative there. It was demonstrated that the enhancement is rewarded by the ratio
$(\Delta \max_t |\partial_t f|)^2/\kappa$, i.e., the peak-to-mean power ratio of the derivative
within Region 2. Smoothness in the middle region was proved to be profitable; had Region 2 been the
straight-line transition that minimizes $\kappa$, the same pulse would have delivered an IE of 40.5 rather
than 193.9.

Two reference cases were quoted for the enhancement values; they should not be confused.
Against the CW sinusoid at $\omega_{\rm rad}$ of the definition (\ref{eq:IE-factor}),
IE is the $153.61$ that the closed form predicts.
Against the physically realizable 5-20-5 multicycle ZAP, which mimics the ideal 1.0 GHz CW signal,
the stunning realized enhancement, 193.9, was confirmed.
The latter was reproduced identically by the full VSWE calculation for two different planar UCAs and
by a wholly independent direct time-domain calculation for two different linear arrays, one of them with more
than a full effective wavelength between its elements.
Neither the dimensionality of the array, nor its size, nor its element spacing moved that number.
Furthermore, the intensity focal spot it produces is more than ten times smaller
in both the axial and the transverse direction than the narrow-band reference spot.
Its own FWHMs are $\lambda_{\rm eff}/52.7$ axially and $\lambda_{\rm eff}/10.5$ transversely.

The IE-ZAP's own intensity focal spot is, moreover, the smaller of the pair it produces; its energy
spot is the wider one. On the other hand, the 1ns single-cycle ZAP achieves an IE of only 1.31 against the
CW sinusoid at its own $\omega_{\rm rad}/2\pi$ = 1.170~GHz, equivalently 1.79 against the 1.0 GHz reference,
and yet its energy focal spot is the narrower one.
The two orderings are opposite because the focal-region waveform of the IE-ZAP is a single smooth hump, for which
$\langle \omega^2 \rangle/\omega_{\rm peak}^2 = 13/20 < 1$, while that of the single-cycle ZAP is
three-lobed, for which the ratio is $65/54 > 1$. A large IE factor and a narrow energy spot are
simply not the same property, and both orderings are in total agreement with the needle-beam
arguments given above.

Taken together, these results identify the temporal waveform, rather than the aperture, as the
design variable that governs localized electromagnetic energy delivery, and the multipole element,
rather than the array size, as the design variable that governs directivity in a
space-constrained volume.
Because a receiver that thresholds an instantaneous field level responds to the intensity beam
while a receiver that correlates over the pulse responds to the energy beam,
the two are also the design variables that govern where information, as distinct from power, is
actually recoverable. Distributed, independently addressable, pulse-driven arrays of
superdirective multipole elements are therefore an attractive route to the intimate control of
electromagnetic power and information transfer that the many applications listed at the outset require.

\section*{Acknowledgments}
\vskip 0.1in

The author was responsible for the directions of this project, the writing, and most of the analytical efforts.
The numerical aspects of these studies were performed primarily by Claude Code, Opus 5 and then assessed
separately by the author. Claude created mirrored codes in both Python and MATLAB; thoroughly checked the
results they produced with both languages; and further confirmed them against a variety of numerical checks it imposed.
Claude offered editing suggestions of the text. Quite unexpectedly, it independently initiated and consummated
the backlobe analysis and their proofs which facilitated the many quantitative, extreme FTBR statements.

The author also extends very sincere thanks to Prof. George Shaker, University of Waterloo, for his
time, efforts, and encouragements in helping me get started using Claude.


\appendices


\section{\textcolor{red}{Derivation of the FTBR at Finite Radius}}
\vskip 0.1in


\subsection{\textcolor{blue}{Endpoint Values of the Angular Functions}}
\vskip 0.1in

\noindent The choice of the phase in the definition of the associated Legendre functions in both Python and MATLAB
is the Condon--Shortley phase, $P_n^{1}(-u) = (-1)^{n+1}P_n^1(u)$.
Setting $\theta = \pi - \delta$ and letting $\delta\to0$,
the endpoint values of the field expressions are:
\begin{equation}\label{eq:endpoints}
\Pi_n(0) = T_n(0) = -\,\frac{n(n+1)}{2},
\qquad
\Pi_n(\pi) = (-1)^{n}\,\frac{n(n+1)}{2},
\qquad
T_n(\pi) = -\,\Pi_n(\pi).
\end{equation}
The relation $\Pi_n = -T_n$ at $\theta = \pi$ holds for every $n$ separately.
Substituting it into (\ref{eq:Eth})--(\ref{eq:Eph}) gives two exact statements that will be used twice below:
\begin{equation}\label{eq:quadrature}
S_\phi (\aleph,\theta=0) \;=\; +j \, S_\theta (\aleph,\theta=0)
\qquad\qquad
S_\phi (\aleph,\theta=\pi) \;=\; - j \, S_\theta (\aleph,\theta=\pi)
\end{equation}
so that $|S_\theta| = |S_\phi|$ on the z-axis in \emph{both} directions, at
\emph{every} radius -- near field as well as far field.
This makes both $D_r(0,\phi)$ and $D_r(\pi,\phi)$ azimuthally flat even though the pattern as a
whole is not.


\subsection{\textcolor{blue}{Exact Radial Identity}}
\vskip 0.1in

\noindent The spherical Hankel function of the second kind can be written in the form:
\begin{equation}\label{eq:un}
h_n^{(2)}(x) \;=\; \frac{e^{-jx}}{x}\,u_n(x),
\qquad
u_n(x) \;=\; j^{\,n+1}\sum_{s=0}^{n}\frac{(n+s)!}{(n-s)!\,s!}
\left(\frac{-j}{2x}\right)^{\!s}.
\end{equation}
i.e., where the sum terminates at $s=n$.
Substituting (\ref{eq:un}) into (\ref{eq:AB}) and using
$\frac{d}{dx}\!\left[x\,h_n^{(2)}\right] = \frac{d}{dx}\!\left[e^{-jx}u_n\right]$
gives $A_n = \frac{e^{-jx}}{x}\left(u_n' - j u_n\right)$ and
$B_n = \frac{e^{-jx}}{x}u_n$, where the prime denotes the derivative with respect to $x$.
Consequently, the two combinations that control the two radial variations become:
\begin{equation}\label{eq:exactid}
A_n + jB_n \;=\; \frac{e^{-jx}}{x}\,u_n'(x),
\qquad\qquad
A_n - jB_n \;=\; \frac{e^{-jx}}{x}\left[\,u_n'(x) - 2j\,u_n(x)\,\right]
\end{equation}
Both are exact.
Note that the forward combination $A_n - jB_n$ is dominated by $-2ju_n = O(1)$;
the backward combination $A_n + jB_n$ retains only $u_n'$. The term $u_n'$ is
\emph{one order smaller in $1/x$ than $u_n - j^{n+1}$} because differentiation of the $s$-th
term of (\ref{eq:un}) converts $x^{-s}$ into $-s\,x^{-s-1}$.

Then, with $\frac{(n+1)!}{(n-1)!\,1!} = n(n+1)$, one has explicitly from (\ref{eq:un}):
\begin{equation}\label{eq:updash}
u_n(x) = j^{\,n+1}\left[1 - j\frac{n(n+1)}{2x} + O(x^{-2})\right],
\qquad
u_n'(x) = j^{\,n+1}\left[\,j\frac{n(n+1)}{2x^{2}} + O(x^{-3})\right],
\end{equation}
and therefore
\begin{equation}\label{eq:leading}
A_n - jB_n = 2\,j^{\,n}\,\frac{e^{-jx}}{x}\left[1+O(x^{-1})\right],
\qquad
A_n + jB_n = -\,j^{\,n}\,\frac{e^{-jx}}{x}\,\frac{n(n+1)}{2x^{2}}
\left[1+O(x^{-1})\right].
\end{equation}
The ratio of the two is $O(x^{-2})$ per degree $n$ -- the $r^{-2}$ in field, $r^{-4}$
in power.
\vskip 0.05in

\noindent It is worth noting \emph{why} the cancellation at $\theta=\pi$ is a
per-degree cancellation rather than an interference between degrees.
Because $b_{n1} = -a_{n1}$, the TM and TE contributions of a \emph{single} $n$
annihilate one another in the backward direction to leading order.
What survives is the failure of $A_n$ and $jB_n$ to be exactly equal and opposite at finite $x$.
This is the reason why the result is so clean: no cross terms between different $n$ ever appear.


\subsection{\textcolor{blue}{Forward and Back Direction Results}}
\vskip 0.1in

\noindent \textbf{In the forward direction,} one has $\Pi_n(0) = T_n(0) = -n(n+1)/2$ in
(\ref{eq:Eth}), then (\ref{eq:leading}) and (\ref{eq:an}), which gives:
\begin{equation}\label{eq:Sfwd}
S_\theta(0) = \sum_{n} a_n\left(-\frac{n(n+1)}{2}\right)\left(A_n - jB_n\right)
= \frac{E_0}{8\pi}\,\frac{e^{-jx}}{x}\sum_{n=1}^{\Nmax}(2n+1)
\left[1+O(x^{-1})\right]
= \frac{E_0}{8\pi}\,\frac{e^{-jx}}{x}\,\Nmax(\Nmax+2),
\end{equation}
where the factors $j^{-n}$ and $j^{n}$ have cancelled degree by degree. {\it This is
exactly the phase alignment that produces the needle effect.}
\vskip 0.05in

\noindent \textbf{In the back direction,} one has $\Pi_n = -T_n$ and, subsequently, the two terms of
(\ref{eq:Eth}) combine into the single product: $T_n(\pi)\left(A_n + jB_n\right)$.
Since
$T_n(\pi) = (-1)^{n+1}n(n+1)/2$ together with (\ref{eq:leading}), (\ref{eq:an}) and the results of the
Lemma in the next subsection for $p=1$, one obtains:
\begin{eqnarray}\label{eq:Sback}
S_\theta(\pi) &=& \sum_{n} a_n\,T_n(\pi)\left(A_n + jB_n\right)
\;=\; \frac{E_0}{8\pi}\,\frac{e^{-jx}}{x}\,\frac{1}{4x^{2}}
\sum_{n=1}^{\Nmax}(-1)^{n+1}(2n+1)\,n(n+1)
\nonumber\\[4pt]
&=& \pm\,\frac{E_0}{8\pi}\,\frac{e^{-jx}}{x}\,
\frac{\Nmax(\Nmax+1)(\Nmax+2)}{4x^{2}}
\end{eqnarray}
\vskip 0.05in

\noindent \textbf{Consequently,} dividing (\ref{eq:Sfwd}) by (\ref{eq:Sback}), the
common prefactor $\frac{E_0}{8\pi}\frac{e^{-jx}}{x}$ drops out and the common factor
$\Nmax(\Nmax+2)$ cancels between numerator and denominator.
The field FTBR becomes
\begin{equation}\label{eq:ftbrfinal}
\ftbr\Big|_{\rm field} = \frac{|S_\theta(0)|}{|S_\theta(\pi)|}
= \frac{\Nmax(\Nmax+2)}
       {\dfrac{\Nmax(\Nmax+1)(\Nmax+2)}{4x^{2}}}
= \frac{4\,(k_0 r)^2}{\Nmax+1}
\end{equation}
The cancellation of $\Nmax(\Nmax+2)$ between the forward sum and the Lemma is
the reason the result is so simple. It is worth remarking that the FTBR
therefore \emph{degrades} as the directivity is raised: doubling $\Nmax$ costs
between 5 and 6~dB of front-to-back ratio at fixed range. Thus,
it is harder, not easier, to maintain the extreme unidirectional behavior of a superdirective
needle at finite ranges.
\vskip 0.05in

\noindent \textbf{Finally, the back-lobe directivity,} follows from (\ref{eq:quadrature}) giving
$|S_\theta| = |S_\phi|$ and, therefore, $|E_\theta|^2+|E_\phi|^2 = 4|S_\theta|^2$
at both poles. {\it There is no residual $\phi$ dependence.}

Since Eq.~(\ref{eq:Directivity}) is a ratio of the same quantity at the
two poles times the same normalisation, one has
$D(\pi, \phi=0)/D(0, \phi=0) = \ftbr|_{\rm field}^{-2}$. Then, with $D(0, \phi=0)\to\Nmax^2+2\Nmax$,
\begin{equation}\label{eq:Dback}
D(\pi, \phi=0) \;=\; \Nmax(\Nmax+2)\,\frac{(\Nmax+1)^2}{16\,x^{4}}
\;=\; \frac{\Nmax(\Nmax+1)^2(\Nmax+2)}{16\,(k_0 r)^4}.
\end{equation}
\vskip 0.05in

\noindent \textbf{A useful by-product} is that the main directivity result: $D(\theta =0, \phi = 0)\to\Nmax^2+2\Nmax$,
can be obtained in closed form from the same ingredients. Furthermore, this confirms the normalisation of (\ref{eq:an})
and of (\ref{eq:P-total}) simultaneously. Then with $|b_{nm}| = |a_{nm}|$ and both $m = \pm1$ contributing,
(\ref{eq:P-total}) gives
\begin{equation}
P_{\rm tot} = \frac{2\pi}{\eta_0 k_0^2}\sum_{n=1}^{\Nmax}
4\,\frac{[n(n+1)]^2}{2n+1}\,|a_n|^2
= \frac{2\pi}{\eta_0 k_0^2}\left(\frac{E_0}{8\pi}\right)^{\!2}
4\sum_{n=1}^{\Nmax}(2n+1)
= \frac{E_0^2\,\Nmax(\Nmax+2)}{8\pi\,\eta_0\,k_0^2},
\end{equation}
while (\ref{eq:Sfwd}) gives
$U(0) = \frac{r^2}{2\eta_0}\cdot4|S_\theta(0)|^2
= \frac{E_0^2\,\Nmax^2(\Nmax+2)^2}{32\pi^2\eta_0 k_0^2}$.
Therefore, $D(\theta=0, \phi=0) = 4\pi U(\theta=0, \phi=0)/P_{\rm tot} = \Nmax(\Nmax+2)$ exactly.


\section{\textcolor{red}{A Family of Exact Integer Identities}}\label{sec:lemma}
\vskip 0.1in

\noindent \textbf{Lemma.} For every integer $p \ge 1$ and every $N \ge 1$,
\begin{equation}\label{eq:lemma}
\sum_{n=1}^{N} (-1)^{n+1}(2n+1)\!\!\prod_{k=1}^{p}(n-k+1)(n+k)
\;=\; (-1)^{N+1}\!\!\prod_{j=-p+1}^{p+1}\!(N+j)
\end{equation}
Thus, the alternating, $(2n+1)$-weighted sum of a product of $2p$ consecutive
integers centered on $n$ equals, up to the sign $(-1)^{N+1}$, the product of the
$2p+1$ consecutive integers from $N-p+1$ to $N+p+1$.
\vskip 0.05in

\noindent \emph{Proof.} Write
$f_p(n) = \prod_{k=1}^{p}(n-k+1)(n+k) = \prod_{i=n-p+1}^{n+p} i$, a product of
$2p$ consecutive integers, and
$F_p(N) = \prod_{i=N-p+1}^{N+p+1} i$, a product of $2p+1$ consecutive integers.
Let $S_p(N)$ denote the left side of (\ref{eq:lemma}).

\emph{Base case $N=1$.} If $p=1$, then $S_1(1) = 3\cdot 1\cdot 2 = 6 = 1\cdot2\cdot3
= F_1(1)$. If $p\ge2$, the factor $(n-1)$ at $n=1$ vanishes and, hence, $S_p(1)=0$. Moreover,
$F_p(1) = \prod_{i=2-p}^{2+p} i$ contains the factor $0$, so $F_p(1)=0$ as well.

\emph{Induction step.} Put $P = f_p(N+1) = \prod_{i=N-p+2}^{N+p+1} i$. Then
$F_p(N) = (N-p+1)\,P$ and $F_p(N+1) = P\,(N+p+2)$ since each term simply appends one
factor at the appropriate end. Therefore,
\begin{equation}
F_p(N+1) + F_p(N) \;=\; P\left[(N+p+2)+(N-p+1)\right] \;=\; (2N+3)\,P
\;=\; (2N+3)\,f_p(N+1).
\end{equation}
Assuming $S_p(N) = (-1)^{N+1}F_p(N)$,
\begin{equation}
S_p(N+1) \;=\; S_p(N) + (-1)^{N+2}(2N+3)f_p(N+1)
\;=\; (-1)^{N+2}\left[(2N+3)f_p(N+1) - F_p(N)\right]
\;=\; (-1)^{N+2}F_p(N+1),
\end{equation}
which is the claim at $N+1$. $\blacksquare$
\vskip 0.05in

\noindent The two cases used above are:
\begin{eqnarray}
p=1:&\quad& \left|\sum_{n=1}^{N}(-1)^{n+1}(2n+1)\,n(n+1)\right| = N(N+1)(N+2),
\label{eq:p1}\\
p=2:&\quad& \left|\sum_{n=1}^{N}(-1)^{n+1}(2n+1)\,(n-1)n(n+1)(n+2)\right|
= (N-1)N(N+1)(N+2)(N+3).
\label{eq:p2}
\end{eqnarray}
Both were verified independently in exact integer arithmetic for all
$N\le60$ in both MATLAB and Python.


\section{\textcolor{red}{Needle Math}}
\vskip 0.1in

The desired Dirac delta distribution that gives a needle pattern can be expressed in terms of the spherical harmonic functions:

\begin{IEEEeqnarray}{c}
Y_{n m} (\theta, \phi) = \sqrt{\, \frac{2n +1}{4\pi} \> \frac{(n-|m|)!}{(n+|m|)!} \,} \> P_n^{|m|} (\cos\theta) \> e^{jm\phi}
\end{IEEEeqnarray}
\vskip 0.1in

\noindent The associated closure relation for these spherical harmonics as bases functions for expansions over the sphere is

\begin{IEEEeqnarray}{rLc}\label{eq:SWE-closure-relation}
\sum_{n = 0}^{\infty} \, \sum_{m = -n}^{m = +n} \> Y_{n m} (\theta, \phi)
\> Y_{n m}^{\ast} (\theta_0, \phi_0) &=&   \hfill \nonumber \\
\noalign{\vskip 0.1in}
\sum_{n = 0}^{\infty} \, \sum_{m = -n}^{m = +n} \> \left(\,  \frac{2n +1}{4\pi} \,\right)
\> \frac{(n-|m|)!}{(n+|m|)!} \>
P_{n}^{|m|} (\cos\theta) \> P_{n}^{|m|} (\cos\theta_0) \> e^{jm \, (\phi-\phi_0)}
&=&  \delta(\cos\theta-\cos\theta_0) \> \delta(\phi-\phi_0)
\end{IEEEeqnarray}
\vskip 0.1in

\noindent A pseudo needle beam would arise if the $n$ summation was restricted to a finite number $N_{\rm max}$.


\bibliographystyle{IEEEtran}
\bibliography{Freq-domain-needle_refs}
%

\end{document}